\documentclass[aps,pra,floatfix,twocolumn,superscriptaddress]{revtex4-2}

\usepackage{mathrsfs}
\usepackage{graphicx}
\usepackage[english]{babel}
\usepackage{amsmath}
\usepackage{amssymb}
\usepackage{xcolor}
\usepackage{cancel}
\usepackage[normalem]{ulem}

\usepackage{relsize}
\usepackage{CJK}
\usepackage{dsfont}
\usepackage{bm}% bold math

\newcommand{\me}{\mathrm{e}}
\newcommand{\mi}{\mathrm{i}}

\newcommand{\dif}{\mathrm{d}}

\newcommand\px{\mathrel{/\mkern-5mu/}}
\allowdisplaybreaks

\begin{document}

\title{Geometry effect of the dynamical quantum phase transitions at finite temperatures}

\author{Jia-Chen Tang}
\affiliation{School of Physics, Southeast University, Jiulonghu Campus, Nanjing 211189, China}

\author{Xu-Yang Hou}
\email{xuyanghouwow@seu.edu.cn}
\affiliation{School of Physics, Southeast University, Jiulonghu Campus, Nanjing 211189, China}

\author{Hao Guo}
\email{guohao.ph@seu.edu.cn}
\affiliation{School of Physics, Southeast University, Jiulonghu Campus, Nanjing 211189, China}
\affiliation{Hefei National Laboratory, Hefei 230088, China}

\begin{abstract}
Dynamical quantum phase transitions (DQPTs) probe the nonequilibrium evolution of quantum systems, unveiling their geometric and topological characteristics. In this study, we introduce the concepts of parallel quench and dynamic geometrical order parameter (DGOP) for non-band models, where these quantities capture the geometric shifts associated with DQPTs. At zero temperature, the DGOP corresponds to the Pancharatnam geometric phase, while at finite temperatures, it extends to the interferometric geometric phase. We further generalize the dynamic topological order parameter (DTOP) to finite-temperature band models, examining its behavior in the Su-Schrieffer-Heeger (SSH) model. Our analysis shows that thermal fluctuations and boundary effects at finite temperatures disrupt the quantization of the DTOP, yet it retains signatures of topological transitions. These findings deepen the understanding of geometric and topological properties in quantum dynamics, illuminating DQPTs across both non-band and band frameworks.
\end{abstract}

\maketitle

\section{Introduction}\label{Sec1}
Recent advancements in the study of ultra-cold atoms in optical lattices have significantly deepened our understanding of nonequilibrium dynamics in isolated quantum systems \cite{DQPTB41,DQPTB4,Zhang_2017,PhysRevApplied.11.044080,PhysRevLett.128.160504}. A key focus in this field is the time evolution of a quantum system following a sudden global quench, a process that can be experimentally realized and theoretically analyzed. The Loschmidt echo, reflecting the overlap between the pre- and post-quench Hamiltonian eigenstates, plays a crucial role in this evolution \cite{Zvyagin16,DQPT13,DQPTreview18}. The formal analogy between the Loschmidt amplitude and the canonical partition function has led to the concept of dynamical quantum phase transitions (DQPTs), which provide a framework for understanding phase transitions in nonequilibrium quantum systems.

The significance of DQPTs lies in their direct connection to observable behavior of quantum many-body systems during quench dynamics \cite{DQPT14,DQPT15,DQPTreview18}.
Specifically, the DQPT occurs at the zeros of  the Loschmidt echo, $\mathcal{L}(t)=|\mathcal{G}(t)|^2$, throughout the nonequilibrium evolution of a quantum system. Here, the Loschmidt amplitude $\mathcal{G}(t)$ quantifies the overlap between the time-evolving state and the initial state. These zeros, known as Fisher zeros \cite{PhysRevLett.118.180601}, reflect the time-evolved critical behaviors of quantum systems, and are thus referred to as critical times~\cite{DQPTreview18}. Additionally, DQPTs can be more clearly illustrated through the concept of rate function, which serves as the ``dynamical analogue'' of free-energy. Its time derivative exhibits cusp-like singularities at critical times, providing a clearer insight into DQPTs.
Another approach to defining DQPTs focuses on the asymptotic late-time behavior of order parameters \cite{Yuzbashyan200696, Barmettler2009102, Eckstein2009103, Sciolla2010105,
 Dziarmaga201059}.
Extensive theoretical research has been conducted on this topic, particularly in systems like XY chains \cite{Vajna201489, PhysRevE.93.052133, Cao202231, Porta.10.1}, Kitaev honeycomb models \cite{Schmitt201592}, non-integrable models \cite{Karrasch201387, Andraschko201489, Heyl2014113, Kriel.90.125106, Sharma.92.104306}, systems with long-range interactions \cite{Halimeh201796, Homrighausen201796,PhysRevB.95.174305, PhysRevE.96.062118, Dutta201796, Bojan2018120, Halimeh.2.033111}, quantum Potts models \cite{PhysRevB.95.075143}, non-Hermitian systems \cite{Zhou201898, Mondal.2022.106, Mondal.2023.107}, Bose-Einstein condensates \cite{Mehdi2019100}, inhomogeneous systems \cite{Yang201796,PhysRevA.97.033624,Mendl2019100,Cao2020102,Modak2021103,Kuliashov.107.094304, Mishra.53.375301}, periodically driven systems \cite{Yang2019100,Zamani2020102, Shirai.101.013809, Zhou202133,Jafari2021103,NC.12.5108,PhysRevB.105.094304,Jafari2022105}, systems in mixed states \cite{Bhattacharya.96.180303, Heyl.96.180304, Lang.98.134310, Bandyopadhyay.8, Hou.102.104305, Kyaw.101.012111, Mera.97.094110, Sedlmayr.97.045147, Link.125.143602, Hou.104.023303},  and others \cite{Heyl.115.140602, Vajna.91.155127, Tatjana.1.1.003, Lang.121.130603, Huang.122.250401, Jafari.99.054302, Khatun.123.160603, Lahiri.99.174311, Liu.99.104307, Srivastav.100.144203, Gulacsi.101.205135, Meibohm_2023.023034, Wong.105.174307, Wrzessniewski.105.094514, Hashizume.4.013250}. Additionally, direct measurements of the nonanalytic behavior in the rate function have been performed using a simulator of the interacting transverse-field Ising model in Ref. \cite{DQPTB41} and in topological nanomechanical systems in Ref.~\cite{Tian19}. Recent demonstrations of DQPTs and their implications have also been observed in various systems, including correlation functions~\cite{DQPTN17a,DQPTN17b}, superconducting qubits~\cite{GuoApplied19}, photonic platforms~\cite{WangPRL19}, NV centers in diamonds~\cite{ChenAPL20}, spinor condensates~\cite{Yang19, Duan20}, and nuclear magnetic resonance quantum simulators~\cite{Nie20}.

Recently, it has been shown that DQPTs can reveal the topological features of real-time quench dynamics, particularly in band models \cite{Budich201693}. In this context, a bulk topological quantum number, the dynamic topological order parameter (DTOP), was introduced. It is defined as the momentum-space winding number of the Pancharatnam geometric phase (PGP) and serves to characterize these topological properties. This raises a compelling question: Can analogous quantities describe the geometric or topological characteristics of non-band models? In this paper, we demonstrate that the PGP itself \cite{Pancharatnam56} can serve this role, exhibiting discrete jumps at DQPTs in certain non-band systems, provided the parallel transport condition is satisfied after a quench. Remarkably, this finding extends to finite temperatures, where the PGP is replaced by the interferometric geometric phase (IGP) \cite{PhysRevLett.85.2845}, contingent on fulfilling the parallel transport condition proposed by Sj\"oqvist et al. As the PGP and IGP capture the geometric features of pure and mixed states at DQPTs, respectively, we designate them the ``dynamic geometrical order parameter" (DGOP) and term the associated quench a ``parallel quench." 
For mixed states, an alternative geometric phase, the Uhlmann phase \cite{Uhlmann86}, emerges from a distinct parallel-transport mechanism, known as the Uhlmann process, which differs from the parallel transport discussed here. Intriguingly, the Uhlmann parallel transport condition can also be integrated into dynamical quench frameworks, as explored in our prior work \cite{PhysRevB.110.134319}. We briefly explain the difference between Uhlmann phase and IGP at the last of Sec.(\ref{QIGPMS})

For band models, we extend the DTOP to finite-temperature scenarios, revealing that thermal fluctuations and boundary effects disrupt its quantization, yet it retains signatures of topological transitions, as exemplified in the Su-Schrieffer-Heeger (SSH) model. These results highlight the interplay between geometric and topological properties in quantum dynamics across both non-band and band systems.

The remainder of the paper is structured as follows. In Sec. \ref{Sec2}, we review the theoretical framework of DQPTs for pure and mixed quantum states. In Sec. \ref{Sec3}, we develop the theory of parallel quench for these two cases. Sec. \ref{Sec4} applies this formalism to non-band models, analyzing two-level and spin-$j$ systems under parallel quenches, and detailing the behavior of the geometric phase and rate function. Sec. \ref{Sec5} extends the analysis to band models, focusing on the DTOP's properties at finite temperatures in the SSH model. Sec. \ref{Sec6} presents possible experimental implication. Finally, Sec. \ref{Sec7} summarizes our conclusions.

\section{Overview of DQPTs of pure and mixed states}\label{Sec2}
For simplicity, we set $c=\hbar=k_B=1$ in the following. Consider a quantum system initially prepared in the ground state $|\psi_0\rangle $ of the Hamiltonian $H_0$. At $t=0^+$, a sudden quench is implemented, changing the Hamiltonian to  $H\neq H_0$. In general, $H_0$ and $H$  do not share common eigenstates. After the quench, the system evolves according to $|\psi(t)\rangle=\me^{-\mi Ht}|\psi_0\rangle$. DQPTs occur whenever the post-quench state $|\psi(t)\rangle$
becomes orthogonal to the pre-quench state $|\psi_0\rangle$. In other words, DQPTs are identified by the zeros $t^*_n$ of the Loschmidt (or return) amplitude~\cite{DQPT13,DQPTreview18}
\begin{align}\label{LA}
\mathcal{G}(t)=\langle\psi(0)|\psi(t)\rangle=\langle\psi(0)|\me^{-\mi Ht}|\psi(0)\rangle,
\end{align}
which measures the deviation of the time-evolved state from the initial condition. Since an ordinary dynamical quench process is not an equilibrium process, rather than the conventional
thermodynamic free-energy, the rate function is introduced as
\begin{align}\label{RF}
r(t)=\lim_{L\rightarrow +\infty}\frac{1}{L}\ln\mathcal{L}(t)
\end{align}
where $L$ is the overall degrees of freedom. Accordingly, $\mathcal{L}(t)$ plays the role of the partition function in thermodynamics~\cite{DQPTreview18}.

The formalism of DQPTs can be directly generalized to finite-temperature scenarios through the pure-state like representation of mixed states \cite{PhysRevB.106.014301}.
For a full-rank density matrix $\rho$ of rank $N$, if it is diagonalized as $\rho=\sum_{n=0}^{N-1}\lambda_n|n\rangle\langle n|$, its purification is given by $W=\sqrt{\rho}U=\sum_{n=0}^{N-1}\sqrt{\lambda_n}|n\rangle\langle n|U$, where $U\in \text{U}(N)$ \cite{Uhlmann86}. Conversely, $\rho=WW^\dag$, indicating that
$W$ serves a role analogous to that of a state vector. This relationship becomes more apparent when $W$ is expressed in the form $|W\rangle=\sum_{n=0}^{N-1}\sqrt{\lambda_n}|n\rangle\otimes U^T|n\rangle$, which is referred to as the purified state representation of $W$. The inner product between two purified states follows the Hilbert-Schmidt product $\langle W_1|W_2\rangle=\text{Tr}(W^\dagger_1 W_2)$.
In this context, the concept of the Loschmidt amplitude can be generalized as follows. If the initial mixed state of a quantum system is represented by $\rho(0)=W(0)W^\dag(0)$, the density matrix evolves according to $\rho(t)=\me^{-\mi Ht}\rho(0)\me^{\mi Ht}$ after a quench governed by $H$. This results in $W(t)=\me^{-\mi Ht}W(0)$. Thus, the Loschmidt amplitude can be obtained by generalizing Eq. (\ref{LA}) to
 \begin{align}\label{Gm}
 \mathcal{G}_\rho(t)
 &=\langle W(0)|W(t)\rangle=\text{Tr}\left[W^\dag(0)W(t)\right]\notag\\
 &=\text{Tr}\left[\rho(0)\me^{-\mi Ht}\right].
\end{align}
Equivalent results can be found in Ref.~\cite{PhysRevB.96.180304,Hou.102.104305}. This formalism generalizes the inner product of quantum states $\langle \psi (0) | \psi (t) \rangle$ to the trace operation. If $\rho$ is the density matrix of a pure state, where $\rho  = | {\psi (0)}\rangle \langle {\psi (0)}|$, this generalized expression exactly recovers the original formulation given in Eq.({\ref{LA}}).

\section{Theoretical Formalism}\label{Sec3}
\subsection{Pure states}

Previous discussion indicates that DQPTs occur when $|\psi(t)\rangle$ is perpendicular to the initial state. In contrast to this, another relation is the parallelity between states \cite{Uhlmann86}: If $|\psi_{1,2}\rangle$ satisfies
  \begin{align}\label{pofstate}
\langle \psi_1|\psi_2\rangle=\langle \psi_2|\psi_1\rangle>0,\end{align} then the two states are said to be parallel to each other. The well-known Berry phase can be considered as a measure of the loss of parallelism during the periodic evolution process.

Consider the dynamical evolution of state after a sudden quench: $|\psi(t)\rangle=\me^{-\mi Ht}|\psi_0\rangle$. If $|\psi(t+\dif t)\rangle$ remains parallel to $|\psi(t)\rangle$ throughout the evolution, the state is said to undergo a parallel transport, and we refer to this type of quench as a parallel quench. Expanding $\langle \psi(t)|\psi(t+\dif t)\rangle>0$ to the first order, and noting that $\langle\psi(t)|\frac{\dif}{\dif t} |\psi(t)\rangle$ is purely imaginary, we obtain the parallel-transport condition
  \begin{align}
 \langle\psi(t)|\frac{\dif}{\dif t} |\psi(t)\rangle=0.
\end{align}
It further leads to
  \begin{align}\label{p1}
 \langle \psi_0|H|\psi_0\rangle=0.
\end{align}
We emphasize that this condition holds true only for quench dynamics. In contrast, for ordinary dynamical evolutions with $H=H_0$, this condition breaks down since $\langle \psi_0|H|\psi_0\rangle=E_0$ is generically non-zero.
Under this condition, the dynamical phase accumulated during the post-quench evolution vanishes:
  \begin{align}
\theta_\text{d}(t)= -\int_0^t\dif t'\langle \psi(t')|H|\psi(t')\rangle =-\langle \psi_0|H|\psi_0\rangle t=0.
\end{align}
This indicates that the argument of the Loschmidt echo is the PGP \cite{MUKUNDA1993205}:
  \begin{align}
\theta_\text{g}(t)=\arg\langle \psi(0)|\psi(t)\rangle-\theta_\text{d}(t)=\arg\langle \psi_0|\me^{-\mi Ht}|\psi_0\rangle.
\end{align}
Interestingly, although this evolution is purely dynamic, only geometric phase is generated after the quench.
In fact, $\theta_\text{g}(t)$ is the negative argument of the Bargmann invariant \cite{MUKUNDA1993205}, which is invariant under the gauge transformation $|\psi(t)\rangle\rightarrow \me^{\mi\chi(t)}|\psi(t)\rangle$.
When DQPTs occur at $t^*_n$ with $\mathcal{G}(t^*_n)=0$, the value of $\theta_\text{g}(t)$ must undergo a discrete jump. This indicates that the geometric properties of the system change as it crosses $t^*_n$.

\subsection{Mixed states}\label{QIGPMS}

The concept of parallel quench can be generalized to mixed states  by employing the parallel-transport condition proposed by Sj\"oqvist et al. \cite{PhysRevLett.85.2845}. Suppose the initial state is described by $\rho(0)$. At $t=0^+$, a sudden quench is implemented, and the Hamiltionian of the system is changed to $H$ such that $[\rho(0),H]\neq 0$. Following this, the density matrix evolves as $ \rho(t)=\me^{-\mi Ht}\rho(0)\me^{\mi Ht}$, which is a unitary evolution. Let the eigenstates of $\rho(0)$ and $\rho(t)$ be $|n\rangle$ and $|n(t)\rangle$, respectively, where $n=0,1,\cdots, N-1$.
By comparing with the optical process in a Mach-Zehnder interferometer, Sj\"oqvist et al suggested that a mixed state acquires the phase 
\begin{align}\label{tht}
\theta(t)=\arg\text{Tr}\left[\rho(0)U(t)\right]
\end{align} 
during a unitary evolution $ \rho(t)=U(t)\rho(0)U^\dag(t)$. 

To ensure that $\rho(t+\dif t)$ is ``in phase'' with $\rho(t)$, the following condition must be satisfied: \cite{PhysRevLett.85.2845}
\begin{align}\label{pfi4}
 \text{Tr}\left[\rho(t)\dot{U}(t)U^\dagger (t)\right]=0,
\end{align}

Following Ref.\cite{PhysRevLett.85.2845}, the dynamical phase accumulated during this evolution is
\begin{align}\label{dfi}
\theta_{\text{d}}(t)=-\int_0^t\dif t' \text{Tr}\left[\rho(t')H\right].
\end{align}
Now consider the quench dynamics, and let $U(t)=\me^{-\mi Ht}$. Using $H=\mi \dot{U}U^\dagger$, the dynamical phase becomes
\begin{align}\label{dfi2}
\theta_{\text{d}}(t)&=-\int_0^t\dif t' \left[\rho(0)U^\dagger(t')\dot{U}(t')\right]\notag\\
&=-\int_0^t\dif t'\text{Tr}\left[\rho(t')\dot{U}(t')U^\dagger(t')\right].
\end{align}
Thus, if the post-quench dynamic evolution satisfies the parallel-transport condition (\ref{pfi4}), no dynamical phase is accumulated, similar to  its pure-state counterpart. In this situation, the argument of the Loschmidt amplitude is the mixed-state geometric phase $\theta_\mathrm{g}(t)$, which is called the interferometric geometric phase (IGP) \cite{PhysRevLett.85.2845,andersson2016geometric}
\begin{align}\label{thg}
\theta_\text{g}(t)&=\arg\mathcal{G}_\rho(t)-\theta_{\text{d}}(t)=\arg\text{Tr}\left[\rho(0)U(t)\right]\notag\\&=\arg\text{Tr}\left[\rho(0)\me^{-\mi Ht}\right],
\end{align}
where Eqs.(\ref{tht}) and (\ref{Gm}) have been applied. In non-band systems, the IGP itself can serve as the dynamical geometric order parameter (DGOP) during parallel quenches.

To more precisely specify the generic evolution $U(t)$, the condition (\ref{pfi4}) can be strengthened as \cite{PhysRevLett.85.2845}
\begin{align}\label{pfi6}
\langle n(t)|\dot{U}(t)U^\dagger(t)|n(t)\rangle=0,\quad n=0,1,\cdots, N-1,
\end{align}
or equivalently
\begin{align}\label{pfi6b}
\langle n|U^\dagger(t)\dot{U}(t)|n\rangle=0,\quad n=0,1,\cdots, N-1
\end{align}
by noting $|n(t)\rangle=\me^{-\mi Ht}|n\rangle$. If a post-quench dynamic evolution satisfies this condition, we also refer to it as the parallel quench. Using $H=\mi U^\dagger  \dot{U}$, the condition (\ref{pfi6b}) becomes
\begin{align}\label{pfi6bc}
\langle n|H|n\rangle=0,\quad n=0,1,\cdots, N-1.
\end{align}
This is in fact a generalization to Eq.(\ref{p1}). If the initial state is prepared in a thermal equilibrium at temperature $T$, then $\rho(0)=\frac{1}{Z}\me^{-\beta H_0}$, where $\beta=\frac{1}{T}$ and $H_0$ is the initial Hamiltonian. Clearly, $\rho(0)$ shares the same eigenstates with $H_0$, which implies that $|0\rangle=|\psi_0\rangle$ is also the ground state of $H_0$. Thus, the condition (\ref{pfi6bc}) covers Eq.(\ref{p1}). In Sec.\ref{Sec4}, we will investigate the case where the initial state is in thermal equilibrium. By taking $T\rightarrow 0$, we can derive the results applicable to pure states.

Another construction of the mixed-state geometric phase, the Uhlmann phase, is essentially different from the IGP.  First, their mathematical foundations diverge: the Uhlmann phase necessarily employs mixed state purification through auxiliary systems, while the IGP originates from quantum interference in operational protocols without requiring such extensions. Second, their geometric constraints exhibit dimensional disparity—the IGP's parallel transport condition operates solely within the system's Hilbert space, whereas Uhlmann's framework  mandates joint evolution in the composite space. Third, while the Uhlmann phase is strictly confined to cyclic processes, the IGP applies universally to arbitrary unitary evolutions. Notably, our recent work reconciled Uhlmann's parallel-transport condition with Hamiltonian dynamics, thereby developing a geometric quantum quench distinct from the one discussed here.\cite{PhysRevB.110.134319}

\subsection{Band models and DTOP}
The preceding discussion is well-suited for non-band models. However, in the case of band models, the condition (\ref{pfi6bc}) generally fails to hold. This arises because both the Hamiltonian $H$ and the state $|n\rangle$ depend on the momentum $k$. As a result, $\langle n(k)|H(k)|n(k)\rangle$ becomes a continuous function of $k$, introducing an additional degree of freedom, and there is no a priori reason for it to vanish. To address this limitation, we will generalize the concept of DTOP to mixed states and investigate whether it can still effectively characterize the topological or geometrical properties of the system's quench dynamics at finite temperatures. For a band model, the Loschmidt amplitude $\mathcal{G}_k(t)$ also depends on $k$. The PGP is then generalized to $\Phi_k^\text{G}(t)=\Phi_k(t)-\Phi^\text{d}_k(t)$ where $\Phi_k(t)$ is the argument of $\mathcal{G}_k(t)$, and the dynamical phase is
 \begin{align}\label{dpms}
\Phi^\text{d}_{k}(t)=-\mathlarger{\int}_0^t\dif t'\text{Tr}\left[\rho_{k}(t')H(k)\right].
\end{align}
The DTOP is defined as
 \begin{align}\label{DTOP}
 \nu(t)=\frac{1}{2\pi}\int_\text{EBZ}\frac{\partial\Phi^\text{G}_k(t)}{\partial k}\dif k,
\end{align}
where the integration is taken over the effetive Brillouin zone (EBZ), provided the system respects certain symmetries. Physically, $\nu(t)$ corresponds to the winding number associated with the PGP. However, it is not necessarily quantized in all systems, such as the quantum Ising chain initiated from BCS-like ground states.

\section{Non-band models}\label{Sec4}

\subsection{Two-level systems}
Let's begin with non-band models undergoing parallel-quench processes. We first examine simple two-level systems.
 Suppose the initial Hamiltonian is parameterized as
\begin{align}\label{H2a}
H_0=\mathbf{R}_0\cdot\boldsymbol{\sigma}=R\begin{pmatrix}\cos\theta_0 & \sin\theta_0\me^{-\mi\phi_0} \\ \sin\theta_0\me^{\mi\phi_0} & -\cos\theta_0
 \end{pmatrix},
\end{align}
where $\mathbf{R}_0=R(\sin\theta_0\cos\phi_0,\sin\theta_0\sin\phi_0,\cos\theta_0)^T$, and $\boldsymbol{\sigma}=(\sigma_x,\sigma_y,\sigma_z)$ are pauli matrices. The two eigenstates are: $|+R\rangle=\left(\begin{array}{c}\cos\frac{\theta_0}{2}\\ \sin\frac{\theta_0}{2}\me^{\mi\phi_0}
 \end{array}\right)$ and $|-R\rangle=\left(\begin{array}{c}
\sin\frac{\theta_0}{2}\\ -\cos\frac{\theta_0}{2}\me^{\mi\phi_0}
 \end{array}\right)$.
At $t=0^+$, a sudden quench is implemented such that the parameter is changed to $\mathbf{R}=R(\sin\theta\cos\phi,\sin\theta\sin\phi,\cos\theta)^T$. Using Eq.(\ref{pfi6bc}), a straightforward evaluation yields:
\begin{align}\label{pcrc}
\langle \pm R|H|\pm R\rangle&=\pm R\left[\cos\theta_0\cos\theta+\sin\theta_0\sin\theta\cos(\phi-\phi_0)\right]\notag\\
&=\pm R\hat{\mathbf{R}}\cdot \hat{\mathbf{R}}_0,
\end{align}
where $\hat{\mathbf{R}}_0=\mathbf{R}_0/R$ and $\hat{\mathbf{R}} =\mathbf{R} /R$.
Therefore, the condition for a parallel quench is $\hat{\mathbf{R}}\cdot \hat{\mathbf{R}}_0=0$. Obviously, this is also true for the pure-state scenarios according to Eq.(\ref{p1}).
Let the system be initially prepared in thermal equilibrium at temperature $T$, with the corresponding density matrix given by \begin{align}
\rho(0)=\frac{\me^{-\beta H_0}}{Z}=\frac{1}{2}\left(\mathbf{1}-\tanh(\beta R)\hat{\mathbf{R}}_0\cdot\boldsymbol{\sigma}\right),
\end{align}
where the partition function $Z$ is invariant under unitary transformations. The post-quench dynamic evolutional operator is
\begin{align}
\me^{-\mi Ht}=\cos(\omega t)\mathbf{1}-\mi\sin(\omega t)\hat{\mathbf{R}}\cdot\boldsymbol{\sigma},
\end{align}
where $\omega=\frac{R}{\hbar}\equiv R$.
Thus, Eq.(\ref{Gm}) leads to
\begin{align}\label{LA2}
\mathcal{G}_\rho(t)=\cos(\omega t)+\mi\sin(\omega t)\tanh(\beta R)\hat{\mathbf{R}}_0\cdot\hat{\mathbf{R}}.
\end{align}
DQPTs occur at $t^*_n=\frac{\left(n+\frac{1}{2}\right)\pi}{\omega}$ with $n$ being a nonnegative integer if $\hat{\mathbf{R}}\cdot \hat{\mathbf{R}}_0=0$. Interestingly, the latter exactly coincides with the parallel-quench condition. Therefore, the argument of $\mathcal{G}_\rho(t)$ is the PGP, which experiences a discrete jump at $t^*_n$.

\subsection{Spin-$j$ systems}

Before presenting our numerical results, we demonstrate that the results of two-level systems can be included into the more general spin-$j$ systems.
Consider an ensemble of spin-$j$ paramagnets influenced by an external magnetic field $\mathbf{B}$. The corresponding Hamiltonian is given by $H=\omega_0\hat{\mathbf{B}}\cdot \mathbf{J}$, where $\omega_0$ is the Larmor frequency, $\hat{\mathbf{B}}=\mathbf{B}/B$ with $B=|\mathbf{B}|$ and $\mathbf{J}$ is the spin angular momentum. Suppose the initial magnetic field is $\mathbf{B}_0=B(\sin\theta_0\cos\phi_0,\sin\theta_0\sin\phi_0,\cos\theta_0)^T$. At $t=0^+$, impose a sudden quench such that $\mathbf{B}_0$ is rotated to $\mathbf{B}=B(\sin\theta\cos\phi,\sin\theta\sin\phi,\cos\theta)^T$. The initial Hamiltonian is
\begin{align}\label{sjH1}
H_0
&=\omega_0\me^{-\mi\phi_0 J_z}\me^{-\mi\theta_0 J_y}J_z\me^{\mi\theta_0 J_y}\me^{\mi\phi_0 J_z}\notag\\
&=\omega_0\mathcal{U}(\theta_0,\phi_0)J_z\mathcal{U}^\dag(\theta_0,\phi_0),
\end{align}
where $\mathcal{U}(\theta_0,\phi_0)=\me^{-\mi\phi_0 J_z}\me^{-\mi\theta_0 J_y}$ is a unitary operator. The eigen-levels of $H_0$ can be constructed by the eigen-levels $|jm\rangle$ of $J_z$:
\begin{align}\label{el0}
|\psi^0_m\rangle=\me^{-\mi\phi_0 (J_z-m)}\me^{-\mi\theta_0 J_y}|jm\rangle
\end{align}
for $m=-j,-j+1,\cdots, j-1,j$. Similarly, the post-quench Hamiltonian is given by: $H=\omega_0\hat{\mathbf{B}}\cdot \mathbf{J}=\omega_0\mathcal{U}(\theta ,\phi)J_z\mathcal{U}^\dag(\theta ,\phi)$. For a parallel-quench, the following condition must be satisfied
\begin{align}\label{el0b}
\langle\psi^0_m|H |\psi^0_m\rangle=0,\quad m=-j,-j+1,\cdots, j-1,j.
\end{align}
Plugging in Eq.(\ref{el0}) and using Eqs.(\ref{het1})-(\ref{het1a}), this condition becomes
\begin{widetext}
\begin{align}\label{pcsj1}
0=&\omega_0\langle jm|\me^{\mi\theta_0 J_y}\me^{-\mi(\phi-\phi_0)J_z}\left(J_x\sin\theta+J_z\cos\theta\right)\me^{\mi(\phi-\phi_0)J_z}\me^{-\mi\theta_0 J_y}|jm\rangle\notag\\
=&\omega_0\langle jm|\me^{\mi\theta_0 J_y}\left[J_x\sin\theta\cos(\phi-\phi_0)+J_y\sin\theta\sin(\phi-\phi_0)+J_z\cos\theta\right]\me^{-\mi\theta_0 J_y}|jm\rangle\notag\\
=&m\omega_0\left[\cos\theta\cos\theta_0+\sin\theta\sin\theta_0\cos(\phi-\phi_0)\right],
\end{align}
\end{widetext}
where we have applied the facts that $\langle jm|J_x|jm\rangle=\langle jm|J_y|jm\rangle=0$ and $\langle jm|J_z|jm\rangle=m$ in the last line. Accordingly, the parallel-quench condition for spin-$j$ model implies
\begin{align}\label{pqsj}
 \hat{\mathbf{B}}_0\cdot\hat{\mathbf{B}}=0.
\end{align}
If $j=\frac{1}{2}$, or equivalently $\mathbf{J}=\frac{1}{2}\boldsymbol{\sigma}$, Eq.(\ref{pqsj}) recovers the parallel-quench condition for two-level systems.

The evaluation of the Loschmidt amplitude needs some algebraic skills. Note the initial mixed state is described by $\rho(0)=\frac{1}{Z(0)}\me^{-\beta H_0}$ with $Z(0)=\text{Tr}\me^{-\beta H_0}$. From this, we further obtain
\begin{align}\label{LAc}
\mathcal{G}_\rho(t)=\text{Tr}\left[\rho(0)\me^{-\mi Ht}\right]\rightarrow Z(0)\mathcal{G}_\rho(t)=\text{Tr}\left(\me^{-\beta H_0}\me^{-\mi Ht}\right).
\end{align}
The trace can be evaluated by means of the representation of the SL$(2,\mathds{C})$ group. Note
\begin{align}
\det\left(\me^{-\beta H_0}\me^{-\mi Ht}\right)=&\det\me^{-\beta H_0}\det\me^{-\mi Ht}\notag\\=&\me^{-\beta\text{Tr}H_0}\me^{-\mi t\text{Tr}H}=1
\end{align}
since $\text{Tr}J_x=\text{Tr}J_y=\text{Tr}J_z=0$. Therefore, $\me^{-\beta H_0}\me^{-\mi Ht}$ is in the ($j,0$)-representation of the SL$(2,\mathds{C})$ group. Basic representation theory of this group tells us that
once the eigenvalues $(\lambda_+,\lambda_-)$ of the $(\frac{1}{2}, 0)$ representation
are known, the eigenvalues for higher $j$ are given by \cite{Jeevanjeebook}
\begin{align}
\lambda^{2j},\lambda^{2j-2},\cdots,\lambda^{-2(j-1)},\lambda^{-2j}
\end{align}
with $\lambda=\lambda_+=\lambda_-^{-1}$. Thus,
\begin{align}\label{G3c}
Z(0)\mathcal{G}(T,t)&=\lambda^{2j}+\lambda^{2j-2}+\cdots+\lambda^{-2j+2}+\lambda^{-2j}\notag\\&=\frac{\lambda^{2j+1}-\lambda^{-2j-1}}{\lambda-\lambda^{-1}}
\end{align}
with
\begin{align}
Z(0)=&\me^{\beta j  \omega_0}+\me^{\beta(j-1)  \omega_0}+\cdots+\me^{-\beta(j-1)  \omega_0}+\me^{-\beta j  \omega_0}\notag\\=&\frac{\sinh\frac{\beta (2j+1)  \omega_0}{2}}{\sinh\frac{\beta  \omega_0}{2}}.
\end{align}

To determine $\lambda_\pm$, we first need to find the results for $j=\frac{1}{2}$ or the $(\frac{1}{2}, 0)$ representation. Using $\mathbf{J}=\frac{1}{2}\boldsymbol{\sigma}$, the initial density matrix and the associated time evolutional operator are respectively given by
\begin{align}\label{sjtmp1}
\rho_{\frac{1}{2}}(0)&=\frac{1}{Z_{\frac{1}{2}}(0)}\me^{-\beta H_0}=\frac{1}{2}\left(\mathbf{1}-\tanh\frac{\beta\omega_0}{2}\hat{\mathbf{B}}_0\cdot\boldsymbol{\sigma}\right),\notag\\
\me^{-\mi Ht}&=\me^{-\mi \frac{\omega_0 t}{2}\hat{\mathbf{B}}\cdot \boldsymbol{\sigma}}=\cos\frac{\omega_0 t}{2}\mathbf{1}-\mi\sin \frac{\omega_0 t}{2}\hat{\mathbf{B}}\cdot \boldsymbol{\sigma},
\end{align}
where $Z_{\frac{1}{2}}(0)=2\cosh\frac{\beta\omega_0}{2}$. Note $Z_{\frac{1}{2}}(0)\rho_{\frac{1}{2}}(0)\me^{-\mi Ht}$ belongs to the $(\frac{1}{2}, 0)$ representation of the SL$(2,\mathds{C})$ group. Its eigenvalues satisfy
$\lambda_+\lambda_-=1$ and
\begin{align}
\lambda_++\lambda_-=\text{Tr}\left[Z_{\frac{1}{2}}(0)\rho_{\frac{1}{2}}(0)\me^{-\mi Ht}\right].
\end{align}
Let
\begin{align}
z&=\frac{1}{2}\text{Tr}\left[Z_{\frac{1}{2}}(0)\rho_{\frac{1}{2}}(0)\me^{-\mi Ht}\right]\notag\\&=\cosh\frac{\beta\omega_0}{2}\cos\frac{\omega_0 t}{2}+\mi\sinh\frac{\beta\omega_0}{2}\sin \frac{\omega_0 t}{2}\hat{\mathbf{B}}_0\cdot \hat{\mathbf{B}}.
\end{align}
It can be found that
\begin{align}
\lambda_+=z+\sqrt{z^2-1}.
\end{align}
Using Eq.(\ref{G3c}), we finally get
\begin{align}\label{rhoeHt2}
\mathcal{G}^j_\rho(T,t)&=\frac{1}{Z(0)}\frac{\left(z+\sqrt{z^2-1}\right)^{2j+1}-\left(z-\sqrt{z^2-1}\right)^{2j+1}}{2\sqrt{z^2-1}}\notag\\
&=\frac{1}{Z_j(0)}U_{2j}(z),
\end{align}
where $U_{2j}(z)$ are the second-kind Chebyshev polynomials. The DQPTs occur at the zeros of $U_{2j}(z)$. If $j=\frac{1}{2}$, $Z_{\frac{1}{2}}(0)=2\cosh\frac{\beta\omega_0}{2}$, and Eq.(\ref{rhoeHt2}) reduces to
\begin{align}\label{rhoeHt3}
\mathcal{G}^{\frac{1}{2}}_\rho(T,t)=\frac{2z}{Z_{\frac{1}{2}}(0)}=\cos\frac{\omega_0 t}{2}+\mi\sin \frac{\omega_0 t}{2}\tanh\frac{\beta\omega_0}{2}\hat{\mathbf{B}}_0\cdot \hat{\mathbf{B}}.
\end{align}
Comparing with Eq.(\ref{LA2}), since $\omega_0=2\omega$, we recover the results of the two-level system. In this case, DQPTs occur at \begin{align}\label{1/2}t^*_n=\frac{2(n+\frac{1}{2})\pi}{\omega_0}\end{align} if $\hat{\mathbf{B}}_0\cdot \hat{\mathbf{B}}=0$. Moreover, by taking $T\rightarrow 0$ and noticing $\lim_{T\rightarrow 0}\tanh\frac{\beta\omega_0}{2}=1$, we can  derive the Loschmidt amplitude for pure states:
\begin{align}\label{psHt3}
\mathcal{G}^{\frac{1}{2}}(t)=\cos\frac{\omega_0 t}{2}+\mi\sin \frac{\omega_0 t}{2}\hat{\mathbf{B}}_0\cdot \hat{\mathbf{B}}.
\end{align}
It can be verified that this expression fully  agrees with the result obtained using Eq.(\ref{LA}).

For generic $j$, $U_{2j}(z)$ has $2j$ roots:
\begin{align}
z^*_k=\cos\frac{k\pi}{2j+1},\text{  for  } k=1,2,\cdots, 2j,
\end{align}
all of which are real. Details can be found in Appendix \ref{app2}. Note DQPTs occur when
\begin{align}
z^*_k=\cosh\frac{\beta\omega_0}{2}\cos\frac{\omega_0 t^{j*}_{n,k}}{2}+\mi\sinh\frac{\beta\omega_0}{2}\sin \frac{\omega_0 t^{j*}_{n,k}}{2}\hat{\mathbf{B}}_0\cdot \hat{\mathbf{B}},
\end{align}
which requires
\begin{align}
\hat{\mathbf{B}}_0\cdot \hat{\mathbf{B}}=0
\end{align}
and
\begin{align}\label{tj}
t^{*}_{n,k}=\frac{2}{\omega_0}\left(2n\pi+\arccos\frac{\cos\frac{k\pi}{2j+1}}{\cosh\frac{\beta\omega_0}{2}}\right),
\end{align}
for $k=1,2,\cdots, 2j$ and arbitrary nonnegative integer $n$. As before, $\hat{\mathbf{B}}_0\cdot \hat{\mathbf{B}}=0$ is also the parallel-quench condition. This means that if DQPT occurs during a post-quench dynamical evolution, the quench must be parallel, and the argument of the Loschmidt amplitude must be the geometric phase. Moreover, for each $n$, there are $2j$ sub-DQPTs according to Eq.(\ref{tj}).

\begin{figure}[th]
\centering
\includegraphics[width=3.5in]{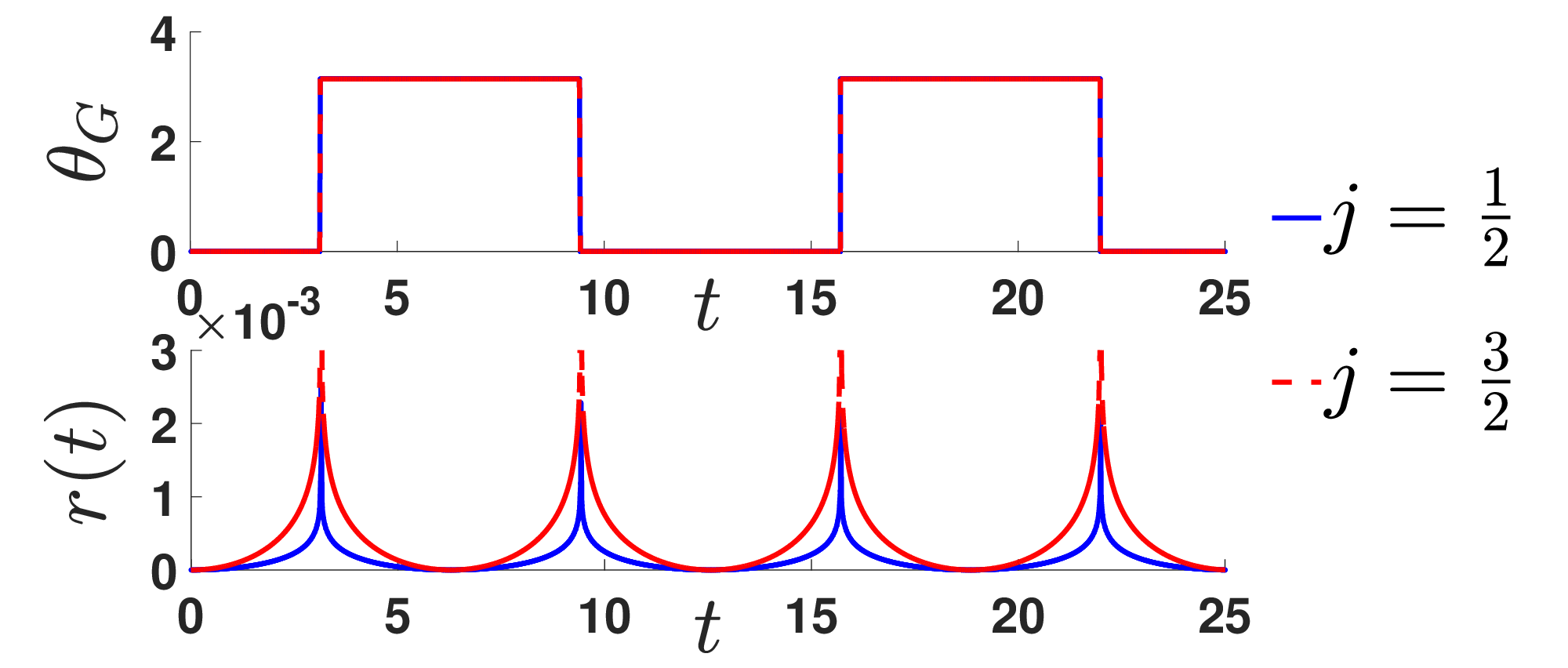}
\caption{(color online) Results for pure states: Geometric phase as a function of $t$ (Top panel), and the rate function as a function of $t$ (Bottom panel). The blue and red lines represent $j=\frac{1}{2}$ and $\frac{3}{2}$ respectively.
}
\label{Fig1}
\end{figure}

In the numerical calculations, we chose $j=\frac{1}{2}$ and $\frac{3}{2}$. In the latter case, a straightforward evaluation yields
 \begin{align}\label{rhoeHt3/2}
\mathcal{G}^{\frac{3}{2}}_\rho(T,t)=\frac{8z^3-4z}{Z_{\frac{3}{2}}(0)},
\end{align}
where $Z_{\frac{3}{2}}(0)=\frac{\sinh(2\beta\omega_0)}{\sinh\frac{\beta\omega_0}{2}}$.
In the zero temperature limit, by using $\sinh\frac{\beta\omega_0}{2}\sim \cosh\frac{\beta\omega_0}{2}\sim \frac{1}{2}\me^{\frac{\beta\omega_0}{2}}$, the Loschmidt amplitude for pure states is given by
\begin{align}\label{psHt3/2}
\mathcal{G}^{\frac{3}{2}}(t)=\left(\cos\frac{\omega_0 t}{2}+\mi\sin \frac{\omega_0 t}{2}\hat{\mathbf{B}}_0\cdot \hat{\mathbf{B}}\right)^3,
\end{align}
which is simply the cube of $\mathcal{G}^{\frac{1}{2}}(t)$ for $j=\frac{1}{2}$ according to Eq.(\ref{psHt3}).
Figure \ref{Fig1} depicts the behavior of the rate function and the corresponding geometric phase for the ground state undergoing parallel quenches as $t$ varies. The blue and red lines represent $j=\frac{1}{2}$ and $\frac{3}{2}$ respectively. It can be found that at each DQPT (the divergences of $r(t)$), the value of $\theta_G(t)$ experiences a $\pm\pi $-jump.
Based on the discussions about Eq.(\ref{pofstate}), when $\theta_G(t)=0$, the evolving state $|\psi(t)\rangle$ is parallel to the initial state. Upon crossing a DQPT point $t^*_n$, where $|\psi(t)\rangle$ becomes perpendicular to the initial state, the value of $\theta_G(t)$ suddenly shifts to $\pi$, indicating that $|\psi(t)\rangle$ is now anti-parallel to the initial state. Thus, the geometrical nature of the post-quench state changes at each $t^*_n$. We refer to $\theta_G$ as the dynamic geometrical order parameter (DGOP), which characterizes the geometric features of the post-quench dynamical evolution.

\begin{figure}[th]
\centering
\includegraphics[width=3.5in]{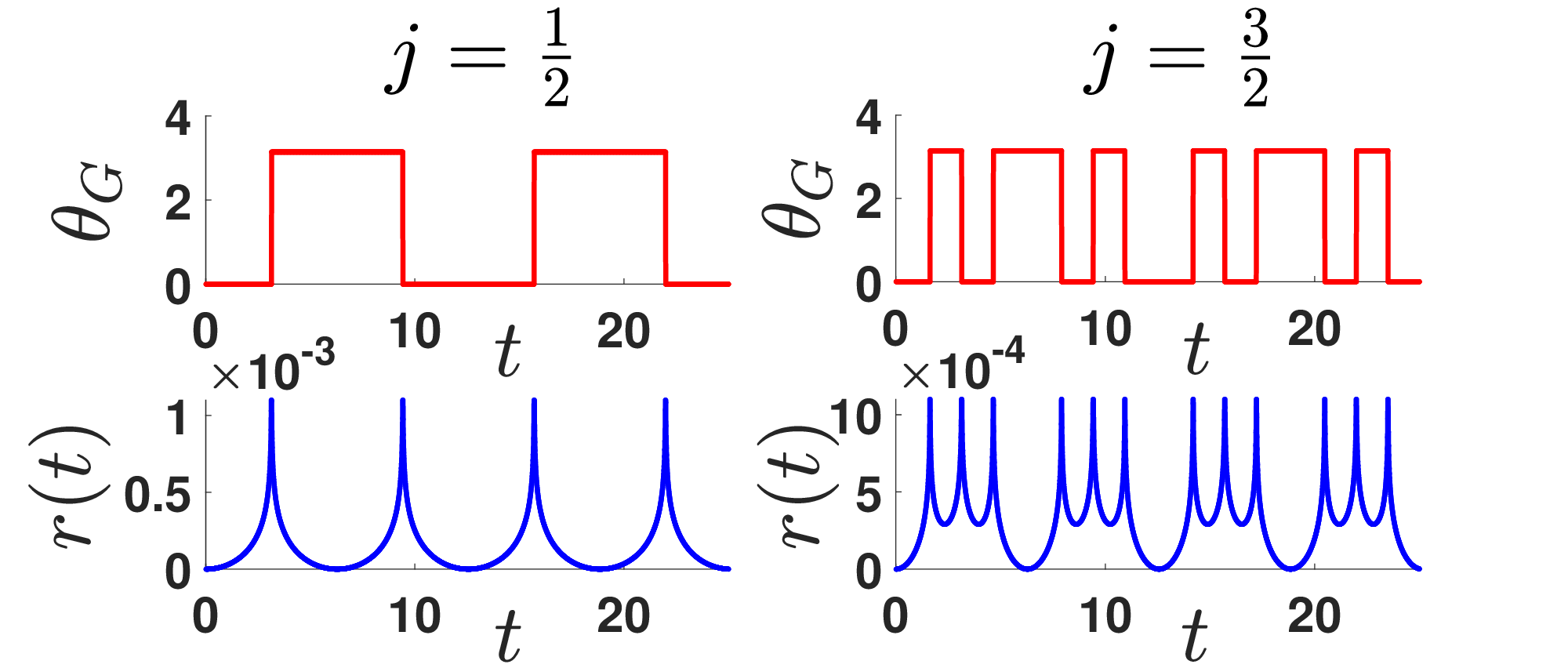}
\caption{(color online) Behaviors of $\theta_G(t)$ and $r(t)$ for mixed states during a parallel quench at $T=2.0\omega_0$. The left and right columns correspond to $j=\frac{1}{2}$ and $\frac{3}{2}$, respectively.
}
\label{Fig2}
\end{figure}

At finite temperatures, these interesting properties of $\theta_G$ are still preserved, as indicated by our previous theoretical analysis.
We visualize our numerical results in Figure \ref{Fig2}, where $\theta_G(t)$ and $r(t)$ are plotted against $t$ for $j=\frac{1}{2}$ and $\frac{3}{2}$ at $T=2.0\omega_0$. According to Eq.(\ref{tj}), at finite temperatures, each DQPT occurring at $t^*_n$ will split into $2j$ sub-DQPTs occurring at $t^{*}_{n,k}$ with $k=1,2,\cdots, 2j$.
For example, the number of divergent peaks of $r(t)$ for $j=\frac{1}{2}$ in Fig.\ref{Fig2} is the same as that in Fig.\ref{Fig1}, while the number of divergent peaks of $r(t)$ for $j=\frac{3}{2}$ is three times that in Fig.\ref{Fig1}. At each DQPT, $\theta_G$ undergoes a $\pm \pi$-jump. If $\theta_G(t)=0$, $\rho(t)$ is in phase with $\rho(0)$. Conversely, if $\theta_G(t)=\pi$, $\rho(t)$ is ``in anti-phase'' with $\rho(0)$. Thus, the geometrical nature of $\rho(t)$ changes at each $t^{*}_{n,k}$.

\begin{figure}[th]
\centering
\includegraphics[width=3.5in]{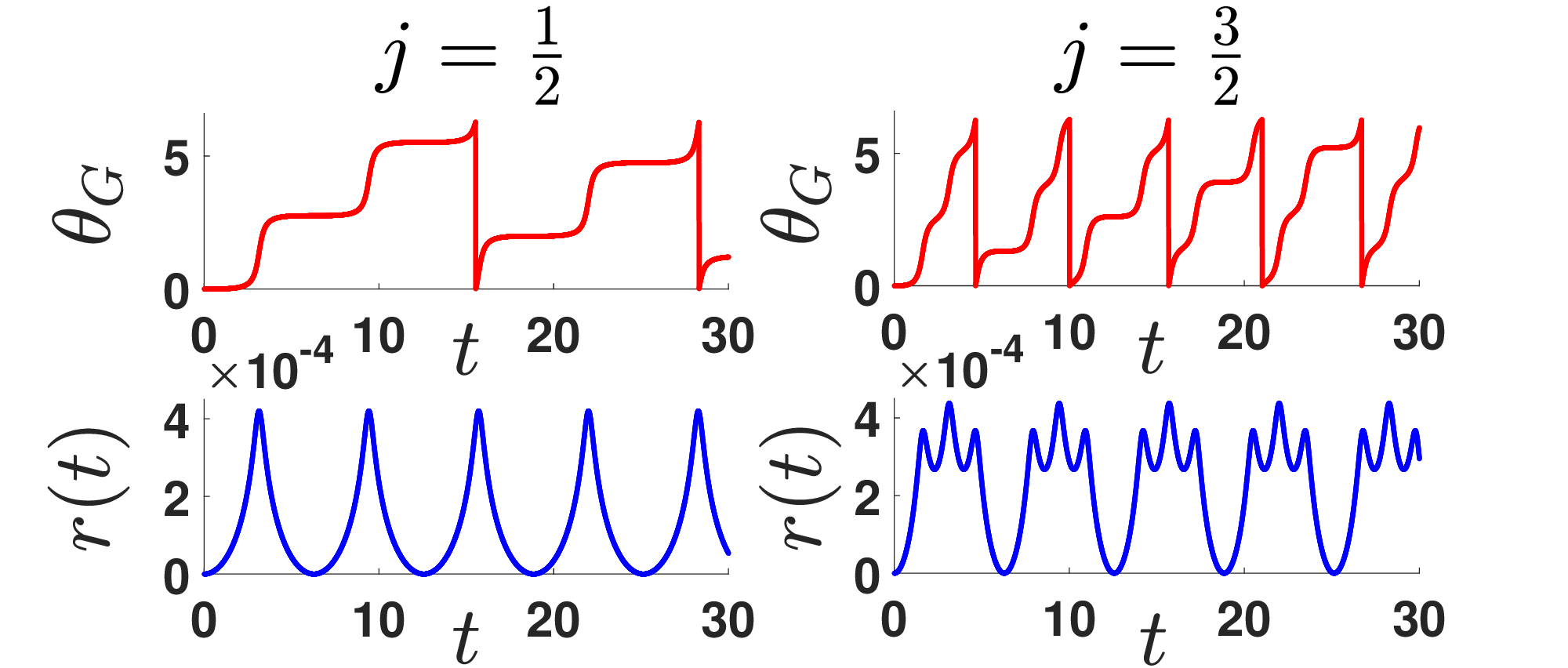}
\caption{(color online) Behaviors of $\theta_G(t)$ and $r(t)$ for mixed states during a non-parallel quench at $T=2.0\omega_0$. The left and right columns correspond to $j=\frac{1}{2}$ and $\frac{3}{2}$, respectively.
}
\label{Fig3}
\end{figure}

As a comparison, we also present numerical results for non-parallel quenches. In Figure \ref{Fig3}, we set $\hat{\mathbf{B}}_0\cdot\hat{\mathbf{B}}=0.5$ and plot both the geometric phase and the rate function as functions of time for $T=2.0\omega_0$. In this case, the dynamical phase does not vanish. Furthermore, since the condition for parallel quenches coincides with the occurrence of DQPTs, no DQPTs are observed during non-parallel quenches. As shown in the figure, the rate function exhibits no singular behavior, and as a result, the geometric phase remains continuous. The apparent $2\pi$-jumps in the figure arise from the fact that we take the principal value of $\theta_G$.

\section{Band models}\label{Sec5}
We consider a generic two-level lattice model, whose dynamics is initially governed by $H^\text{i}(k)=\mathbf{R}^\text{i}(k)\cdot \boldsymbol{\sigma}$. We parameterize the initial Bloch vector as $\mathbf{R}^\text{i}(k)=R^\text{i}_k(\sin\theta^\text{i}_{k}\cos\phi^\text{i}_{k},\sin\theta^\text{i}_{k}\sin\phi^\text{i}_{k},\cos\theta^\text{i}_{k})^T$, the eigen-levels of $H^\text{i}(k)$ are $|+R^\text{i}_k\rangle=\left(\begin{array}{c}\cos\frac{\theta^\text{i}_{k}}{2} \\\sin\frac{\theta^\text{i}_{k}}{2}\me^{\mi\phi^\text{i}_{k}}
 \end{array}\right)$ and $
|-R^\text{i}_k\rangle=\left(\begin{array}{c}
\sin\frac{\theta^\text{i}_{k}}{2}\\ -\cos\frac{\theta^\text{i}_{k}}{2}\me^{\mi\phi^\text{i}_{k}}.
 \end{array}\right)$. 
The initial system is prepared in a thermal equilibrium state at the temperature $T$, which is characterized by the density matrix $\rho(0)=\prod_k\otimes\rho_k(0)$ with
\begin{align}\label{rho0}
\rho_k(0)=\frac{1}{2}\left[\mathbf{1}-\tanh(\beta R^\text{i}_k)\hat{\mathbf{R}}^\text{i}(k)\cdot\boldsymbol{\sigma}\right].
\end{align}
At $t=0^+$, the Hamiltonian is suddenly switched to $H^\text{f}(k)=\mathbf{R}^\text{f}(k)\cdot\boldsymbol{\sigma}$. 
Similar to Eq.(\ref{pcrc}), it is straightforward to verify the parallel-quench condition is
\begin{align}\label{pcr1b}
0=\langle\pm R^\text{i}_k|H^\text{f}(k)|\pm R^\text{i}_k\rangle=\pm R^\text{i}_k\hat{\mathbf{R}}^\text{i}(k)\cdot\hat{\mathbf{R}}^\text{f}(k).
\end{align}
According to Eq.(\ref{dpms}), the accumulated dynamical phase is 
\begin{align}\label{dpms2}
\Phi^\text{d}_{k}(t)=&-\text{Tr}\left[\rho_{k}(0)H^\text{f}(k)\right]t\notag\\=&\tanh(\beta R^\text{i}_{k})\hat{\mathbf{R}}^\text{i}(k)\cdot\hat{\mathbf{R}}^\text{f}(k)R^\text{f}_kt.
\end{align}
The time-evolution operator following the quench is given by $U(t)=\prod_k\otimes U_k(t)$ with $U_k(t)=\me^{-\mi H^\text{f}(k)t}=\cos(R_k^\text{f} t)\mathbf{1}-\mi\sin(R_k^\text{f} t)\hat{\mathbf{R}}^\text{f}(k)\cdot\boldsymbol{\sigma}$. 
Thus, the Loschmidt amplitude is $\mathcal{G}_\rho(t)=\prod_k\mathcal{G}_k(t)$ with
\begin{align}\label{Gms}
&\mathcal{G}_k(t)=\text{Tr}\left[\rho_k(0)U_k(t)\right]=\text{Tr}\left[\rho_k(0)\me^{-\mi H^\text{f}(k)t}\right]\notag\\
=& \left[\cos(R_k^\text{f} t)+\mi\sin(R_k^\text{f} t)\tanh(\beta R^\text{i}_k)\hat{\mathbf{R}}^\text{i}(k)\cdot\hat{\mathbf{R}}^\text{f}(k)\right].
\end{align}
DQPTs occur at
 \begin{align}\label{tcnms}
t^*_n=\frac{\left(n+\frac{1}{2}\right)\pi}{R^\text{f}_{k_c}}.
\end{align}
where $k_c$ is the critical momentum such that  \begin{align}\hat{\mathbf{R}}^\text{i}(k_c)\cdot\hat{\mathbf{R}}^\text{f}(k_c)=0.\end{align}
Comparing with Eqs.(\ref{pcr1b}) and (\ref{dpms2}), it is evident that the parallel-quench condition is satisfied only at the critical momentum.  
Moreover, the dynamical phase vanishes at this point, yielding $\Phi^\text{G}_{k_c}(t)=\arg \mathcal{G}_{k_c}(t)$. 
For generic momentum, the parallel-quench condition is not respected, and we need to focus on the DTOP instead.

To illustrate the behaviors of DTOP at finite temperatures, we consider Su-Schrieffer-Heeger(SSH) model. Its Bloch Hamiltonian is given by 
$H_k=-(J_1+J_2\cos k)\sigma_x+J_2\sin k\sigma_y$. Introducing $m=\frac{J_1}{J_2}$, $H_k$ can be cast into the form  
$H_k=J_2\left[-(m+\cos k)\sigma_x+\sin k\sigma_y\right]$, which corresponds to the Bloch vector $\mathbf{R}=J_2(-m-\cos k,\sin k,0)^T$. This model respects the particle-hole symmetry (PHS):  $\mathcal{C}H(k)\mathcal{C}^{-1}=-H^*(-k)$ with $\mathcal{C}=\sigma_z$. At zero temperature, the system exhibits two distinct topological phases:
$m<1$ corresponds to a topologically nontrivial phase, while $m>1$ corresponds to a trivial phase. 
In addition to $k_c$, there are other types of momenta that require careful consideration. 
Note the PHS introduces the momentum inversion $k\rightarrow -k$. 
The first type of momenta consists of fixed points under this operation, satisfying $k\equiv -k \mod 2\pi$. Clearly, these momenta are $k_\mathcal{C}=0$, $\pi$.
%Thus, the Brillouin zone reduces to the EBZ $[0,\pi]$
The second type refers to the momenta $k^*$ at which the initial and final normalized Bloch vectors are either parallel or anti-parallel, satisfying $\hat{\mathbf{R}}^\text{i}(k^*)\px \pm\hat{\mathbf{R}}^\text{f}(k^*)$, or equivalently, $\hat{\mathbf{R}}^\text{i}(k^*)\cdot\hat{\mathbf{R}}^\text{f}(k^*)=\pm1$.
When $T=0$, the Loschmidt amplitude at $k^*$ simplifies to $\mathcal{G}_{k^*}(t)=\me^{\pm\mi R_{k^*}^\text{f} t}$ since $\lim_{\beta\rightarrow +\infty}\tanh(\beta R^\text{i}_{k^*})=1$.
This leads to
 \begin{align}\Phi_{k^*}(t)=\arg\mathcal{G}_{k^*}(t)=\pm R^\text{f}_{k^*}t=\Phi^\text{d}_{k^*}(t)\end{align} according to Eqs.(\ref{dpms2}). Consequently, the PGP is pinned to zero at $k^*$ when $T=0$.
Here, we consider a quench which switches $m=0.5$ in the initial Hamiltonian to $m=2.0$ in the final Hamiltonian, thereby connecting two topologically distinct phases. In this case, it is straightforward to verify that $\hat{\mathbf{R}}^\text{i}(0)\cdot\hat{\mathbf{R}}^\text{f}(0)=1$ and $\hat{\mathbf{R}}^\text{i}(\pi)\cdot\hat{\mathbf{R}}^\text{f}(\pi)=-1$. Thus, the momenta $k^*$ exactly coincide with $k_\mathcal{C}$, leading to  $\Phi^\text{G}_k(t,T=0)=0$ at $k=k_\mathcal{C}$. Due to the PHS, the Brillouin zone reduces to the EBZ $[0,\pi]$, which is further endowed with the topology of $S^1$ by identifying the fixed points. Similarly, the image of the mapping $\text{exp}(\mi \Phi^\text{G}_k(t,T=0))$  defined on the EBZ also inherits the topology of $U(1)\sim S^1$. In fact, $\me^{\mi \Phi^\text{G}_k(t)}$ is a homotopic mapping at $T=0$, and its topological class is characterized by the DTOP
 \begin{align}\label{DTOPk}
 \nu(t,T=0)=\frac{1}{2\pi}\oint_0^{\pi}\frac{\partial\Phi^\text{G}_k(t,T=0)}{\partial k}\dif k,
\end{align}
which is integer-quantized.
%In the case of the SSH model, these two types of momenta coincide with each other.

\begin{figure}[th]
\centering
\includegraphics[width=3.6in]{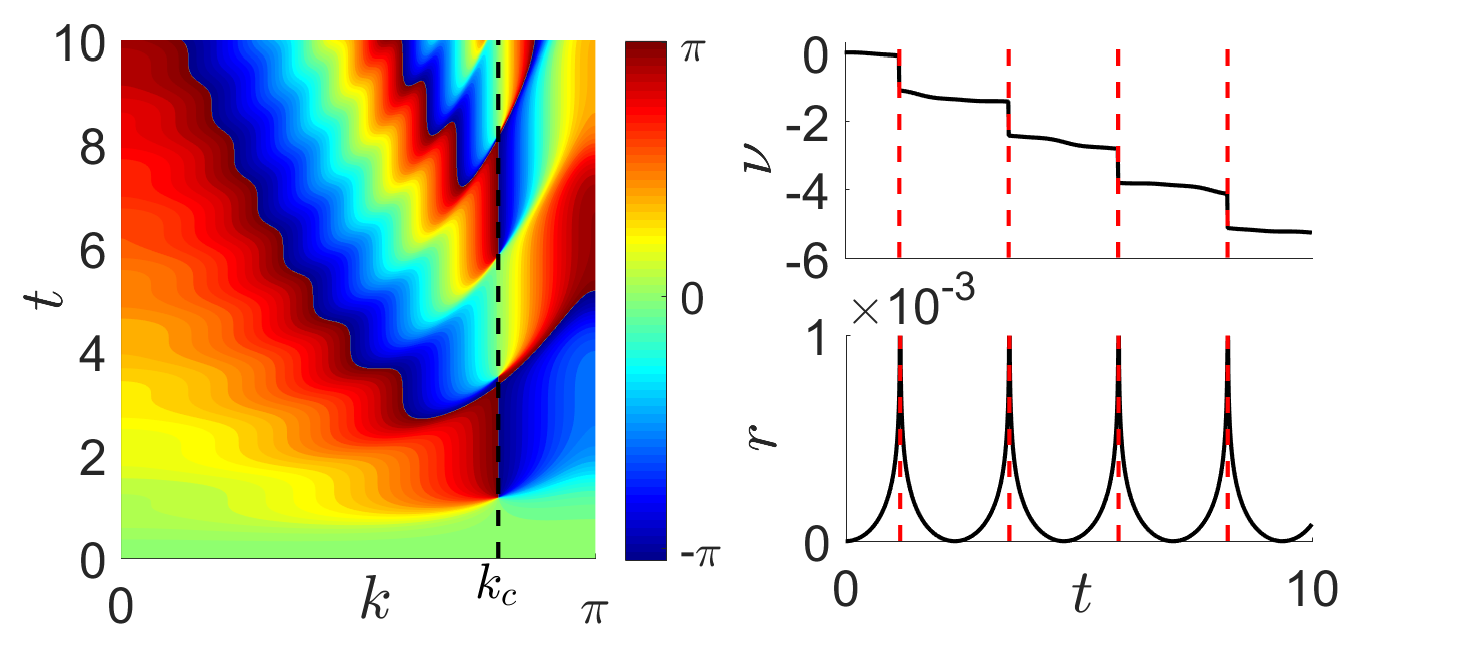}
\caption{(color online) Left panel: Contour plot of the PGP $\Phi^\text{G}_k(t)$ at $T=1.0 J_2$ for quenches from $m=0.5$ to $m=2.0$ in the SSH model. Right panel: DTOP (top) and the rate function (bottom) as functions of time.
}
\label{Fig4}
\end{figure}

At finite temperatures, the situation is substantially different. It can be verified that $\Phi^\text{G}_0(t)\neq \Phi^\text{G}_\pi(t) \mod 2\pi$ when $\beta>0$. As a result, the image of $\text{exp}(\mi \Phi^\text{G}_k(t,T>0))$ becomes an open line segment, which is topologically different from $S^1$. Accordingly, $\text{exp}(\mi \Phi^\text{G}_k(t,T>0))$ is no longer a homotopic mapping. To understand its physical origin, we notice that at zero temperature, the PGP satisfies the inversion symmetry
 \begin{align}\label{IS}
\Phi^\text{G}_k(t)=-\Phi^\text{G}_{-k}(t) \mod 2\pi
\end{align}
at $k_\mathcal{C}$. Thus, $ \nu(t,T=0)$ is a symmetry-protected topological invariant. At finite temperatures, this symmetry is broken because $0<\tanh(\beta R^\text{i}_{k^*})<1$ due to the emergence of thermal fluctuations. Therefore, $\nu(t,T>0)$ is no longer quantized as an integer. Nevertheless, at the two end points of EBZ, $\hat{\mathbf{R}}^\text{i}(0)\px \hat{\mathbf{R}}^\text{f}(0)$ while $\hat{\mathbf{R}}^\text{i}(\pi)\px -\hat{\mathbf{R}}^\text{f}(\pi)$, and there must exist a critical momentum $k_c\in [0,\pi]$ such that $\hat{\mathbf{R}}^\text{i}(k_c)\bot \hat{\mathbf{R}}^\text{f}(k_c)$. Thus, $\nu(t,T>0)$ should still be able to reflect this topological difference between the two end points. In fact, it can be proved that
 \begin{align}
 \Delta\nu(t^*_n)=&\lim_{\tau\rightarrow 0^+}\left[\nu(t^*_n+\tau)-\nu(t^*_n-\tau)\right]\notag\\=&\text{sgn}\left[\partial_kf(k)\right]|_{k_c},
\end{align}
where $f(k)=\tanh(\beta R^\text{i}_k)\hat{\mathbf{R}}^\text{i}(k)\cdot\hat{\mathbf{R}}^\text{f}(k)$. 
This means that at DQPTs, the DTOP still experiences a unit jump, which is a direct generalization of the result in Ref.\cite{Budich201693}. The detailed proof can be found in Appendix \ref{appp}. Evaluating the integral in Eq.(\ref{DTOPk}), the DTOP is given by \begin{align}\label{nuD3}
 \nu(t)=\frac{\Phi^\text{G}_{\pi}(t)-\Phi^\text{G}_{0}(t)}{2\pi}+\mathcal{N},
\end{align}
where the first term represents  the ``boundary effect'', capturing the ``non-closure" of the map $\text{exp}(\mi \Phi^\text{G}_k(t,T>0))$, and $\mathcal{N}$ denotes the
number of times that $\Phi^\text{G}_{\pi}(t)$ is folded into its principal angle value. Explicitly, $\mathcal{N}$ can be determined by generalizing the idea of relative Bloch sphere in Ref.\cite{Budich201693}. We consider $\tilde{\mathbf{R}}^\text{i}(k)\equiv\tanh(\beta R^\text{i}_k)\hat{\mathbf{R}}^\text{i}(k)$ as the reference vector, which points to the north-pole of the relative Bloch sphere defined at $k$. In fact, $\tilde{\mathbf{R}}^\text{i}(k)$ is the Bloch vector of the initial density matrix $\rho_k(0)$, whose end point lies inside the relative Bloch sphere since $\tanh(\beta R^\text{i}_k)<1$ for $\beta>0$. Note that $f(k)=\tilde{\mathbf{R}}^\text{i}(k)\cdot\hat{\mathbf{R}}^\text{f}(k)$, and $\Delta\nu(t^*_n)$ is then determined by the direction of $\hat{\mathbf{R}}^\text{f}(k)$ relative to $\tilde{\mathbf{R}}^\text{i}(k)$, or the north-pole direction. When $k$ increases from $k_c-0^+$ to $k_c+0^+$, if $\hat{\mathbf{R}}^\text{f}(k)$ traverses the equator of the relative Bloch sphere from the northern to the southern hemisphere, then $\Delta\nu(t^*_n)=-1$. Otherwise, $\Delta\nu(t^*_n)=1$.

\begin{figure}[th]
\centering
\includegraphics[width=3.6in]{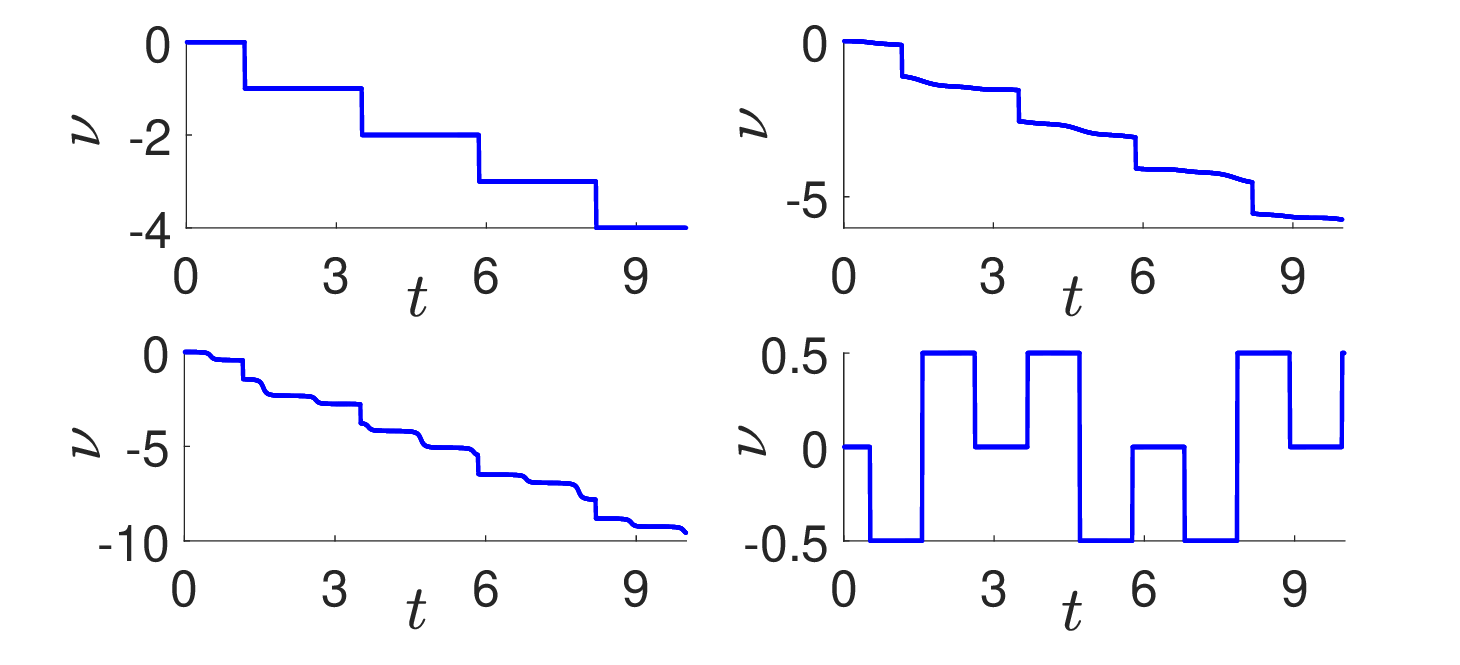}
\caption{ DTOP as a function of $t$ at $T=0.01J_2$ (top-left), $1.25J_2$ (top-right), $10.0J_2$ (bottom-left) and $+\infty$ (bottom-right).
}
\label{Fig5}
\end{figure}

In Figure \ref{Fig4}, we show the contour plot of the PGP $\Phi^\text{G}_k(t)$ as a function of $k$ and $t$ at $T=1.0J_2$ (left panel). 
It is clear that at the critical momentum $k_c$, the PGP undergoes a $\pm \pi$ jump at the DQPTs. 
The corresponding DTOP and rate function are shown as functions of time in the right panels. 
As expected, DTOP exhibits unit jumps at $t^*_n$, however, its behavior is, to some extent, distorted by boundary effects.
To illustrate the temperature dependence of the DTOP in greater detail, we present the results for $T=0.01J_2$, $1.25J_2$, $10.0J_2$, and $+\infty$ in Fig.~\ref{Fig5}. At very low temperatures, the DTOP remains almost integer-quantized, and its behavior closely resembles that of the pure-state case. As the temperature increases, boundary effects progressively become more significant and eventually dominate the behavior of the DTOP. At $T=10.0J_2$, the DTOP exhibits rather irregular behavior; however, its discrete jumps at the DQPTs are still discernible. In the infinite-temperature limit, the behavior changes completely. Since $\tanh(\beta R^\text{i}_k) = 0$ when $T = +\infty$, the initial density matrix becomes fully mixed, $\rho_k(0) = \frac{1}{2} \mathbf{1}$, and its corresponding Bloch vector is $\tilde{\mathbf{R}}^\text{i}(k) = \mathbf{0}$. Consequently, $f(k) = 0$, and $\Delta \nu(t^*_n)$ is no longer well-defined. In fact, $\nu(t)$ loses its capacity to characterize DQPTs.
To understand this, we note that at infinite temperature, $\Phi^\text{G}_k(t) = \arg[\cos(R_k^\text{f} t)]$ according to Eqs.~(\ref{dpms}) and (\ref{Gms}). As a result, $\text{exp}(\mi \Phi^\text{G}_k(t, T=+\infty))$ maps the EBZ to the discrete set of end points $\{1, -1\}$ rather than a line segment. Therefore, in this limit, $\nu(t)$ only captures the boundary effects, given by \begin{align}\label{nuD4} \nu(t, T=+\infty) = \frac{\arg[\cos(R_\pi^\text{f} t)] - \arg[\cos(R_0^\text{f} t)]}{2\pi}. \end{align}
As shown in the bottom-right panel of Fig.~\ref{Fig5}, although $\nu(t, T=+\infty)$ still exhibits finite jumps, these jumps are unrelated to DQPTs and instead reflect the periodicities at the boundary points. Consequently, the DTOP loses its original physical meaning at infinite temperature.

\section{Experimental implication}\label{Sec6}

Experimentally, the spin-$j$ state $|jm\rangle$ can be achieved by using the hyperfine states of cold atoms~\cite{FootBook,PethickBook}, which enables precise control over preparation, manipulation, and measurement. Moreover, as discussed in Sec.\ref{Sec1}, DQPTs have already been realized in several types of physical systems, including nuclear magnetic resonance (NMR) quantum simulators~\cite{Nie20}.
The IGP has also been measured via various experimental techniques~\cite{PhysRevLett.91.100403,PhysRevLett.94.050401,PhysRevLett.101.150404,GHOSH200627}, also including the NMR interferometry \cite{PhysRevLett.91.100403}. Therefore, the dynamical parallel-quench in the spin-$j$ system is experimentally feasible, and our findings contribute to enhancing control over such quantum systems in future experiments.

\section{Conclusion}\label{Sec7}

In conclusion, this study provides deep insights into the geometric and topological properties of quantum systems undergoing dynamical quantum phase transitions (DQPTs). For non-band models, we have shown that the Pancharatnam geometric phase (PGP) at zero temperature, and its finite-temperature counterpart, the interferometric geometric phase (IGP), serve as dynamic geometric order parameters (DGOPs), effectively capturing the geometric changes during DQPTs when the parallel-transport condition is satisfied. This highlights the power of these geometric phases as accessible tools for probing complex quantum dynamics. By extending our analysis to band models, we generalized the dynamic topological order parameter (DTOP) to finite-temperature scenarios, using the SSH model as a case study. Our results reveal that thermal fluctuations and boundary effects disrupt the DTOP's quantization, yet it retains the ability to signal topological transitions, broadening its applicability. Collectively, these findings enhance the theoretical framework of DQPTs, bridging geometric and topological perspectives, and lay the foundation for future studies of quantum dynamical evolution across a variety of systems.

\section{Acknowledgments}
H.G. was supported by the Innovation Program for Quantum
Science and Technology (Grant No. 2021ZD0301904) and the National Natural Science
Foundation of China (Grant No. 12074064). 
X. Y. H. was supported by the National Natural Science
Foundation of China (Grant No. 12405008) and the Jiangsu Funding Program for Excellent Postdoctoral Talent (Grant No. 2023ZB611).

\appendix
\section{Some derivations}\label{app1}
By using the Campbell-Baker-Hausdorff formula $\me^{\hat{A}}\hat{B}\me^{-\hat{A}}=\hat{B}+[\hat{A},\hat{B}]+\frac{1}{2!}[\hat{A},[\hat{A},\hat{B}]]+\cdots$, the following results can be obtained
\begin{align}
\me^{-\mi\theta J_y}J_z\me^{\mi\theta J_y}&
%J_z+\frac{-\mi\theta}{1! }[J_y,J_z]+\frac{(-\mi\theta)^2}{2! }[J_y,[J_y,J_z]]+\frac{(-\mi\theta)^3}{3! }[J_y,[J_y,[J_y,J_z]]+\cdots\notag\\
%&=J_x(\theta-\frac{\theta^3}{3!}+\frac{\theta^5}{5!}+\cdots)+J_z(1-\frac{\theta^2}{2!}+\frac{\theta^4}{4!}+\cdots)\notag\\&
=J_x\sin\theta+J_z\cos\theta, \label{het1}\\
\me^{-\mi\phi J_z}J_x\me^{\mi\phi J_z}&=J_x\cos\phi+J_y\sin\phi, \label{het2}\\
\me^{-\mi\phi J_z}J_y\me^{\mi\phi J_z}&=-J_x\sin\phi+J_y\cos\phi, \label{het2a}\\
\me^{-\mi\theta J_y}J_x\me^{\mi\theta J_y}&=-J_z\sin\theta+J_x\cos\theta.\label{het1a}
\end{align}

%\section{Two-band model}\label{appc}

\section{Root of the second-kind Chebyshev polynomials}\label{app2}

The expression for the second-kind Chebyshev polynomials is
\begin{align}\label{Un}
U_n(z)=\frac{\left(z+\sqrt{z^2-1}\right)^{n+1}-\left(z-\sqrt{z^2-1}\right)^{n+1}}{2\sqrt{z^2-1}}.
\end{align}
It is straightforward to verify that $U_0(z)=1$, $U_1(z)=2z$, and the recurrence relation
\begin{align}\label{Uni}
2zU_n(z)-U_{n-1}(z)=U_{n+1}(z).
\end{align}
Using the recursion relation (\ref{Uni}) and the method of induction,  we can show that $U_n(z)$ is a polynomial of degree $n$ in $z$.

To find the roots of $U_n(z)$, we set $z=\cos x$. It can be shown that $U_0(\cos x)=1=\frac{\sin(0+1)x}{\sin x}$, $U_1(\cos x)=2\cos x=\frac{\sin(1+1)x}{\sin x}$. This suggests the reasonable conjecture that $U_n(\cos x)=\frac{\sin(n+1)x}{\sin x}$. This can be verified by the method of induction. Suppose the proposition is valid for $n\le N$. For $n=N+1$, using Eq.(\ref{Uni}), we have
 \begin{align}\label{Uni2}
U_{N+1}(\cos x)&=2\cos x\frac{\sin(N+1)x}{\sin x}-\frac{\sin Nx}{\sin x}\notag\\
&=\frac{\sin(N+2)x}{\sin x},
\end{align}
which confirms our initial assertion. The roots of $U_n(\cos x)=\frac{\sin(n+1)x}{\sin x}$ satisfy $\sin(n+1)x=0$, leading to $x=\frac{k\pi}{n+1}$ with $k=1,2,\cdots, n$ (Note that if $k=0$, then $U_n(\cos \frac{k\pi}{n+1})=n+1$). Therefore, $U_n(z)$ has $n$ roots
 \begin{align}\label{Uroot}
z_k=\cos\frac{k\pi}{n+1},\quad k=1,2,\cdots, n.
\end{align}
Since $U_n(z)$ is a polynomial of degree $n$,  all of its roots, as shown in Eq.(\ref{Uroot}), are real.

\section{Dynamical phase for non-parallel quenches}\label{appd}
For spin-$j$ systems, the dynamical phase during a non-parallel quench is evaluated according to Eq.(\ref{dfi}):
\begin{align}\label{dfij}
\theta_{\text{d}}(t)&=-\int_0^t\dif t' \text{Tr}\left[\rho(t')H\right]\notag\\
&=-\int_0^t\dif t'\text{Tr}\left[\rho(0)H\right]\notag\\
&=-\sum_{m=-j}^j\frac{\me^{-\beta m\omega_0}}{Z_j(0)}\langle \psi_m^0|H|\psi^0_m\rangle t,
\end{align}
where we have applied the fact that $[U(t), H]=0$. Applying the Eq.(\ref{pcsj1}), we finally obtain
\begin{align}\label{dfij2}
\theta_{\text{d}}(t)=-\sum_{m=-j}^j\frac{\me^{-\beta m\omega_0}m\omega_0 t}{Z_j(0)}\hat{\mathbf{B}}_0\cdot \hat{\mathbf{B}}.
\end{align}
For pure states (ground state), the dynamical phase is evaluated as
\begin{align}\label{dfips}
\theta_{\text{d}}(t)&=-\int_0^t\dif t'\langle \psi^0_{-j}(t)|H|\psi^0_{-j}(t)\rangle\notag\\
&=-\int_0^t\dif t'\langle \psi^0_{-j}|H|\psi^0_{-j}\rangle\notag\\
&=j\omega_0 t\hat{\mathbf{B}}_0\cdot \hat{\mathbf{B}}.
\end{align}
In the zero temperature limit, Eq.(\ref{dfij2}) exactly reduces to Eq.(\ref{dfips}).

\section{Proof of the formula for the winding number}\label{appp}
The winding number is given by Eq.(\ref{DTOP}), where
 \begin{align}
\Phi^\text{G}_k(t)=&\Phi_k(t)-\Phi^\text{dyn}_{k}(t)\notag\\=&\arg \mathcal{G}_{k}(t)+\int_0^t\dif t'\text{Tr}\left[\rho_{k}(t')H^\text{f}(k)\right]\notag\\
=&\arg \mathcal{G}_{k}(t)+\text{Tr}\left[\rho_{k}(0)H^\text{f}(k)\right]t.
\end{align}
Here the fact $\rho_{k}(t')=\me^{-\mi H^\text{f}(k) t'}\rho(0)\me^{\mi H^\text{f}(k) t'}$ has been applied. For a generic two-band model, the Loschmidt amplitude is given by Eq.(\ref{Gms}). 
Note the principal value of the argument of $\mathcal{G}_k(t)$ is given by
\begin{widetext}
\begin{align}\label{Ud9}
\Phi_k(t)
&=\left\{\begin{array}{cc}\arctan\left[\tan (R_k^\text{f} t)f(k)\right], & \text{if } R_k^\text{f} t \in  \left(2n\pi-\frac{\pi}{2},2n\pi+\frac{\pi}{2}\right);\\
\pm\frac{\pi}{2}\text{sgn}[f(k)], & \text{if  } R_k^\text{f} t=2n\pi\pm\frac{\pi}{2};\\
\arctan\left[\tan (R_k^\text{f} t)f(k)\right]+\pi\text{sgn}[f(k)], & \text{if  } R_k^\text{f} t \in \left(2n\pi+\frac{\pi}{2},2n\pi+\pi\right);\\
\arctan\left[\tan (R_k^\text{f} t)f(k)\right]-\pi\text{sgn}[f(k)], & \text{if  } R_k^\text{f} t \in \big[2n\pi-\pi,2n\pi-\frac{\pi}{2}\big).
\end{array}\right.
\end{align}
\end{widetext}
Plugging in $H^\text{f}(k)=\mathbf{R}^\text{f}(k)\cdot\boldsymbol{\sigma}$ and Eq.(\ref{rho0}), the dynamical phase is evaluated as
\begin{align}
\Phi^\text{dyn}_{k}(t)=\tanh(\beta R^\text{i}_{k})\hat{\mathbf{R}}^\text{i}(k)\cdot\hat{\mathbf{R}}^\text{f}(k)R^\text{f}_kt,
\end{align}
which is a continuous function of both $k$ and $t$. Hence,
\begin{align}
 \nu(t)=\frac{1}{2\pi}\int_0^\pi\frac{\partial\Phi_k(t)}{\partial k}\dif k-\frac{\Phi^\text{dyn}_{\pi}(t)-\Phi^\text{dyn}_{0}(t)}{2\pi}.
\end{align}
Taking the principal value of the integral, we further have
\begin{align}\label{nuD3b}
 \nu(t)=\frac{\Phi^\text{G}_{\pi}(t)-\Phi^\text{G}_{0}(t)}{2\pi}+\lim_{\epsilon\rightarrow 0^+}\int_{k_c^-}^{k_c^+}\frac{\partial\Phi_k(t)}{\partial k}\dif k,
\end{align}
where $k_c^\pm=k_c\pm \epsilon$. For simplicity, we have assumed that there is only one $k_c\in [0,\pi]$ such that $\hat{\mathbf{R}}^\text{i}(k_c)\cdot\hat{\mathbf{R}}^\text{f}(k_c)=0$. Note $\Phi_k(t)$ is singular at $t^*_n$. Define the jump of $\nu$ at $t^*_n$ as $\Delta\nu(t^*_n)=\lim_{\tau\rightarrow 0^+}\left[\nu(t^*_n+\tau)-\nu(t^*_n-\tau)\right]$ with $\tau>0$. Using Eq.(\ref{nuD3b}), we have
\begin{align}
\Delta\nu(t^*_n)=&\lim_{\tau\rightarrow 0^+}\lim_{\epsilon\rightarrow 0^+}\Big[\Phi_{k^+_c}(t^{*+}_n)-\Phi_{k^-_c}(t^{*+}_n)\notag\\&-\Phi_{k^+_c}(t^{*-}_n)+\Phi_{k^-_c}(t^{*-}_n)\Big].
\end{align}
where $t^{*\pm}_n=t^*_n\pm \tau$. We further assume that $R^\text{f}(k)$ is a monotonically increasing function of $k$ for simplicity. Next, we will analyze the situation based on different cases.

If $t^*_n$ satisfies $R^\text{f}_{k_c}t^*_n=2n\pi+\frac{\pi}{2}$
 then $R^\text{f}_{k^+_c}t^*_n\in \left(2n\pi+\frac{\pi}{2},2n\pi+\pi\right)$ and $R^\text{f}_{k^-_c}t^*_n\in \left(2n\pi-\frac{\pi}{2},2n\pi+\frac{\pi}{2}\right)$ since $R^\text{f}(k)$ monotonically increases with $k$. Thus, we must have
 $R^\text{f}_{k^+_c}t^{*+}_n\in \left(2n\pi+\frac{\pi}{2},2n\pi+\pi\right)$ and $R^\text{f}_{k^-_c}t^{*-}_n\in \left(2n\pi-\frac{\pi}{2},2n\pi+\frac{\pi}{2}\right)$. Note
 \begin{align}
R^\text{f}_{k^+_c}t^{*-}_n&=(R^\text{f}_{k_c}+\partial_k R^\text{f}_k\big|_{k_c}\epsilon )(t^*_n-\tau)\notag\\&=2n\pi+\frac{\pi}{2}-\tau R^\text{f}_{k_c}+\epsilon \partial_k R^\text{f}_k\big|_{k_c} t^*_n,\notag\\
R^\text{f}_{k^-_c}t^{*+}_n&=(R^\text{f}_{k_c}-\partial_k R^\text{f}_k\big|_{k_c}\epsilon )(t^*_n+\tau)\notag\\&=2n\pi+\frac{\pi}{2}+\tau R^\text{f}_{k_c}-\epsilon \partial_k R^\text{f}_k\big|_{k_c} t^*_n.
\end{align}
Note the limit $\epsilon\rightarrow 0^+$ is taken before $\tau\rightarrow 0^+$, then we must have $\tau R^\text{f}_{k_c}-\epsilon \partial_k R^\text{f}_k\big|_{k_c} t^*_n>0$. This implies
 \begin{align}
&R^\text{f}_{k^+_c}t^{*-}_n\in \left(2n\pi-\frac{\pi}{2},2n\pi+\frac{\pi}{2}\right),\notag\\&R^\text{f}_{k^-_c}t^{*+}_n\in \left(2n\pi+\frac{\pi}{2},2n\pi+\pi\right).
\end{align}
According to Eq.(\ref{Ud9}), we have
\begin{align}\label{Ud9b}
\Phi_{k_c^+}(t^{*+}_n)=&\arctan\left[\tan (R_{k_c^+}^\text{f} t^{*+}_n)f({k_c^+})\right]+\pi\text{sgn}[f(k_c^+)],\notag\\
\Phi_{k_c^-}(t^{*+}_n)=&\arctan\left[\tan (R_{k_c^-}^\text{f} t^{*+}_n)f({k_c^-})\right]+\pi\text{sgn}[f(k_c^-)],\notag\\
\Phi_{k_c^+}(t^{*-}_n)=&\arctan\left[\tan (R_{k_c^+}^\text{f} t^{*-}_n)f({k_c^+})\right],\notag\\
\Phi_{k_c^-}(t^{*-}_n)=&\arctan\left[\tan (R_{k_c^-}^\text{f} t^{*-}_n)f({k_c^-})\right].
\end{align}
Note
\begin{align}\label{ta0}
\text{sgn}[f(k_c^\pm)]=\pm\text{sgn}\left[\partial_k f(k)\big|_{k_c}\right]
\end{align}
where we have applied the facts $f(k_c)=0$ and $\text{sgn}(\pm x\epsilon)=\pm \text{sgn}(x)$ due to $\epsilon>0$.
Expanding the functions in Eq.(\ref{Ud9b}) to the first order of $\epsilon$ and $\tau$ at $(k_c, t^*_n)$, we have
\begin{align}\label{ta1}
&\tan (R_{k_c^\pm}^\text{f} t^{*+}_n)f({k_c^\pm})\notag\\=&\tan \left[R_{k_c}^\text{f} t^*_n+\tau R^\text{f}_{k_c}\pm \partial_k R^\text{f}_k\big|_{k_c}\epsilon t^*_n\right]\left[f({k_c})\pm\partial_k f(k)\big|_{k_c}\epsilon\right]\notag\\
%=&\tan\left [2n\pi+\frac{\pi}{2}+\tau R^\text{f}_{k_c}\pm \partial_k R^\text{f}_k\big|_{k_c}\epsilon t^*_n\right]\partial_k f(k)\big|_{k_c}(\pm\epsilon)\notag\\
=&-\text{ctan}\left(\tau R^\text{f}_{k_c}\pm \partial_k R^\text{f}_k\big|_{k_c}\epsilon t^*_n\right) \partial_k f(k)\big|_{k_c}(\pm\epsilon)\notag\\
%=&-\frac{\cos\left(\partial_k R^\text{f}_k\big|_{k_c}\epsilon t^*_n\pm\tau R^\text{f}_{k_c}\right)}{\sin\left(\partial_k R^\text{f}_k\big|_{k_c}\epsilon t^*_n\pm \tau R^\text{f}_{k_c}\right)}\partial_k f(k)\big|_{k_c}\epsilon\notag\\
=&-\frac{\partial_k f(k)\big|_{k_c}\epsilon}{\partial_k R^\text{f}_k\big|_{k_c} t^*_n\epsilon\pm \tau R^\text{f}_{k_c}},
\end{align}
where we have applied the fact $\tan(\frac{\pi}{2}+x)=-\text{ctan}x$.
Using Eqs.(\ref{Ud9b}), (\ref{ta0}) and (\ref{ta1}) and taking the limits with respect to $\epsilon$ and $\tau$ in order, we have
\begin{align}\label{ta2}
\lim_{\tau\rightarrow 0}\lim_{\epsilon\rightarrow 0}\left[\Phi_{k^+_c}(t^{*+}_n)-\Phi_{k^-_c}(t^{*+}_n)\right]=2\pi \text{sgn}\left[\partial_k f(k)\big|_{k_c}\right].
\end{align}
since $\lim_{\tau\rightarrow 0}\lim_{\epsilon\rightarrow 0}\frac{\partial_k f(k)|_{k_c}\epsilon}{\partial_k R^\text{f}_k|_{k_c} t^*_n\epsilon\pm \tau R^\text{f}_{k_c}}=0$. Similarly,
\begin{align}\label{ta1b}
\tan (R_{k_c^\pm}^\text{f} t^{*-}_n)f({k_c^\pm})=-\frac{\partial_k f(k)\big|_{k_c}\epsilon}{\partial_k R^\text{f}_k\big|_{k_c} t^*_n\epsilon\mp \tau R^\text{f}_{k_c}},
\end{align}
and accordingly
\begin{align}\label{ta2b}
\lim_{\tau\rightarrow 0}\lim_{\epsilon\rightarrow 0}\left[\Phi_{k^+_c}(t^{*-}_n)-\Phi_{k^-_c}(t^{*-}_n)\right]=0.
\end{align}
Based on the above results, we can conclude that
\begin{align}
\Delta \nu(t^*_n)=\text{sgn}\left[\partial_k f(k)\big|_{k_c}\right].
\end{align}

If $t^*_n$ satisfies $R^\text{f}_{k_c}t^*_n=2n\pi-\frac{\pi}{2}$
 then
 \begin{align}
&R^\text{f}_{k^\pm_c}t^{*+}_n\in \left(2n\pi-\frac{\pi}{2},2n\pi+\frac{\pi}{2}\right),\notag\\&R^\text{f}_{k^\pm_c}t^{*-}_n\in \left(2n\pi-\pi,2n\pi-\frac{\pi}{2}\right).
\end{align}
  According to Eq.(\ref{Ud9}), we have
 \begin{align}\label{Ud9c}
\Phi_{k_c^+}(t^{*+}_n)=&\arctan\left[\tan (R_{k_c^+}^\text{f} t^{*+}_n)f({k_c^+})\right],\notag\\
\Phi_{k_c^+}(t^{*-}_n)=&\arctan\left[\tan (R_{k_c^+}^\text{f} t^{*-}_n)f({k_c^+})\right]-\pi\text{sgn}[f(k_c^+)],\notag\\
\Phi_{k_c^-}(t^{*+}_n)=&\arctan\left[\tan (R_{k_c^-}^\text{f} t^{*+}_n)f({k_c^-})\right],\notag\\
\Phi_{k_c^-}(t^{*-}_n)=&\arctan\left[\tan (R_{k_c^-}^\text{f} t^{*-}_n)f({k_c^-})\right]-\pi\text{sgn}[f(k_c^-)].
\end{align}
Following the same steps, Eqs.(\ref{ta2}) and (\ref{ta2b}) respectively become
\begin{align}\label{ta2c}
\lim_{\tau\rightarrow 0}\lim_{\epsilon\rightarrow 0}\left[\Phi_{k^+_c}(t^{*+}_n)-\Phi_{k^-_c}(t^{*+}_n)\right]=\pi \text{sgn}\left[\partial_k f(k)\big|_{k_c}\right],
\end{align}
and
 \begin{align}\label{ta2d}
\lim_{\tau\rightarrow 0}\lim_{\epsilon\rightarrow 0}\left[\Phi_{k^+_c}(t^{*-}_n)-\Phi_{k^-_c}(t^{*-}_n)\right]=\pi \text{sgn}\left[\partial_k f(k)\big|_{k_c}\right].
\end{align}
Thus, we also get
\begin{align}
 \Delta\nu(t^*_n)=\text{sgn}\left[\partial_k f(k)\big|_{k_c}\right].
\end{align}

The same conclusion can be obtained for other cases through similar derivations.

\bibliography{Review1,dqpt,Review}

\providecommand{\noopsort}[1]{}\providecommand{\singleletter}[1]{#1}%
\begin{thebibliography}{103}%
\makeatletter
\providecommand \@ifxundefined [1]{%
 \@ifx{#1\undefined}
}%
\providecommand \@ifnum [1]{%
 \ifnum #1\expandafter \@firstoftwo
 \else \expandafter \@secondoftwo
 \fi
}%
\providecommand \@ifx [1]{%
 \ifx #1\expandafter \@firstoftwo
 \else \expandafter \@secondoftwo
 \fi
}%
\providecommand \natexlab [1]{#1}%
\providecommand \enquote  [1]{``#1''}%
\providecommand \bibnamefont  [1]{#1}%
\providecommand \bibfnamefont [1]{#1}%
\providecommand \citenamefont [1]{#1}%
\providecommand \href@noop [0]{\@secondoftwo}%
\providecommand \href [0]{\begingroup \@sanitize@url \@href}%
\providecommand \@href[1]{\@@startlink{#1}\@@href}%
\providecommand \@@href[1]{\endgroup#1\@@endlink}%
\providecommand \@sanitize@url [0]{\catcode `\\12\catcode `\$12\catcode
  `\&12\catcode `\#12\catcode `\^12\catcode `\_12\catcode `\%12\relax}%
\providecommand \@@startlink[1]{}%
\providecommand \@@endlink[0]{}%
\providecommand \url  [0]{\begingroup\@sanitize@url \@url }%
\providecommand \@url [1]{\endgroup\@href {#1}{\urlprefix }}%
\providecommand \urlprefix  [0]{URL }%
\providecommand \Eprint [0]{\href }%
\providecommand \doibase [0]{https://doi.org/}%
\providecommand \selectlanguage [0]{\@gobble}%
\providecommand \bibinfo  [0]{\@secondoftwo}%
\providecommand \bibfield  [0]{\@secondoftwo}%
\providecommand \translation [1]{[#1]}%
\providecommand \BibitemOpen [0]{}%
\providecommand \bibitemStop [0]{}%
\providecommand \bibitemNoStop [0]{.\EOS\space}%
\providecommand \EOS [0]{\spacefactor3000\relax}%
\providecommand \BibitemShut  [1]{\csname bibitem#1\endcsname}%
\let\auto@bib@innerbib\@empty
%</preamble>
\bibitem [{\citenamefont {Jurcevic}\ \emph {et~al.}()\citenamefont {Jurcevic},
  \citenamefont {Shen}, \citenamefont {Hauke}, \citenamefont {Maier},
  \citenamefont {Brydges}, \citenamefont {Hempel}, \citenamefont {Lanyon},
  \citenamefont {Heyl}, \citenamefont {Blatt},\ and\ \citenamefont
  {Roos}}]{DQPTB41}%
  \BibitemOpen
  \bibfield  {author} {\bibinfo {author} {\bibfnamefont {P.}~\bibnamefont
  {Jurcevic}}, \bibinfo {author} {\bibfnamefont {H.}~\bibnamefont {Shen}},
  \bibinfo {author} {\bibfnamefont {P.}~\bibnamefont {Hauke}}, \bibinfo
  {author} {\bibfnamefont {C.}~\bibnamefont {Maier}}, \bibinfo {author}
  {\bibfnamefont {T.}~\bibnamefont {Brydges}}, \bibinfo {author} {\bibfnamefont
  {C.}~\bibnamefont {Hempel}}, \bibinfo {author} {\bibfnamefont {B.~P.}\
  \bibnamefont {Lanyon}}, \bibinfo {author} {\bibfnamefont {M.}~\bibnamefont
  {Heyl}}, \bibinfo {author} {\bibfnamefont {R.}~\bibnamefont {Blatt}},\ and\
  \bibinfo {author} {\bibfnamefont {C.~F.}\ \bibnamefont {Roos}},\ }\bibfield
  {title} {\bibinfo {title} {Direct observation of dynamical quantum phase
  transitions in an interacting many-body system},\ }\href@noop {} {\
  }\BibitemShut {NoStop}%
\bibitem [{\citenamefont {Fl$\ddot{\textrm{a}}$schner}\ \emph
  {et~al.}(2018)\citenamefont {Fl$\ddot{\textrm{a}}$schner}, \citenamefont
  {Vogel}, \citenamefont {Tarnowski}, \citenamefont {Rem}, \citenamefont
  {Luhmann}, \citenamefont {Heyl}, \citenamefont {Budich}, \citenamefont
  {Mathe}, \citenamefont {Sengstock},\ and\ \citenamefont
  {Weitenberg}}]{DQPTB4}%
  \BibitemOpen
  \bibfield  {author} {\bibinfo {author} {\bibfnamefont {N.}~\bibnamefont
  {Fl$\ddot{\textrm{a}}$schner}}, \bibinfo {author} {\bibfnamefont
  {D.}~\bibnamefont {Vogel}}, \bibinfo {author} {\bibfnamefont
  {M.}~\bibnamefont {Tarnowski}}, \bibinfo {author} {\bibfnamefont {B.~S.}\
  \bibnamefont {Rem}}, \bibinfo {author} {\bibfnamefont {D.~S.}\ \bibnamefont
  {Luhmann}}, \bibinfo {author} {\bibfnamefont {M.}~\bibnamefont {Heyl}},
  \bibinfo {author} {\bibfnamefont {J.~C.}\ \bibnamefont {Budich}}, \bibinfo
  {author} {\bibfnamefont {Y.~L.}\ \bibnamefont {Mathe}}, \bibinfo {author}
  {\bibfnamefont {K.}~\bibnamefont {Sengstock}},\ and\ \bibinfo {author}
  {\bibfnamefont {C.}~\bibnamefont {Weitenberg}},\ }\bibfield  {title}
  {\bibinfo {title} {Observation of dynamical vortices after quenches in a
  system with topology},\ }\href@noop {} {\bibfield  {journal} {\bibinfo
  {journal} {Nat. Phys.}\ }\textbf {\bibinfo {volume} {14}},\ \bibinfo {pages}
  {265} (\bibinfo {year} {2018})}\BibitemShut {NoStop}%
\bibitem [{\citenamefont {Zhang}\ \emph
  {et~al.}(2017{\natexlab{a}})\citenamefont {Zhang}, \citenamefont {Pagano},
  \citenamefont {Hess}, \citenamefont {Kyprianidis}, \citenamefont {Becker},
  \citenamefont {Kaplan}, \citenamefont {Gorshkov}, \citenamefont {Gong},\ and\
  \citenamefont {Monroe}}]{Zhang_2017}%
  \BibitemOpen
  \bibfield  {author} {\bibinfo {author} {\bibfnamefont {J.}~\bibnamefont
  {Zhang}}, \bibinfo {author} {\bibfnamefont {G.}~\bibnamefont {Pagano}},
  \bibinfo {author} {\bibfnamefont {P.~W.}\ \bibnamefont {Hess}}, \bibinfo
  {author} {\bibfnamefont {A.}~\bibnamefont {Kyprianidis}}, \bibinfo {author}
  {\bibfnamefont {P.}~\bibnamefont {Becker}}, \bibinfo {author} {\bibfnamefont
  {H.}~\bibnamefont {Kaplan}}, \bibinfo {author} {\bibfnamefont {A.~V.}\
  \bibnamefont {Gorshkov}}, \bibinfo {author} {\bibfnamefont {Z.-X.}\
  \bibnamefont {Gong}},\ and\ \bibinfo {author} {\bibfnamefont
  {C.}~\bibnamefont {Monroe}},\ }\bibfield  {title} {\bibinfo {title}
  {Observation of a many-body dynamical phase transition with a 53-qubit
  quantum simulator},\ }\href {https://doi.org/10.1038/nature24654} {\bibfield
  {journal} {\bibinfo  {journal} {Nature}\ }\textbf {\bibinfo {volume} {551}},\
  \bibinfo {pages} {601–604} (\bibinfo {year}
  {2017}{\natexlab{a}})}\BibitemShut {NoStop}%
\bibitem [{\citenamefont {Guo}\ \emph {et~al.}(2019{\natexlab{a}})\citenamefont
  {Guo}, \citenamefont {Yang}, \citenamefont {Zeng}, \citenamefont {Peng},
  \citenamefont {Li}, \citenamefont {Deng}, \citenamefont {Jin}, \citenamefont
  {Chen}, \citenamefont {Zheng},\ and\ \citenamefont
  {Fan}}]{PhysRevApplied.11.044080}%
  \BibitemOpen
  \bibfield  {author} {\bibinfo {author} {\bibfnamefont {X.-Y.}\ \bibnamefont
  {Guo}}, \bibinfo {author} {\bibfnamefont {C.}~\bibnamefont {Yang}}, \bibinfo
  {author} {\bibfnamefont {Y.}~\bibnamefont {Zeng}}, \bibinfo {author}
  {\bibfnamefont {Y.}~\bibnamefont {Peng}}, \bibinfo {author} {\bibfnamefont
  {H.-K.}\ \bibnamefont {Li}}, \bibinfo {author} {\bibfnamefont
  {H.}~\bibnamefont {Deng}}, \bibinfo {author} {\bibfnamefont {Y.-R.}\
  \bibnamefont {Jin}}, \bibinfo {author} {\bibfnamefont {S.}~\bibnamefont
  {Chen}}, \bibinfo {author} {\bibfnamefont {D.}~\bibnamefont {Zheng}},\ and\
  \bibinfo {author} {\bibfnamefont {H.}~\bibnamefont {Fan}},\ }\bibfield
  {title} {\bibinfo {title} {Observation of a dynamical quantum phase
  transition by a superconducting qubit simulation},\ }\href
  {https://doi.org/10.1103/PhysRevApplied.11.044080} {\bibfield  {journal}
  {\bibinfo  {journal} {Phys. Rev. Appl.}\ }\textbf {\bibinfo {volume} {11}},\
  \bibinfo {pages} {044080} (\bibinfo {year} {2019}{\natexlab{a}})}\BibitemShut
  {NoStop}%
\bibitem [{\citenamefont {Mei}\ \emph {et~al.}(2022)\citenamefont {Mei},
  \citenamefont {Li}, \citenamefont {Wu}, \citenamefont {Cai}, \citenamefont
  {Wang}, \citenamefont {Yao}, \citenamefont {Zhou},\ and\ \citenamefont
  {Duan}}]{PhysRevLett.128.160504}%
  \BibitemOpen
  \bibfield  {author} {\bibinfo {author} {\bibfnamefont {Q.-X.}\ \bibnamefont
  {Mei}}, \bibinfo {author} {\bibfnamefont {B.-W.}\ \bibnamefont {Li}},
  \bibinfo {author} {\bibfnamefont {Y.-K.}\ \bibnamefont {Wu}}, \bibinfo
  {author} {\bibfnamefont {M.-L.}\ \bibnamefont {Cai}}, \bibinfo {author}
  {\bibfnamefont {Y.}~\bibnamefont {Wang}}, \bibinfo {author} {\bibfnamefont
  {L.}~\bibnamefont {Yao}}, \bibinfo {author} {\bibfnamefont {Z.-C.}\
  \bibnamefont {Zhou}},\ and\ \bibinfo {author} {\bibfnamefont {L.-M.}\
  \bibnamefont {Duan}},\ }\bibfield  {title} {\bibinfo {title} {Experimental
  realization of the rabi-hubbard model with trapped ions},\ }\href
  {https://doi.org/10.1103/PhysRevLett.128.160504} {\bibfield  {journal}
  {\bibinfo  {journal} {Phys. Rev. Lett.}\ }\textbf {\bibinfo {volume} {128}},\
  \bibinfo {pages} {160504} (\bibinfo {year} {2022})}\BibitemShut {NoStop}%
\bibitem [{\citenamefont {Zvyagin}(2016)}]{Zvyagin16}%
  \BibitemOpen
  \bibfield  {author} {\bibinfo {author} {\bibfnamefont {A.~A.}\ \bibnamefont
  {Zvyagin}},\ }\bibfield  {title} {\bibinfo {title} {Dynamical quantum phase
  transitions (review article)},\ }\href@noop {} {\bibfield  {journal}
  {\bibinfo  {journal} {Low Temp. Phys.}\ }\textbf {\bibinfo {volume} {42}},\
  \bibinfo {pages} {971} (\bibinfo {year} {2016})}\BibitemShut {NoStop}%
\bibitem [{\citenamefont {Heyl}\ \emph {et~al.}(2013)\citenamefont {Heyl},
  \citenamefont {Polkovnikov},\ and\ \citenamefont {Kehrein}}]{DQPT13}%
  \BibitemOpen
  \bibfield  {author} {\bibinfo {author} {\bibfnamefont {M.}~\bibnamefont
  {Heyl}}, \bibinfo {author} {\bibfnamefont {A.}~\bibnamefont {Polkovnikov}},\
  and\ \bibinfo {author} {\bibfnamefont {S.}~\bibnamefont {Kehrein}},\
  }\bibfield  {title} {\bibinfo {title} {Dynamical quantum phase transitions in
  the transverse-field ising model},\ }\href@noop {} {\bibfield  {journal}
  {\bibinfo  {journal} {Phys. Rev. Lett.}\ }\textbf {\bibinfo {volume} {110}},\
  \bibinfo {pages} {135704} (\bibinfo {year} {2013})}\BibitemShut {NoStop}%
\bibitem [{\citenamefont {Heyl}(2018)}]{DQPTreview18}%
  \BibitemOpen
  \bibfield  {author} {\bibinfo {author} {\bibfnamefont {M.}~\bibnamefont
  {Heyl}},\ }\bibfield  {title} {\bibinfo {title} {Dynamical quantum phase
  transitions: a review},\ }\href@noop {} {\bibfield  {journal} {\bibinfo
  {journal} {Rep. Prog. Phys.}\ }\textbf {\bibinfo {volume} {81}},\ \bibinfo
  {pages} {054001} (\bibinfo {year} {2018})}\BibitemShut {NoStop}%
\bibitem [{\citenamefont {Heyl}(2014{\natexlab{a}})}]{DQPT14}%
  \BibitemOpen
  \bibfield  {author} {\bibinfo {author} {\bibfnamefont {M.}~\bibnamefont
  {Heyl}},\ }\bibfield  {title} {\bibinfo {title} {Dynamical quantum phase
  transitions in systems with broken-symmetry phases},\ }\href@noop {}
  {\bibfield  {journal} {\bibinfo  {journal} {Phys. Rev. Lett.}\ }\textbf
  {\bibinfo {volume} {113}},\ \bibinfo {pages} {205701} (\bibinfo {year}
  {2014}{\natexlab{a}})}\BibitemShut {NoStop}%
\bibitem [{\citenamefont {Heyl}(2015{\natexlab{a}})}]{DQPT15}%
  \BibitemOpen
  \bibfield  {author} {\bibinfo {author} {\bibfnamefont {M.}~\bibnamefont
  {Heyl}},\ }\bibfield  {title} {\bibinfo {title} {Scaling and universality at
  dynamical quantum phase transitions},\ }\href@noop {} {\bibfield  {journal}
  {\bibinfo  {journal} {Phys. Rev. Lett.}\ }\textbf {\bibinfo {volume} {115}},\
  \bibinfo {pages} {140602} (\bibinfo {year} {2015}{\natexlab{a}})}\BibitemShut
  {NoStop}%
\bibitem [{\citenamefont {Brandner}\ \emph {et~al.}(2017)\citenamefont
  {Brandner}, \citenamefont {Maisi}, \citenamefont {Pekola}, \citenamefont
  {Garrahan},\ and\ \citenamefont {Flindt}}]{PhysRevLett.118.180601}%
  \BibitemOpen
  \bibfield  {author} {\bibinfo {author} {\bibfnamefont {K.}~\bibnamefont
  {Brandner}}, \bibinfo {author} {\bibfnamefont {V.~F.}\ \bibnamefont {Maisi}},
  \bibinfo {author} {\bibfnamefont {J.~P.}\ \bibnamefont {Pekola}}, \bibinfo
  {author} {\bibfnamefont {J.~P.}\ \bibnamefont {Garrahan}},\ and\ \bibinfo
  {author} {\bibfnamefont {C.}~\bibnamefont {Flindt}},\ }\bibfield  {title}
  {\bibinfo {title} {Experimental determination of dynamical lee-yang zeros},\
  }\href {https://doi.org/10.1103/PhysRevLett.118.180601} {\bibfield  {journal}
  {\bibinfo  {journal} {Phys. Rev. Lett.}\ }\textbf {\bibinfo {volume} {118}},\
  \bibinfo {pages} {180601} (\bibinfo {year} {2017})}\BibitemShut {NoStop}%
\bibitem [{\citenamefont {Yuzbashyan}\ \emph {et~al.}(2006)\citenamefont
  {Yuzbashyan}, \citenamefont {Tsyplyatyev},\ and\ \citenamefont
  {Altshuler}}]{Yuzbashyan200696}%
  \BibitemOpen
  \bibfield  {author} {\bibinfo {author} {\bibfnamefont {E.~A.}\ \bibnamefont
  {Yuzbashyan}}, \bibinfo {author} {\bibfnamefont {O.}~\bibnamefont
  {Tsyplyatyev}},\ and\ \bibinfo {author} {\bibfnamefont {B.~L.}\ \bibnamefont
  {Altshuler}},\ }\bibfield  {title} {\bibinfo {title} {Relaxation and
  persistent oscillations of the order parameter in fermionic condensates},\
  }\href {https://doi.org/10.1103/PhysRevLett.96.097005} {\bibfield  {journal}
  {\bibinfo  {journal} {Phys. Rev. Lett.}\ }\textbf {\bibinfo {volume} {96}},\
  \bibinfo {pages} {097005} (\bibinfo {year} {2006})}\BibitemShut {NoStop}%
\bibitem [{\citenamefont {Barmettler}\ \emph {et~al.}(2009)\citenamefont
  {Barmettler}, \citenamefont {Punk}, \citenamefont {Gritsev}, \citenamefont
  {Demler},\ and\ \citenamefont {Altman}}]{Barmettler2009102}%
  \BibitemOpen
  \bibfield  {author} {\bibinfo {author} {\bibfnamefont {P.}~\bibnamefont
  {Barmettler}}, \bibinfo {author} {\bibfnamefont {M.}~\bibnamefont {Punk}},
  \bibinfo {author} {\bibfnamefont {V.}~\bibnamefont {Gritsev}}, \bibinfo
  {author} {\bibfnamefont {E.}~\bibnamefont {Demler}},\ and\ \bibinfo {author}
  {\bibfnamefont {E.}~\bibnamefont {Altman}},\ }\bibfield  {title} {\bibinfo
  {title} {Relaxation of antiferromagnetic order in spin-$1/2$ chains following
  a quantum quench},\ }\href {https://doi.org/10.1103/PhysRevLett.102.130603}
  {\bibfield  {journal} {\bibinfo  {journal} {Phys. Rev. Lett.}\ }\textbf
  {\bibinfo {volume} {102}},\ \bibinfo {pages} {130603} (\bibinfo {year}
  {2009})}\BibitemShut {NoStop}%
\bibitem [{\citenamefont {Eckstein}\ \emph {et~al.}(2009)\citenamefont
  {Eckstein}, \citenamefont {Kollar},\ and\ \citenamefont
  {Werner}}]{Eckstein2009103}%
  \BibitemOpen
  \bibfield  {author} {\bibinfo {author} {\bibfnamefont {M.}~\bibnamefont
  {Eckstein}}, \bibinfo {author} {\bibfnamefont {M.}~\bibnamefont {Kollar}},\
  and\ \bibinfo {author} {\bibfnamefont {P.}~\bibnamefont {Werner}},\
  }\bibfield  {title} {\bibinfo {title} {Thermalization after an interaction
  quench in the hubbard model},\ }\href
  {https://doi.org/10.1103/PhysRevLett.103.056403} {\bibfield  {journal}
  {\bibinfo  {journal} {Phys. Rev. Lett.}\ }\textbf {\bibinfo {volume} {103}},\
  \bibinfo {pages} {056403} (\bibinfo {year} {2009})}\BibitemShut {NoStop}%
\bibitem [{\citenamefont {Sciolla}\ and\ \citenamefont
  {Biroli}(2010)}]{Sciolla2010105}%
  \BibitemOpen
  \bibfield  {author} {\bibinfo {author} {\bibfnamefont {B.}~\bibnamefont
  {Sciolla}}\ and\ \bibinfo {author} {\bibfnamefont {G.}~\bibnamefont
  {Biroli}},\ }\bibfield  {title} {\bibinfo {title} {Quantum quenches and
  off-equilibrium dynamical transition in the infinite-dimensional bose-hubbard
  model},\ }\href {https://doi.org/10.1103/PhysRevLett.105.220401} {\bibfield
  {journal} {\bibinfo  {journal} {Phys. Rev. Lett.}\ }\textbf {\bibinfo
  {volume} {105}},\ \bibinfo {pages} {220401} (\bibinfo {year}
  {2010})}\BibitemShut {NoStop}%
\bibitem [{\citenamefont {Dziarmaga}(2010)}]{Dziarmaga201059}%
  \BibitemOpen
  \bibfield  {author} {\bibinfo {author} {\bibfnamefont {J.}~\bibnamefont
  {Dziarmaga}},\ }\bibfield  {title} {\bibinfo {title} {Dynamics of a quantum
  phase transition and relaxation to a steady state},\ }\href
  {https://doi.org/10.1080/00018732.2010.514702} {\bibfield  {journal}
  {\bibinfo  {journal} {Advances in Physics}\ }\textbf {\bibinfo {volume}
  {59}},\ \bibinfo {pages} {1063} (\bibinfo {year} {2010})}\BibitemShut
  {NoStop}%
\bibitem [{\citenamefont {Vajna}\ and\ \citenamefont
  {D\'ora}(2014)}]{Vajna201489}%
  \BibitemOpen
  \bibfield  {author} {\bibinfo {author} {\bibfnamefont {S.}~\bibnamefont
  {Vajna}}\ and\ \bibinfo {author} {\bibfnamefont {B.}~\bibnamefont {D\'ora}},\
  }\bibfield  {title} {\bibinfo {title} {Disentangling dynamical phase
  transitions from equilibrium phase transitions},\ }\href
  {https://doi.org/10.1103/PhysRevB.89.161105} {\bibfield  {journal} {\bibinfo
  {journal} {Phys. Rev. B}\ }\textbf {\bibinfo {volume} {89}},\ \bibinfo
  {pages} {161105} (\bibinfo {year} {2014})}\BibitemShut {NoStop}%
\bibitem [{\citenamefont {Divakaran}\ \emph {et~al.}(2016)\citenamefont
  {Divakaran}, \citenamefont {Sharma},\ and\ \citenamefont
  {Dutta}}]{PhysRevE.93.052133}%
  \BibitemOpen
  \bibfield  {author} {\bibinfo {author} {\bibfnamefont {U.}~\bibnamefont
  {Divakaran}}, \bibinfo {author} {\bibfnamefont {S.}~\bibnamefont {Sharma}},\
  and\ \bibinfo {author} {\bibfnamefont {A.}~\bibnamefont {Dutta}},\ }\bibfield
   {title} {\bibinfo {title} {Tuning the presence of dynamical phase
  transitions in a generalized $xy$ spin chain},\ }\href
  {https://doi.org/10.1103/PhysRevE.93.052133} {\bibfield  {journal} {\bibinfo
  {journal} {Phys. Rev. E}\ }\textbf {\bibinfo {volume} {93}},\ \bibinfo
  {pages} {052133} (\bibinfo {year} {2016})}\BibitemShut {NoStop}%
\bibitem [{\citenamefont {Cao}\ \emph {et~al.}(2022)\citenamefont {Cao},
  \citenamefont {Zhong},\ and\ \citenamefont {Tong}}]{Cao202231}%
  \BibitemOpen
  \bibfield  {author} {\bibinfo {author} {\bibfnamefont {K.}~\bibnamefont
  {Cao}}, \bibinfo {author} {\bibfnamefont {M.}~\bibnamefont {Zhong}},\ and\
  \bibinfo {author} {\bibfnamefont {P.}~\bibnamefont {Tong}},\ }\bibfield
  {title} {\bibinfo {title} {Dynamical quantum phase transition in {XY} chains
  with the dzyaloshinskii-moriya and {XZY}{\textendash}{YZX} three-site
  interactions},\ }\href {https://doi.org/10.1088/1674-1056/ac4a6e} {\bibfield
  {journal} {\bibinfo  {journal} {Chinese Physics B}\ }\textbf {\bibinfo
  {volume} {31}},\ \bibinfo {pages} {060505} (\bibinfo {year}
  {2022})}\BibitemShut {NoStop}%
\bibitem [{\citenamefont {Porta}\ \emph {et~al.}(2020)\citenamefont {Porta},
  \citenamefont {Cavaliere}, \citenamefont {Sassetti},\ and\ \citenamefont
  {Ziani}}]{Porta.10.1}%
  \BibitemOpen
  \bibfield  {author} {\bibinfo {author} {\bibfnamefont {S.}~\bibnamefont
  {Porta}}, \bibinfo {author} {\bibfnamefont {F.}~\bibnamefont {Cavaliere}},
  \bibinfo {author} {\bibfnamefont {M.}~\bibnamefont {Sassetti}},\ and\
  \bibinfo {author} {\bibfnamefont {N.~T.}\ \bibnamefont {Ziani}},\ }\bibfield
  {title} {\bibinfo {title} {Topological classification of dynamical quantum
  phase transitions in the xy chain},\ }\bibfield  {journal} {\bibinfo
  {journal} {Scientific Reports}\ }\textbf {\bibinfo {volume} {10}},\ \href
  {https://doi.org/10.1038/s41598-020-69621-8} {10.1038/s41598-020-69621-8}
  (\bibinfo {year} {2020})\BibitemShut {NoStop}%
\bibitem [{\citenamefont {Schmitt}\ and\ \citenamefont
  {Kehrein}(2015)}]{Schmitt201592}%
  \BibitemOpen
  \bibfield  {author} {\bibinfo {author} {\bibfnamefont {M.}~\bibnamefont
  {Schmitt}}\ and\ \bibinfo {author} {\bibfnamefont {S.}~\bibnamefont
  {Kehrein}},\ }\bibfield  {title} {\bibinfo {title} {Dynamical quantum phase
  transitions in the kitaev honeycomb model},\ }\href
  {https://doi.org/10.1103/PhysRevB.92.075114} {\bibfield  {journal} {\bibinfo
  {journal} {Phys. Rev. B}\ }\textbf {\bibinfo {volume} {92}},\ \bibinfo
  {pages} {075114} (\bibinfo {year} {2015})}\BibitemShut {NoStop}%
\bibitem [{\citenamefont {Karrasch}\ and\ \citenamefont
  {Schuricht}(2013)}]{Karrasch201387}%
  \BibitemOpen
  \bibfield  {author} {\bibinfo {author} {\bibfnamefont {C.}~\bibnamefont
  {Karrasch}}\ and\ \bibinfo {author} {\bibfnamefont {D.}~\bibnamefont
  {Schuricht}},\ }\bibfield  {title} {\bibinfo {title} {Dynamical phase
  transitions after quenches in nonintegrable models},\ }\href
  {https://doi.org/10.1103/PhysRevB.87.195104} {\bibfield  {journal} {\bibinfo
  {journal} {Phys. Rev. B}\ }\textbf {\bibinfo {volume} {87}},\ \bibinfo
  {pages} {195104} (\bibinfo {year} {2013})}\BibitemShut {NoStop}%
\bibitem [{\citenamefont {Andraschko}\ and\ \citenamefont
  {Sirker}(2014)}]{Andraschko201489}%
  \BibitemOpen
  \bibfield  {author} {\bibinfo {author} {\bibfnamefont {F.}~\bibnamefont
  {Andraschko}}\ and\ \bibinfo {author} {\bibfnamefont {J.}~\bibnamefont
  {Sirker}},\ }\bibfield  {title} {\bibinfo {title} {Dynamical quantum phase
  transitions and the loschmidt echo: A transfer matrix approach},\ }\href
  {https://doi.org/10.1103/PhysRevB.89.125120} {\bibfield  {journal} {\bibinfo
  {journal} {Phys. Rev. B}\ }\textbf {\bibinfo {volume} {89}},\ \bibinfo
  {pages} {125120} (\bibinfo {year} {2014})}\BibitemShut {NoStop}%
\bibitem [{\citenamefont {Heyl}(2014{\natexlab{b}})}]{Heyl2014113}%
  \BibitemOpen
  \bibfield  {author} {\bibinfo {author} {\bibfnamefont {M.}~\bibnamefont
  {Heyl}},\ }\bibfield  {title} {\bibinfo {title} {Dynamical quantum phase
  transitions in systems with broken-symmetry phases},\ }\href
  {https://doi.org/10.1103/PhysRevLett.113.205701} {\bibfield  {journal}
  {\bibinfo  {journal} {Phys. Rev. Lett.}\ }\textbf {\bibinfo {volume} {113}},\
  \bibinfo {pages} {205701} (\bibinfo {year} {2014}{\natexlab{b}})}\BibitemShut
  {NoStop}%
\bibitem [{\citenamefont {Kriel}\ \emph {et~al.}(2014)\citenamefont {Kriel},
  \citenamefont {Karrasch},\ and\ \citenamefont {Kehrein}}]{Kriel.90.125106}%
  \BibitemOpen
  \bibfield  {author} {\bibinfo {author} {\bibfnamefont {J.~N.}\ \bibnamefont
  {Kriel}}, \bibinfo {author} {\bibfnamefont {C.}~\bibnamefont {Karrasch}},\
  and\ \bibinfo {author} {\bibfnamefont {S.}~\bibnamefont {Kehrein}},\
  }\bibfield  {title} {\bibinfo {title} {Dynamical quantum phase transitions in
  the axial next-nearest-neighbor ising chain},\ }\href
  {https://doi.org/10.1103/PhysRevB.90.125106} {\bibfield  {journal} {\bibinfo
  {journal} {Phys. Rev. B}\ }\textbf {\bibinfo {volume} {90}},\ \bibinfo
  {pages} {125106} (\bibinfo {year} {2014})}\BibitemShut {NoStop}%
\bibitem [{\citenamefont {Sharma}\ \emph {et~al.}(2015)\citenamefont {Sharma},
  \citenamefont {Suzuki},\ and\ \citenamefont {Dutta}}]{Sharma.92.104306}%
  \BibitemOpen
  \bibfield  {author} {\bibinfo {author} {\bibfnamefont {S.}~\bibnamefont
  {Sharma}}, \bibinfo {author} {\bibfnamefont {S.}~\bibnamefont {Suzuki}},\
  and\ \bibinfo {author} {\bibfnamefont {A.}~\bibnamefont {Dutta}},\ }\bibfield
   {title} {\bibinfo {title} {Quenches and dynamical phase transitions in a
  nonintegrable quantum ising model},\ }\href
  {https://doi.org/10.1103/PhysRevB.92.104306} {\bibfield  {journal} {\bibinfo
  {journal} {Phys. Rev. B}\ }\textbf {\bibinfo {volume} {92}},\ \bibinfo
  {pages} {104306} (\bibinfo {year} {2015})}\BibitemShut {NoStop}%
\bibitem [{\citenamefont {Halimeh}\ and\ \citenamefont
  {Zauner-Stauber}(2017)}]{Halimeh201796}%
  \BibitemOpen
  \bibfield  {author} {\bibinfo {author} {\bibfnamefont {J.~C.}\ \bibnamefont
  {Halimeh}}\ and\ \bibinfo {author} {\bibfnamefont {V.}~\bibnamefont
  {Zauner-Stauber}},\ }\bibfield  {title} {\bibinfo {title} {Dynamical phase
  diagram of quantum spin chains with long-range interactions},\ }\href
  {https://doi.org/10.1103/PhysRevB.96.134427} {\bibfield  {journal} {\bibinfo
  {journal} {Phys. Rev. B}\ }\textbf {\bibinfo {volume} {96}},\ \bibinfo
  {pages} {134427} (\bibinfo {year} {2017})}\BibitemShut {NoStop}%
\bibitem [{\citenamefont {Homrighausen}\ \emph {et~al.}(2017)\citenamefont
  {Homrighausen}, \citenamefont {Abeling}, \citenamefont {Zauner-Stauber},\
  and\ \citenamefont {Halimeh}}]{Homrighausen201796}%
  \BibitemOpen
  \bibfield  {author} {\bibinfo {author} {\bibfnamefont {I.}~\bibnamefont
  {Homrighausen}}, \bibinfo {author} {\bibfnamefont {N.~O.}\ \bibnamefont
  {Abeling}}, \bibinfo {author} {\bibfnamefont {V.}~\bibnamefont
  {Zauner-Stauber}},\ and\ \bibinfo {author} {\bibfnamefont {J.~C.}\
  \bibnamefont {Halimeh}},\ }\bibfield  {title} {\bibinfo {title} {Anomalous
  dynamical phase in quantum spin chains with long-range interactions},\ }\href
  {https://doi.org/10.1103/PhysRevB.96.104436} {\bibfield  {journal} {\bibinfo
  {journal} {Phys. Rev. B}\ }\textbf {\bibinfo {volume} {96}},\ \bibinfo
  {pages} {104436} (\bibinfo {year} {2017})}\BibitemShut {NoStop}%
\bibitem [{\citenamefont {Obuchi}\ \emph {et~al.}(2017)\citenamefont {Obuchi},
  \citenamefont {Suzuki},\ and\ \citenamefont
  {Takahashi}}]{PhysRevB.95.174305}%
  \BibitemOpen
  \bibfield  {author} {\bibinfo {author} {\bibfnamefont {T.}~\bibnamefont
  {Obuchi}}, \bibinfo {author} {\bibfnamefont {S.}~\bibnamefont {Suzuki}},\
  and\ \bibinfo {author} {\bibfnamefont {K.}~\bibnamefont {Takahashi}},\
  }\bibfield  {title} {\bibinfo {title} {Complex semiclassical analysis of the
  loschmidt amplitude and dynamical quantum phase transitions},\ }\href
  {https://doi.org/10.1103/PhysRevB.95.174305} {\bibfield  {journal} {\bibinfo
  {journal} {Phys. Rev. B}\ }\textbf {\bibinfo {volume} {95}},\ \bibinfo
  {pages} {174305} (\bibinfo {year} {2017})}\BibitemShut {NoStop}%
\bibitem [{\citenamefont {Zauner-Stauber}\ and\ \citenamefont
  {Halimeh}(2017)}]{PhysRevE.96.062118}%
  \BibitemOpen
  \bibfield  {author} {\bibinfo {author} {\bibfnamefont {V.}~\bibnamefont
  {Zauner-Stauber}}\ and\ \bibinfo {author} {\bibfnamefont {J.~C.}\
  \bibnamefont {Halimeh}},\ }\bibfield  {title} {\bibinfo {title} {Probing the
  anomalous dynamical phase in long-range quantum spin chains through
  fisher-zero lines},\ }\href {https://doi.org/10.1103/PhysRevE.96.062118}
  {\bibfield  {journal} {\bibinfo  {journal} {Phys. Rev. E}\ }\textbf {\bibinfo
  {volume} {96}},\ \bibinfo {pages} {062118} (\bibinfo {year}
  {2017})}\BibitemShut {NoStop}%
\bibitem [{\citenamefont {Dutta}\ and\ \citenamefont
  {Dutta}(2017)}]{Dutta201796}%
  \BibitemOpen
  \bibfield  {author} {\bibinfo {author} {\bibfnamefont {A.}~\bibnamefont
  {Dutta}}\ and\ \bibinfo {author} {\bibfnamefont {A.}~\bibnamefont {Dutta}},\
  }\bibfield  {title} {\bibinfo {title} {Probing the role of long-range
  interactions in the dynamics of a long-range kitaev chain},\ }\href
  {https://doi.org/10.1103/PhysRevB.96.125113} {\bibfield  {journal} {\bibinfo
  {journal} {Phys. Rev. B}\ }\textbf {\bibinfo {volume} {96}},\ \bibinfo
  {pages} {125113} (\bibinfo {year} {2017})}\BibitemShut {NoStop}%
\bibitem [{\citenamefont {\ifmmode \check{Z}\else
  \v{Z}\fi{}unkovi\ifmmode~\check{c}\else \v{c}\fi{}}\ \emph
  {et~al.}(2018)\citenamefont {\ifmmode \check{Z}\else
  \v{Z}\fi{}unkovi\ifmmode~\check{c}\else \v{c}\fi{}}, \citenamefont {Heyl},
  \citenamefont {Knap},\ and\ \citenamefont {Silva}}]{Bojan2018120}%
  \BibitemOpen
  \bibfield  {author} {\bibinfo {author} {\bibfnamefont {B.}~\bibnamefont
  {\ifmmode \check{Z}\else \v{Z}\fi{}unkovi\ifmmode~\check{c}\else
  \v{c}\fi{}}}, \bibinfo {author} {\bibfnamefont {M.}~\bibnamefont {Heyl}},
  \bibinfo {author} {\bibfnamefont {M.}~\bibnamefont {Knap}},\ and\ \bibinfo
  {author} {\bibfnamefont {A.}~\bibnamefont {Silva}},\ }\bibfield  {title}
  {\bibinfo {title} {Dynamical quantum phase transitions in spin chains with
  long-range interactions: Merging different concepts of nonequilibrium
  criticality},\ }\href {https://doi.org/10.1103/PhysRevLett.120.130601}
  {\bibfield  {journal} {\bibinfo  {journal} {Phys. Rev. Lett.}\ }\textbf
  {\bibinfo {volume} {120}},\ \bibinfo {pages} {130601} (\bibinfo {year}
  {2018})}\BibitemShut {NoStop}%
\bibitem [{\citenamefont {Halimeh}\ \emph {et~al.}(2020)\citenamefont
  {Halimeh}, \citenamefont {Van~Damme}, \citenamefont {Zauner-Stauber},\ and\
  \citenamefont {Vanderstraeten}}]{Halimeh.2.033111}%
  \BibitemOpen
  \bibfield  {author} {\bibinfo {author} {\bibfnamefont {J.~C.}\ \bibnamefont
  {Halimeh}}, \bibinfo {author} {\bibfnamefont {M.}~\bibnamefont {Van~Damme}},
  \bibinfo {author} {\bibfnamefont {V.}~\bibnamefont {Zauner-Stauber}},\ and\
  \bibinfo {author} {\bibfnamefont {L.}~\bibnamefont {Vanderstraeten}},\
  }\bibfield  {title} {\bibinfo {title} {Quasiparticle origin of dynamical
  quantum phase transitions},\ }\href
  {https://doi.org/10.1103/PhysRevResearch.2.033111} {\bibfield  {journal}
  {\bibinfo  {journal} {Phys. Rev. Res.}\ }\textbf {\bibinfo {volume} {2}},\
  \bibinfo {pages} {033111} (\bibinfo {year} {2020})}\BibitemShut {NoStop}%
\bibitem [{\citenamefont {Karrasch}\ and\ \citenamefont
  {Schuricht}(2017)}]{PhysRevB.95.075143}%
  \BibitemOpen
  \bibfield  {author} {\bibinfo {author} {\bibfnamefont {C.}~\bibnamefont
  {Karrasch}}\ and\ \bibinfo {author} {\bibfnamefont {D.}~\bibnamefont
  {Schuricht}},\ }\bibfield  {title} {\bibinfo {title} {Dynamical quantum phase
  transitions in the quantum potts chain},\ }\href
  {https://doi.org/10.1103/PhysRevB.95.075143} {\bibfield  {journal} {\bibinfo
  {journal} {Phys. Rev. B}\ }\textbf {\bibinfo {volume} {95}},\ \bibinfo
  {pages} {075143} (\bibinfo {year} {2017})}\BibitemShut {NoStop}%
\bibitem [{\citenamefont {Zhou}\ \emph {et~al.}(2018)\citenamefont {Zhou},
  \citenamefont {Wang}, \citenamefont {Wang},\ and\ \citenamefont
  {Gong}}]{Zhou201898}%
  \BibitemOpen
  \bibfield  {author} {\bibinfo {author} {\bibfnamefont {L.}~\bibnamefont
  {Zhou}}, \bibinfo {author} {\bibfnamefont {Q.-h.}\ \bibnamefont {Wang}},
  \bibinfo {author} {\bibfnamefont {H.}~\bibnamefont {Wang}},\ and\ \bibinfo
  {author} {\bibfnamefont {J.}~\bibnamefont {Gong}},\ }\bibfield  {title}
  {\bibinfo {title} {Dynamical quantum phase transitions in non-hermitian
  lattices},\ }\href {https://doi.org/10.1103/PhysRevA.98.022129} {\bibfield
  {journal} {\bibinfo  {journal} {Phys. Rev. A}\ }\textbf {\bibinfo {volume}
  {98}},\ \bibinfo {pages} {022129} (\bibinfo {year} {2018})}\BibitemShut
  {NoStop}%
\bibitem [{\citenamefont {Mondal}\ and\ \citenamefont
  {Nag}(2022)}]{Mondal.2022.106}%
  \BibitemOpen
  \bibfield  {author} {\bibinfo {author} {\bibfnamefont {D.}~\bibnamefont
  {Mondal}}\ and\ \bibinfo {author} {\bibfnamefont {T.}~\bibnamefont {Nag}},\
  }\bibfield  {title} {\bibinfo {title} {Anomaly in the dynamical quantum phase
  transition in a non-hermitian system with extended gapless phases},\ }\href
  {https://doi.org/10.1103/PhysRevB.106.054308} {\bibfield  {journal} {\bibinfo
   {journal} {Phys. Rev. B}\ }\textbf {\bibinfo {volume} {106}},\ \bibinfo
  {pages} {054308} (\bibinfo {year} {2022})}\BibitemShut {NoStop}%
\bibitem [{\citenamefont {Mondal}\ and\ \citenamefont
  {Nag}(2023)}]{Mondal.2023.107}%
  \BibitemOpen
  \bibfield  {author} {\bibinfo {author} {\bibfnamefont {D.}~\bibnamefont
  {Mondal}}\ and\ \bibinfo {author} {\bibfnamefont {T.}~\bibnamefont {Nag}},\
  }\bibfield  {title} {\bibinfo {title} {Finite-temperature dynamical quantum
  phase transition in a non-hermitian system},\ }\href
  {https://doi.org/10.1103/PhysRevB.107.184311} {\bibfield  {journal} {\bibinfo
   {journal} {Phys. Rev. B}\ }\textbf {\bibinfo {volume} {107}},\ \bibinfo
  {pages} {184311} (\bibinfo {year} {2023})}\BibitemShut {NoStop}%
\bibitem [{\citenamefont {Abdi}(2019)}]{Mehdi2019100}%
  \BibitemOpen
  \bibfield  {author} {\bibinfo {author} {\bibfnamefont {M.}~\bibnamefont
  {Abdi}},\ }\bibfield  {title} {\bibinfo {title} {Dynamical quantum phase
  transition in bose-einstein condensates},\ }\href
  {https://doi.org/10.1103/PhysRevB.100.184310} {\bibfield  {journal} {\bibinfo
   {journal} {Phys. Rev. B}\ }\textbf {\bibinfo {volume} {100}},\ \bibinfo
  {pages} {184310} (\bibinfo {year} {2019})}\BibitemShut {NoStop}%
\bibitem [{\citenamefont {Yang}\ \emph {et~al.}(2017)\citenamefont {Yang},
  \citenamefont {Wang}, \citenamefont {Wang}, \citenamefont {Gao},\ and\
  \citenamefont {Chen}}]{Yang201796}%
  \BibitemOpen
  \bibfield  {author} {\bibinfo {author} {\bibfnamefont {C.}~\bibnamefont
  {Yang}}, \bibinfo {author} {\bibfnamefont {Y.}~\bibnamefont {Wang}}, \bibinfo
  {author} {\bibfnamefont {P.}~\bibnamefont {Wang}}, \bibinfo {author}
  {\bibfnamefont {X.}~\bibnamefont {Gao}},\ and\ \bibinfo {author}
  {\bibfnamefont {S.}~\bibnamefont {Chen}},\ }\bibfield  {title} {\bibinfo
  {title} {Dynamical signature of localization-delocalization transition in a
  one-dimensional incommensurate lattice},\ }\href
  {https://doi.org/10.1103/PhysRevB.95.184201} {\bibfield  {journal} {\bibinfo
  {journal} {Phys. Rev. B}\ }\textbf {\bibinfo {volume} {95}},\ \bibinfo
  {pages} {184201} (\bibinfo {year} {2017})}\BibitemShut {NoStop}%
\bibitem [{\citenamefont {Yin}\ \emph {et~al.}(2018)\citenamefont {Yin},
  \citenamefont {Chen}, \citenamefont {Gao},\ and\ \citenamefont
  {Wang}}]{PhysRevA.97.033624}%
  \BibitemOpen
  \bibfield  {author} {\bibinfo {author} {\bibfnamefont {H.}~\bibnamefont
  {Yin}}, \bibinfo {author} {\bibfnamefont {S.}~\bibnamefont {Chen}}, \bibinfo
  {author} {\bibfnamefont {X.}~\bibnamefont {Gao}},\ and\ \bibinfo {author}
  {\bibfnamefont {P.}~\bibnamefont {Wang}},\ }\bibfield  {title} {\bibinfo
  {title} {Zeros of loschmidt echo in the presence of anderson localization},\
  }\href {https://doi.org/10.1103/PhysRevA.97.033624} {\bibfield  {journal}
  {\bibinfo  {journal} {Phys. Rev. A}\ }\textbf {\bibinfo {volume} {97}},\
  \bibinfo {pages} {033624} (\bibinfo {year} {2018})}\BibitemShut {NoStop}%
\bibitem [{\citenamefont {Mendl}\ and\ \citenamefont
  {Budich}(2019)}]{Mendl2019100}%
  \BibitemOpen
  \bibfield  {author} {\bibinfo {author} {\bibfnamefont {C.~B.}\ \bibnamefont
  {Mendl}}\ and\ \bibinfo {author} {\bibfnamefont {J.~C.}\ \bibnamefont
  {Budich}},\ }\bibfield  {title} {\bibinfo {title} {Stability of dynamical
  quantum phase transitions in quenched topological insulators: From multiband
  to disordered systems},\ }\href {https://doi.org/10.1103/PhysRevB.100.224307}
  {\bibfield  {journal} {\bibinfo  {journal} {Phys. Rev. B}\ }\textbf {\bibinfo
  {volume} {100}},\ \bibinfo {pages} {224307} (\bibinfo {year}
  {2019})}\BibitemShut {NoStop}%
\bibitem [{\citenamefont {Cao}\ \emph {et~al.}(2020)\citenamefont {Cao},
  \citenamefont {Li}, \citenamefont {Zhong},\ and\ \citenamefont
  {Tong}}]{Cao2020102}%
  \BibitemOpen
  \bibfield  {author} {\bibinfo {author} {\bibfnamefont {K.}~\bibnamefont
  {Cao}}, \bibinfo {author} {\bibfnamefont {W.}~\bibnamefont {Li}}, \bibinfo
  {author} {\bibfnamefont {M.}~\bibnamefont {Zhong}},\ and\ \bibinfo {author}
  {\bibfnamefont {P.}~\bibnamefont {Tong}},\ }\bibfield  {title} {\bibinfo
  {title} {Influence of weak disorder on the dynamical quantum phase
  transitions in the anisotropic xy chain},\ }\href
  {https://doi.org/10.1103/PhysRevB.102.014207} {\bibfield  {journal} {\bibinfo
   {journal} {Phys. Rev. B}\ }\textbf {\bibinfo {volume} {102}},\ \bibinfo
  {pages} {014207} (\bibinfo {year} {2020})}\BibitemShut {NoStop}%
\bibitem [{\citenamefont {Modak}\ and\ \citenamefont
  {Rakshit}(2021)}]{Modak2021103}%
  \BibitemOpen
  \bibfield  {author} {\bibinfo {author} {\bibfnamefont {R.}~\bibnamefont
  {Modak}}\ and\ \bibinfo {author} {\bibfnamefont {D.}~\bibnamefont
  {Rakshit}},\ }\bibfield  {title} {\bibinfo {title} {Many-body dynamical phase
  transition in a quasiperiodic potential},\ }\href
  {https://doi.org/10.1103/PhysRevB.103.224310} {\bibfield  {journal} {\bibinfo
   {journal} {Phys. Rev. B}\ }\textbf {\bibinfo {volume} {103}},\ \bibinfo
  {pages} {224310} (\bibinfo {year} {2021})}\BibitemShut {NoStop}%
\bibitem [{\citenamefont {Kuliashov}\ \emph {et~al.}(2023)\citenamefont
  {Kuliashov}, \citenamefont {Markov},\ and\ \citenamefont
  {Rubtsov}}]{Kuliashov.107.094304}%
  \BibitemOpen
  \bibfield  {author} {\bibinfo {author} {\bibfnamefont {O.~N.}\ \bibnamefont
  {Kuliashov}}, \bibinfo {author} {\bibfnamefont {A.~A.}\ \bibnamefont
  {Markov}},\ and\ \bibinfo {author} {\bibfnamefont {A.~N.}\ \bibnamefont
  {Rubtsov}},\ }\bibfield  {title} {\bibinfo {title} {Dynamical quantum phase
  transition without an order parameter},\ }\href
  {https://doi.org/10.1103/PhysRevB.107.094304} {\bibfield  {journal} {\bibinfo
   {journal} {Phys. Rev. B}\ }\textbf {\bibinfo {volume} {107}},\ \bibinfo
  {pages} {094304} (\bibinfo {year} {2023})}\BibitemShut {NoStop}%
\bibitem [{\citenamefont {Mishra}\ \emph {et~al.}(2020)\citenamefont {Mishra},
  \citenamefont {Jafari},\ and\ \citenamefont {Akbari}}]{Mishra.53.375301}%
  \BibitemOpen
  \bibfield  {author} {\bibinfo {author} {\bibfnamefont {U.}~\bibnamefont
  {Mishra}}, \bibinfo {author} {\bibfnamefont {R.}~\bibnamefont {Jafari}},\
  and\ \bibinfo {author} {\bibfnamefont {A.}~\bibnamefont {Akbari}},\
  }\bibfield  {title} {\bibinfo {title} {Disordered kitaev chain with
  long-range pairing: Loschmidt echo revivals and dynamical phase
  transitions},\ }\href {https://doi.org/10.1088/1751-8121/ab97de} {\bibfield
  {journal} {\bibinfo  {journal} {Journal of Physics A: Mathematical and
  Theoretical}\ }\textbf {\bibinfo {volume} {53}},\ \bibinfo {pages} {375301}
  (\bibinfo {year} {2020})}\BibitemShut {NoStop}%
\bibitem [{\citenamefont {Yang}\ \emph
  {et~al.}(2019{\natexlab{a}})\citenamefont {Yang}, \citenamefont {Zhou},
  \citenamefont {Ma}, \citenamefont {Kong}, \citenamefont {Wang}, \citenamefont
  {Qin}, \citenamefont {Rong}, \citenamefont {Wang}, \citenamefont {Shi},
  \citenamefont {Gong},\ and\ \citenamefont {Du}}]{Yang2019100}%
  \BibitemOpen
  \bibfield  {author} {\bibinfo {author} {\bibfnamefont {K.}~\bibnamefont
  {Yang}}, \bibinfo {author} {\bibfnamefont {L.}~\bibnamefont {Zhou}}, \bibinfo
  {author} {\bibfnamefont {W.}~\bibnamefont {Ma}}, \bibinfo {author}
  {\bibfnamefont {X.}~\bibnamefont {Kong}}, \bibinfo {author} {\bibfnamefont
  {P.}~\bibnamefont {Wang}}, \bibinfo {author} {\bibfnamefont {X.}~\bibnamefont
  {Qin}}, \bibinfo {author} {\bibfnamefont {X.}~\bibnamefont {Rong}}, \bibinfo
  {author} {\bibfnamefont {Y.}~\bibnamefont {Wang}}, \bibinfo {author}
  {\bibfnamefont {F.}~\bibnamefont {Shi}}, \bibinfo {author} {\bibfnamefont
  {J.}~\bibnamefont {Gong}},\ and\ \bibinfo {author} {\bibfnamefont
  {J.}~\bibnamefont {Du}},\ }\bibfield  {title} {\bibinfo {title} {Floquet
  dynamical quantum phase transitions},\ }\href
  {https://doi.org/10.1103/PhysRevB.100.085308} {\bibfield  {journal} {\bibinfo
   {journal} {Phys. Rev. B}\ }\textbf {\bibinfo {volume} {100}},\ \bibinfo
  {pages} {085308} (\bibinfo {year} {2019}{\natexlab{a}})}\BibitemShut
  {NoStop}%
\bibitem [{\citenamefont {Zamani}\ \emph {et~al.}(2020)\citenamefont {Zamani},
  \citenamefont {Jafari},\ and\ \citenamefont {Langari}}]{Zamani2020102}%
  \BibitemOpen
  \bibfield  {author} {\bibinfo {author} {\bibfnamefont {S.}~\bibnamefont
  {Zamani}}, \bibinfo {author} {\bibfnamefont {R.}~\bibnamefont {Jafari}},\
  and\ \bibinfo {author} {\bibfnamefont {A.}~\bibnamefont {Langari}},\
  }\bibfield  {title} {\bibinfo {title} {Floquet dynamical quantum phase
  transition in the extended xy model: Nonadiabatic to adiabatic topological
  transition},\ }\href {https://doi.org/10.1103/PhysRevB.102.144306} {\bibfield
   {journal} {\bibinfo  {journal} {Phys. Rev. B}\ }\textbf {\bibinfo {volume}
  {102}},\ \bibinfo {pages} {144306} (\bibinfo {year} {2020})}\BibitemShut
  {NoStop}%
\bibitem [{\citenamefont {Shirai}\ \emph {et~al.}(2020)\citenamefont {Shirai},
  \citenamefont {Todo},\ and\ \citenamefont {Miyashita}}]{Shirai.101.013809}%
  \BibitemOpen
  \bibfield  {author} {\bibinfo {author} {\bibfnamefont {T.}~\bibnamefont
  {Shirai}}, \bibinfo {author} {\bibfnamefont {S.}~\bibnamefont {Todo}},\ and\
  \bibinfo {author} {\bibfnamefont {S.}~\bibnamefont {Miyashita}},\ }\bibfield
  {title} {\bibinfo {title} {Dynamical phase transition in floquet optical
  bistable systems: An approach from finite-size quantum systems},\ }\href
  {https://doi.org/10.1103/PhysRevA.101.013809} {\bibfield  {journal} {\bibinfo
   {journal} {Phys. Rev. A}\ }\textbf {\bibinfo {volume} {101}},\ \bibinfo
  {pages} {013809} (\bibinfo {year} {2020})}\BibitemShut {NoStop}%
\bibitem [{\citenamefont {Zhou}\ and\ \citenamefont {Du}(2021)}]{Zhou202133}%
  \BibitemOpen
  \bibfield  {author} {\bibinfo {author} {\bibfnamefont {L.}~\bibnamefont
  {Zhou}}\ and\ \bibinfo {author} {\bibfnamefont {Q.}~\bibnamefont {Du}},\
  }\bibfield  {title} {\bibinfo {title} {Floquet dynamical quantum phase
  transitions in periodically quenched systems},\ }\href
  {https://doi.org/10.1088/1361-648x/ac0b60} {\bibfield  {journal} {\bibinfo
  {journal} {Journal of Physics: Condensed Matter}\ }\textbf {\bibinfo {volume}
  {33}},\ \bibinfo {pages} {345403} (\bibinfo {year} {2021})}\BibitemShut
  {NoStop}%
\bibitem [{\citenamefont {Jafari}\ and\ \citenamefont
  {Akbari}(2021)}]{Jafari2021103}%
  \BibitemOpen
  \bibfield  {author} {\bibinfo {author} {\bibfnamefont {R.}~\bibnamefont
  {Jafari}}\ and\ \bibinfo {author} {\bibfnamefont {A.}~\bibnamefont
  {Akbari}},\ }\bibfield  {title} {\bibinfo {title} {Floquet dynamical phase
  transition and entanglement spectrum},\ }\href
  {https://doi.org/10.1103/PhysRevA.103.012204} {\bibfield  {journal} {\bibinfo
   {journal} {Phys. Rev. A}\ }\textbf {\bibinfo {volume} {103}},\ \bibinfo
  {pages} {012204} (\bibinfo {year} {2021})}\BibitemShut {NoStop}%
\bibitem [{\citenamefont {Hamazaki}(2021)}]{NC.12.5108}%
  \BibitemOpen
  \bibfield  {author} {\bibinfo {author} {\bibfnamefont {R.}~\bibnamefont
  {Hamazaki}},\ }\bibfield  {title} {\bibinfo {title} {Exceptional dynamical
  quantum phase transitions in periodically driven systems},\ }\href
  {https://doi.org/10.1038/s41467-021-25355-3} {\bibfield  {journal} {\bibinfo
  {journal} {Nature Communications}\ }\textbf {\bibinfo {volume} {12}},\
  \bibinfo {pages} {5108} (\bibinfo {year} {2021})}\BibitemShut {NoStop}%
\bibitem [{\citenamefont {Zamani}\ \emph {et~al.}(2022)\citenamefont {Zamani},
  \citenamefont {Jafari},\ and\ \citenamefont {Langari}}]{PhysRevB.105.094304}%
  \BibitemOpen
  \bibfield  {author} {\bibinfo {author} {\bibfnamefont {S.}~\bibnamefont
  {Zamani}}, \bibinfo {author} {\bibfnamefont {R.}~\bibnamefont {Jafari}},\
  and\ \bibinfo {author} {\bibfnamefont {A.}~\bibnamefont {Langari}},\
  }\bibfield  {title} {\bibinfo {title} {Out-of-time-order correlations and
  floquet dynamical quantum phase transition},\ }\href
  {https://doi.org/10.1103/PhysRevB.105.094304} {\bibfield  {journal} {\bibinfo
   {journal} {Phys. Rev. B}\ }\textbf {\bibinfo {volume} {105}},\ \bibinfo
  {pages} {094304} (\bibinfo {year} {2022})}\BibitemShut {NoStop}%
\bibitem [{\citenamefont {Jafari}\ \emph {et~al.}(2022)\citenamefont {Jafari},
  \citenamefont {Akbari}, \citenamefont {Mishra},\ and\ \citenamefont
  {Johannesson}}]{Jafari2022105}%
  \BibitemOpen
  \bibfield  {author} {\bibinfo {author} {\bibfnamefont {R.}~\bibnamefont
  {Jafari}}, \bibinfo {author} {\bibfnamefont {A.}~\bibnamefont {Akbari}},
  \bibinfo {author} {\bibfnamefont {U.}~\bibnamefont {Mishra}},\ and\ \bibinfo
  {author} {\bibfnamefont {H.}~\bibnamefont {Johannesson}},\ }\bibfield
  {title} {\bibinfo {title} {Floquet dynamical quantum phase transitions under
  synchronized periodic driving},\ }\href
  {https://doi.org/10.1103/PhysRevB.105.094311} {\bibfield  {journal} {\bibinfo
   {journal} {Phys. Rev. B}\ }\textbf {\bibinfo {volume} {105}},\ \bibinfo
  {pages} {094311} (\bibinfo {year} {2022})}\BibitemShut {NoStop}%
\bibitem [{\citenamefont {Bhattacharya}\ \emph {et~al.}(2017)\citenamefont
  {Bhattacharya}, \citenamefont {Bandyopadhyay},\ and\ \citenamefont
  {Dutta}}]{Bhattacharya.96.180303}%
  \BibitemOpen
  \bibfield  {author} {\bibinfo {author} {\bibfnamefont {U.}~\bibnamefont
  {Bhattacharya}}, \bibinfo {author} {\bibfnamefont {S.}~\bibnamefont
  {Bandyopadhyay}},\ and\ \bibinfo {author} {\bibfnamefont {A.}~\bibnamefont
  {Dutta}},\ }\bibfield  {title} {\bibinfo {title} {Mixed state dynamical
  quantum phase transitions},\ }\href
  {https://doi.org/10.1103/PhysRevB.96.180303} {\bibfield  {journal} {\bibinfo
  {journal} {Phys. Rev. B}\ }\textbf {\bibinfo {volume} {96}},\ \bibinfo
  {pages} {180303} (\bibinfo {year} {2017})}\BibitemShut {NoStop}%
\bibitem [{\citenamefont {Heyl}\ and\ \citenamefont
  {Budich}(2017{\natexlab{a}})}]{Heyl.96.180304}%
  \BibitemOpen
  \bibfield  {author} {\bibinfo {author} {\bibfnamefont {M.}~\bibnamefont
  {Heyl}}\ and\ \bibinfo {author} {\bibfnamefont {J.~C.}\ \bibnamefont
  {Budich}},\ }\bibfield  {title} {\bibinfo {title} {Dynamical topological
  quantum phase transitions for mixed states},\ }\href
  {https://doi.org/10.1103/PhysRevB.96.180304} {\bibfield  {journal} {\bibinfo
  {journal} {Phys. Rev. B}\ }\textbf {\bibinfo {volume} {96}},\ \bibinfo
  {pages} {180304} (\bibinfo {year} {2017}{\natexlab{a}})}\BibitemShut
  {NoStop}%
\bibitem [{\citenamefont {Lang}\ \emph
  {et~al.}(2018{\natexlab{a}})\citenamefont {Lang}, \citenamefont {Chen},
  \citenamefont {Hong},\ and\ \citenamefont {Fan}}]{Lang.98.134310}%
  \BibitemOpen
  \bibfield  {author} {\bibinfo {author} {\bibfnamefont {H.}~\bibnamefont
  {Lang}}, \bibinfo {author} {\bibfnamefont {Y.}~\bibnamefont {Chen}}, \bibinfo
  {author} {\bibfnamefont {Q.}~\bibnamefont {Hong}},\ and\ \bibinfo {author}
  {\bibfnamefont {H.}~\bibnamefont {Fan}},\ }\bibfield  {title} {\bibinfo
  {title} {Dynamical quantum phase transition for mixed states in open
  systems},\ }\href {https://doi.org/10.1103/PhysRevB.98.134310} {\bibfield
  {journal} {\bibinfo  {journal} {Phys. Rev. B}\ }\textbf {\bibinfo {volume}
  {98}},\ \bibinfo {pages} {134310} (\bibinfo {year}
  {2018}{\natexlab{a}})}\BibitemShut {NoStop}%
\bibitem [{\citenamefont {Bandyopadhyay}\ \emph {et~al.}(2018)\citenamefont
  {Bandyopadhyay}, \citenamefont {Laha}, \citenamefont {Bhattacharya},\ and\
  \citenamefont {Dutta}}]{Bandyopadhyay.8}%
  \BibitemOpen
  \bibfield  {author} {\bibinfo {author} {\bibfnamefont {S.}~\bibnamefont
  {Bandyopadhyay}}, \bibinfo {author} {\bibfnamefont {S.}~\bibnamefont {Laha}},
  \bibinfo {author} {\bibfnamefont {U.}~\bibnamefont {Bhattacharya}},\ and\
  \bibinfo {author} {\bibfnamefont {A.}~\bibnamefont {Dutta}},\ }\bibfield
  {title} {\bibinfo {title} {Exploring the possibilities of dynamical quantum
  phase transitions in the presence of a markovian bath},\ }\bibfield
  {journal} {\bibinfo  {journal} {Scientific Reports}\ }\textbf {\bibinfo
  {volume} {8}},\ \href {https://doi.org/10.1038/s41598-018-30377-x}
  {10.1038/s41598-018-30377-x} (\bibinfo {year} {2018})\BibitemShut {NoStop}%
\bibitem [{\citenamefont {Hou}\ \emph {et~al.}(2020)\citenamefont {Hou},
  \citenamefont {Gao}, \citenamefont {Guo}, \citenamefont {He}, \citenamefont
  {Liu},\ and\ \citenamefont {Chien}}]{Hou.102.104305}%
  \BibitemOpen
  \bibfield  {author} {\bibinfo {author} {\bibfnamefont {X.-Y.}\ \bibnamefont
  {Hou}}, \bibinfo {author} {\bibfnamefont {Q.-C.}\ \bibnamefont {Gao}},
  \bibinfo {author} {\bibfnamefont {H.}~\bibnamefont {Guo}}, \bibinfo {author}
  {\bibfnamefont {Y.}~\bibnamefont {He}}, \bibinfo {author} {\bibfnamefont
  {T.}~\bibnamefont {Liu}},\ and\ \bibinfo {author} {\bibfnamefont {C.-C.}\
  \bibnamefont {Chien}},\ }\bibfield  {title} {\bibinfo {title} {Ubiquity of
  zeros of the loschmidt amplitude for mixed states in different physical
  processes and its implication},\ }\href
  {https://doi.org/10.1103/PhysRevB.102.104305} {\bibfield  {journal} {\bibinfo
   {journal} {Phys. Rev. B}\ }\textbf {\bibinfo {volume} {102}},\ \bibinfo
  {pages} {104305} (\bibinfo {year} {2020})}\BibitemShut {NoStop}%
\bibitem [{\citenamefont {Kyaw}\ \emph {et~al.}(2020)\citenamefont {Kyaw},
  \citenamefont {Bastidas}, \citenamefont {Tangpanitanon}, \citenamefont
  {Romero},\ and\ \citenamefont {Kwek}}]{Kyaw.101.012111}%
  \BibitemOpen
  \bibfield  {author} {\bibinfo {author} {\bibfnamefont {T.~H.}\ \bibnamefont
  {Kyaw}}, \bibinfo {author} {\bibfnamefont {V.~M.}\ \bibnamefont {Bastidas}},
  \bibinfo {author} {\bibfnamefont {J.}~\bibnamefont {Tangpanitanon}}, \bibinfo
  {author} {\bibfnamefont {G.}~\bibnamefont {Romero}},\ and\ \bibinfo {author}
  {\bibfnamefont {L.-C.}\ \bibnamefont {Kwek}},\ }\bibfield  {title} {\bibinfo
  {title} {Dynamical quantum phase transitions and non-markovian dynamics},\
  }\href {https://doi.org/10.1103/PhysRevA.101.012111} {\bibfield  {journal}
  {\bibinfo  {journal} {Phys. Rev. A}\ }\textbf {\bibinfo {volume} {101}},\
  \bibinfo {pages} {012111} (\bibinfo {year} {2020})}\BibitemShut {NoStop}%
\bibitem [{\citenamefont {Mera}\ \emph {et~al.}(2018)\citenamefont {Mera},
  \citenamefont {Vlachou}, \citenamefont {Paunkovi\ifmmode~\acute{c}\else
  \'{c}\fi{}}, \citenamefont {Vieira},\ and\ \citenamefont
  {Viyuela}}]{Mera.97.094110}%
  \BibitemOpen
  \bibfield  {author} {\bibinfo {author} {\bibfnamefont {B.}~\bibnamefont
  {Mera}}, \bibinfo {author} {\bibfnamefont {C.}~\bibnamefont {Vlachou}},
  \bibinfo {author} {\bibfnamefont {N.}~\bibnamefont
  {Paunkovi\ifmmode~\acute{c}\else \'{c}\fi{}}}, \bibinfo {author}
  {\bibfnamefont {V.~R.}\ \bibnamefont {Vieira}},\ and\ \bibinfo {author}
  {\bibfnamefont {O.}~\bibnamefont {Viyuela}},\ }\bibfield  {title} {\bibinfo
  {title} {Dynamical phase transitions at finite temperature from fidelity and
  interferometric loschmidt echo induced metrics},\ }\href
  {https://doi.org/10.1103/PhysRevB.97.094110} {\bibfield  {journal} {\bibinfo
  {journal} {Phys. Rev. B}\ }\textbf {\bibinfo {volume} {97}},\ \bibinfo
  {pages} {094110} (\bibinfo {year} {2018})}\BibitemShut {NoStop}%
\bibitem [{\citenamefont {Sedlmayr}\ \emph {et~al.}(2018)\citenamefont
  {Sedlmayr}, \citenamefont {Fleischhauer},\ and\ \citenamefont
  {Sirker}}]{Sedlmayr.97.045147}%
  \BibitemOpen
  \bibfield  {author} {\bibinfo {author} {\bibfnamefont {N.}~\bibnamefont
  {Sedlmayr}}, \bibinfo {author} {\bibfnamefont {M.}~\bibnamefont
  {Fleischhauer}},\ and\ \bibinfo {author} {\bibfnamefont {J.}~\bibnamefont
  {Sirker}},\ }\bibfield  {title} {\bibinfo {title} {Fate of dynamical phase
  transitions at finite temperatures and in open systems},\ }\href
  {https://doi.org/10.1103/PhysRevB.97.045147} {\bibfield  {journal} {\bibinfo
  {journal} {Phys. Rev. B}\ }\textbf {\bibinfo {volume} {97}},\ \bibinfo
  {pages} {045147} (\bibinfo {year} {2018})}\BibitemShut {NoStop}%
\bibitem [{\citenamefont {Link}\ and\ \citenamefont
  {Strunz}(2020)}]{Link.125.143602}%
  \BibitemOpen
  \bibfield  {author} {\bibinfo {author} {\bibfnamefont {V.}~\bibnamefont
  {Link}}\ and\ \bibinfo {author} {\bibfnamefont {W.~T.}\ \bibnamefont
  {Strunz}},\ }\bibfield  {title} {\bibinfo {title} {Dynamical phase
  transitions in dissipative quantum dynamics with quantum optical
  realization},\ }\href {https://doi.org/10.1103/PhysRevLett.125.143602}
  {\bibfield  {journal} {\bibinfo  {journal} {Phys. Rev. Lett.}\ }\textbf
  {\bibinfo {volume} {125}},\ \bibinfo {pages} {143602} (\bibinfo {year}
  {2020})}\BibitemShut {NoStop}%
\bibitem [{\citenamefont {Hou}\ \emph {et~al.}(2021)\citenamefont {Hou},
  \citenamefont {Guo},\ and\ \citenamefont {Chien}}]{Hou.104.023303}%
  \BibitemOpen
  \bibfield  {author} {\bibinfo {author} {\bibfnamefont {X.-Y.}\ \bibnamefont
  {Hou}}, \bibinfo {author} {\bibfnamefont {H.}~\bibnamefont {Guo}},\ and\
  \bibinfo {author} {\bibfnamefont {C.-C.}\ \bibnamefont {Chien}},\ }\bibfield
  {title} {\bibinfo {title} {Finite-temperature topological phase transitions
  of spin-$j$ systems in uhlmann processes: General formalism and experimental
  protocols},\ }\href {https://doi.org/10.1103/PhysRevA.104.023303} {\bibfield
  {journal} {\bibinfo  {journal} {Phys. Rev. A}\ }\textbf {\bibinfo {volume}
  {104}},\ \bibinfo {pages} {023303} (\bibinfo {year} {2021})}\BibitemShut
  {NoStop}%
\bibitem [{\citenamefont {Heyl}(2015{\natexlab{b}})}]{Heyl.115.140602}%
  \BibitemOpen
  \bibfield  {author} {\bibinfo {author} {\bibfnamefont {M.}~\bibnamefont
  {Heyl}},\ }\bibfield  {title} {\bibinfo {title} {Scaling and universality at
  dynamical quantum phase transitions},\ }\href
  {https://doi.org/10.1103/PhysRevLett.115.140602} {\bibfield  {journal}
  {\bibinfo  {journal} {Phys. Rev. Lett.}\ }\textbf {\bibinfo {volume} {115}},\
  \bibinfo {pages} {140602} (\bibinfo {year} {2015}{\natexlab{b}})}\BibitemShut
  {NoStop}%
\bibitem [{\citenamefont {Vajna}\ and\ \citenamefont
  {D\'ora}(2015)}]{Vajna.91.155127}%
  \BibitemOpen
  \bibfield  {author} {\bibinfo {author} {\bibfnamefont {S.}~\bibnamefont
  {Vajna}}\ and\ \bibinfo {author} {\bibfnamefont {B.}~\bibnamefont {D\'ora}},\
  }\bibfield  {title} {\bibinfo {title} {Topological classification of
  dynamical phase transitions},\ }\href
  {https://doi.org/10.1103/PhysRevB.91.155127} {\bibfield  {journal} {\bibinfo
  {journal} {Phys. Rev. B}\ }\textbf {\bibinfo {volume} {91}},\ \bibinfo
  {pages} {155127} (\bibinfo {year} {2015})}\BibitemShut {NoStop}%
\bibitem [{\citenamefont {Puskarov}\ and\ \citenamefont
  {Schuricht}(2016)}]{Tatjana.1.1.003}%
  \BibitemOpen
  \bibfield  {author} {\bibinfo {author} {\bibfnamefont {T.}~\bibnamefont
  {Puskarov}}\ and\ \bibinfo {author} {\bibfnamefont {D.}~\bibnamefont
  {Schuricht}},\ }\bibfield  {title} {\bibinfo {title} {{Time evolution during
  and after finite-time quantum quenches in the transverse-field Ising
  chain}},\ }\href {https://doi.org/10.21468/SciPostPhys.1.1.003} {\bibfield
  {journal} {\bibinfo  {journal} {SciPost Phys.}\ }\textbf {\bibinfo {volume}
  {1}},\ \bibinfo {pages} {003} (\bibinfo {year} {2016})}\BibitemShut {NoStop}%
\bibitem [{\citenamefont {Lang}\ \emph
  {et~al.}(2018{\natexlab{b}})\citenamefont {Lang}, \citenamefont {Frank},\
  and\ \citenamefont {Halimeh}}]{Lang.121.130603}%
  \BibitemOpen
  \bibfield  {author} {\bibinfo {author} {\bibfnamefont {J.}~\bibnamefont
  {Lang}}, \bibinfo {author} {\bibfnamefont {B.}~\bibnamefont {Frank}},\ and\
  \bibinfo {author} {\bibfnamefont {J.~C.}\ \bibnamefont {Halimeh}},\
  }\bibfield  {title} {\bibinfo {title} {Dynamical quantum phase transitions: A
  geometric picture},\ }\href {https://doi.org/10.1103/PhysRevLett.121.130603}
  {\bibfield  {journal} {\bibinfo  {journal} {Phys. Rev. Lett.}\ }\textbf
  {\bibinfo {volume} {121}},\ \bibinfo {pages} {130603} (\bibinfo {year}
  {2018}{\natexlab{b}})}\BibitemShut {NoStop}%
\bibitem [{\citenamefont {Huang}\ \emph {et~al.}(2019)\citenamefont {Huang},
  \citenamefont {Banerjee},\ and\ \citenamefont {Heyl}}]{Huang.122.250401}%
  \BibitemOpen
  \bibfield  {author} {\bibinfo {author} {\bibfnamefont {Y.-P.}\ \bibnamefont
  {Huang}}, \bibinfo {author} {\bibfnamefont {D.}~\bibnamefont {Banerjee}},\
  and\ \bibinfo {author} {\bibfnamefont {M.}~\bibnamefont {Heyl}},\ }\bibfield
  {title} {\bibinfo {title} {Dynamical quantum phase transitions in u(1)
  quantum link models},\ }\href
  {https://doi.org/10.1103/PhysRevLett.122.250401} {\bibfield  {journal}
  {\bibinfo  {journal} {Phys. Rev. Lett.}\ }\textbf {\bibinfo {volume} {122}},\
  \bibinfo {pages} {250401} (\bibinfo {year} {2019})}\BibitemShut {NoStop}%
\bibitem [{\citenamefont {Jafari}\ \emph {et~al.}(2019)\citenamefont {Jafari},
  \citenamefont {Johannesson}, \citenamefont {Langari},\ and\ \citenamefont
  {Martin-Delgado}}]{Jafari.99.054302}%
  \BibitemOpen
  \bibfield  {author} {\bibinfo {author} {\bibfnamefont {R.}~\bibnamefont
  {Jafari}}, \bibinfo {author} {\bibfnamefont {H.}~\bibnamefont {Johannesson}},
  \bibinfo {author} {\bibfnamefont {A.}~\bibnamefont {Langari}},\ and\ \bibinfo
  {author} {\bibfnamefont {M.~A.}\ \bibnamefont {Martin-Delgado}},\ }\bibfield
  {title} {\bibinfo {title} {Quench dynamics and zero-energy modes: The case of
  the creutz model},\ }\href {https://doi.org/10.1103/PhysRevB.99.054302}
  {\bibfield  {journal} {\bibinfo  {journal} {Phys. Rev. B}\ }\textbf {\bibinfo
  {volume} {99}},\ \bibinfo {pages} {054302} (\bibinfo {year}
  {2019})}\BibitemShut {NoStop}%
\bibitem [{\citenamefont {Khatun}\ and\ \citenamefont
  {Bhattacharjee}(2019)}]{Khatun.123.160603}%
  \BibitemOpen
  \bibfield  {author} {\bibinfo {author} {\bibfnamefont {A.}~\bibnamefont
  {Khatun}}\ and\ \bibinfo {author} {\bibfnamefont {S.~M.}\ \bibnamefont
  {Bhattacharjee}},\ }\bibfield  {title} {\bibinfo {title} {Boundaries and
  unphysical fixed points in dynamical quantum phase transitions},\ }\href
  {https://doi.org/10.1103/PhysRevLett.123.160603} {\bibfield  {journal}
  {\bibinfo  {journal} {Phys. Rev. Lett.}\ }\textbf {\bibinfo {volume} {123}},\
  \bibinfo {pages} {160603} (\bibinfo {year} {2019})}\BibitemShut {NoStop}%
\bibitem [{\citenamefont {Lahiri}\ and\ \citenamefont
  {Bera}(2019)}]{Lahiri.99.174311}%
  \BibitemOpen
  \bibfield  {author} {\bibinfo {author} {\bibfnamefont {A.}~\bibnamefont
  {Lahiri}}\ and\ \bibinfo {author} {\bibfnamefont {S.}~\bibnamefont {Bera}},\
  }\bibfield  {title} {\bibinfo {title} {Dynamical quantum phase transitions in
  weyl semimetals},\ }\href {https://doi.org/10.1103/PhysRevB.99.174311}
  {\bibfield  {journal} {\bibinfo  {journal} {Phys. Rev. B}\ }\textbf {\bibinfo
  {volume} {99}},\ \bibinfo {pages} {174311} (\bibinfo {year}
  {2019})}\BibitemShut {NoStop}%
\bibitem [{\citenamefont {Liu}\ and\ \citenamefont
  {Guo}(2019)}]{Liu.99.104307}%
  \BibitemOpen
  \bibfield  {author} {\bibinfo {author} {\bibfnamefont {T.}~\bibnamefont
  {Liu}}\ and\ \bibinfo {author} {\bibfnamefont {H.}~\bibnamefont {Guo}},\
  }\bibfield  {title} {\bibinfo {title} {Dynamical quantum phase transitions on
  cross-stitch flat band networks},\ }\href
  {https://doi.org/10.1103/PhysRevB.99.104307} {\bibfield  {journal} {\bibinfo
  {journal} {Phys. Rev. B}\ }\textbf {\bibinfo {volume} {99}},\ \bibinfo
  {pages} {104307} (\bibinfo {year} {2019})}\BibitemShut {NoStop}%
\bibitem [{\citenamefont {Srivastav}\ \emph {et~al.}(2019)\citenamefont
  {Srivastav}, \citenamefont {Bhattacharya},\ and\ \citenamefont
  {Dutta}}]{Srivastav.100.144203}%
  \BibitemOpen
  \bibfield  {author} {\bibinfo {author} {\bibfnamefont {V.}~\bibnamefont
  {Srivastav}}, \bibinfo {author} {\bibfnamefont {U.}~\bibnamefont
  {Bhattacharya}},\ and\ \bibinfo {author} {\bibfnamefont {A.}~\bibnamefont
  {Dutta}},\ }\bibfield  {title} {\bibinfo {title} {Dynamical quantum phase
  transitions in extended toric-code models},\ }\href
  {https://doi.org/10.1103/PhysRevB.100.144203} {\bibfield  {journal} {\bibinfo
   {journal} {Phys. Rev. B}\ }\textbf {\bibinfo {volume} {100}},\ \bibinfo
  {pages} {144203} (\bibinfo {year} {2019})}\BibitemShut {NoStop}%
\bibitem [{\citenamefont {Gul\'acsi}\ \emph {et~al.}(2020)\citenamefont
  {Gul\'acsi}, \citenamefont {Heyl},\ and\ \citenamefont
  {D\'ora}}]{Gulacsi.101.205135}%
  \BibitemOpen
  \bibfield  {author} {\bibinfo {author} {\bibfnamefont {B.}~\bibnamefont
  {Gul\'acsi}}, \bibinfo {author} {\bibfnamefont {M.}~\bibnamefont {Heyl}},\
  and\ \bibinfo {author} {\bibfnamefont {B.}~\bibnamefont {D\'ora}},\
  }\bibfield  {title} {\bibinfo {title} {Geometrical quench and dynamical
  quantum phase transition in the $\ensuremath{\alpha}\ensuremath{-}{T}_{3}$
  lattice},\ }\href {https://doi.org/10.1103/PhysRevB.101.205135} {\bibfield
  {journal} {\bibinfo  {journal} {Phys. Rev. B}\ }\textbf {\bibinfo {volume}
  {101}},\ \bibinfo {pages} {205135} (\bibinfo {year} {2020})}\BibitemShut
  {NoStop}%
\bibitem [{\citenamefont {Meibohm}\ and\ \citenamefont
  {Esposito}(2023)}]{Meibohm_2023.023034}%
  \BibitemOpen
  \bibfield  {author} {\bibinfo {author} {\bibfnamefont {J.}~\bibnamefont
  {Meibohm}}\ and\ \bibinfo {author} {\bibfnamefont {M.}~\bibnamefont
  {Esposito}},\ }\bibfield  {title} {\bibinfo {title} {Landau theory for
  finite-time dynamical phase transitions},\ }\href
  {https://doi.org/10.1088/1367-2630/acbc41} {\bibfield  {journal} {\bibinfo
  {journal} {New Journal of Physics}\ }\textbf {\bibinfo {volume} {25}},\
  \bibinfo {pages} {023034} (\bibinfo {year} {2023})}\BibitemShut {NoStop}%
\bibitem [{\citenamefont {Wong}\ and\ \citenamefont
  {Yu}(2022)}]{Wong.105.174307}%
  \BibitemOpen
  \bibfield  {author} {\bibinfo {author} {\bibfnamefont {C.~Y.}\ \bibnamefont
  {Wong}}\ and\ \bibinfo {author} {\bibfnamefont {W.~C.}\ \bibnamefont {Yu}},\
  }\bibfield  {title} {\bibinfo {title} {Loschmidt amplitude spectrum in
  dynamical quantum phase transitions},\ }\href
  {https://doi.org/10.1103/PhysRevB.105.174307} {\bibfield  {journal} {\bibinfo
   {journal} {Phys. Rev. B}\ }\textbf {\bibinfo {volume} {105}},\ \bibinfo
  {pages} {174307} (\bibinfo {year} {2022})}\BibitemShut {NoStop}%
\bibitem [{\citenamefont {Wrze\ifmmode~\acute{s}\else \'{s}\fi{}niewski}\ \emph
  {et~al.}(2022)\citenamefont {Wrze\ifmmode~\acute{s}\else \'{s}\fi{}niewski},
  \citenamefont {Weymann}, \citenamefont {Sedlmayr},\ and\ \citenamefont
  {Doma\ifmmode~\acute{n}\else \'{n}\fi{}ski}}]{Wrzessniewski.105.094514}%
  \BibitemOpen
  \bibfield  {author} {\bibinfo {author} {\bibfnamefont {K.}~\bibnamefont
  {Wrze\ifmmode~\acute{s}\else \'{s}\fi{}niewski}}, \bibinfo {author}
  {\bibfnamefont {I.}~\bibnamefont {Weymann}}, \bibinfo {author} {\bibfnamefont
  {N.}~\bibnamefont {Sedlmayr}},\ and\ \bibinfo {author} {\bibfnamefont
  {T.}~\bibnamefont {Doma\ifmmode~\acute{n}\else \'{n}\fi{}ski}},\ }\bibfield
  {title} {\bibinfo {title} {Dynamical quantum phase transitions in a
  mesoscopic superconducting system},\ }\href
  {https://doi.org/10.1103/PhysRevB.105.094514} {\bibfield  {journal} {\bibinfo
   {journal} {Phys. Rev. B}\ }\textbf {\bibinfo {volume} {105}},\ \bibinfo
  {pages} {094514} (\bibinfo {year} {2022})}\BibitemShut {NoStop}%
\bibitem [{\citenamefont {Hashizume}\ \emph {et~al.}(2022)\citenamefont
  {Hashizume}, \citenamefont {McCulloch},\ and\ \citenamefont
  {Halimeh}}]{Hashizume.4.013250}%
  \BibitemOpen
  \bibfield  {author} {\bibinfo {author} {\bibfnamefont {T.}~\bibnamefont
  {Hashizume}}, \bibinfo {author} {\bibfnamefont {I.~P.}\ \bibnamefont
  {McCulloch}},\ and\ \bibinfo {author} {\bibfnamefont {J.~C.}\ \bibnamefont
  {Halimeh}},\ }\bibfield  {title} {\bibinfo {title} {Dynamical phase
  transitions in the two-dimensional transverse-field ising model},\ }\href
  {https://doi.org/10.1103/PhysRevResearch.4.013250} {\bibfield  {journal}
  {\bibinfo  {journal} {Phys. Rev. Res.}\ }\textbf {\bibinfo {volume} {4}},\
  \bibinfo {pages} {013250} (\bibinfo {year} {2022})}\BibitemShut {NoStop}%
\bibitem [{\citenamefont {Tian}\ \emph {et~al.}(2019)\citenamefont {Tian},
  \citenamefont {Ke}, \citenamefont {Zhang}, \citenamefont {Lin}, \citenamefont
  {Shi}, \citenamefont {Huang}, \citenamefont {Lee},\ and\ \citenamefont
  {Du}}]{Tian19}%
  \BibitemOpen
  \bibfield  {author} {\bibinfo {author} {\bibfnamefont {T.}~\bibnamefont
  {Tian}}, \bibinfo {author} {\bibfnamefont {Y.}~\bibnamefont {Ke}}, \bibinfo
  {author} {\bibfnamefont {L.}~\bibnamefont {Zhang}}, \bibinfo {author}
  {\bibfnamefont {S.}~\bibnamefont {Lin}}, \bibinfo {author} {\bibfnamefont
  {Z.}~\bibnamefont {Shi}}, \bibinfo {author} {\bibfnamefont {P.}~\bibnamefont
  {Huang}}, \bibinfo {author} {\bibfnamefont {C.}~\bibnamefont {Lee}},\ and\
  \bibinfo {author} {\bibfnamefont {J.}~\bibnamefont {Du}},\ }\bibfield
  {title} {\bibinfo {title} {Observation of dynamical phase transitions in a
  topological nanomechanical system},\ }\href@noop {} {\bibfield  {journal}
  {\bibinfo  {journal} {Phys. Rev. B}\ }\textbf {\bibinfo {volume} {100}},\
  \bibinfo {pages} {024310} (\bibinfo {year} {2019})}\BibitemShut {NoStop}%
\bibitem [{\citenamefont {Bernien}\ \emph {et~al.}(2017)\citenamefont
  {Bernien}, \citenamefont {Schwartz}, \citenamefont {Keesling}, \citenamefont
  {Levine}, \citenamefont {Omran}, \citenamefont {Pichler}, \citenamefont
  {Choi}, \citenamefont {Zibrov}, \citenamefont {Endres}, \citenamefont
  {Greiner}, \citenamefont {Vuleti$\acute{\text{ c}}$},\ and\ \citenamefont
  {Lukin}}]{DQPTN17a}%
  \BibitemOpen
  \bibfield  {author} {\bibinfo {author} {\bibfnamefont {H.}~\bibnamefont
  {Bernien}}, \bibinfo {author} {\bibfnamefont {S.}~\bibnamefont {Schwartz}},
  \bibinfo {author} {\bibfnamefont {A.}~\bibnamefont {Keesling}}, \bibinfo
  {author} {\bibfnamefont {H.}~\bibnamefont {Levine}}, \bibinfo {author}
  {\bibfnamefont {A.}~\bibnamefont {Omran}}, \bibinfo {author} {\bibfnamefont
  {H.}~\bibnamefont {Pichler}}, \bibinfo {author} {\bibfnamefont
  {S.}~\bibnamefont {Choi}}, \bibinfo {author} {\bibfnamefont {A.~S.}\
  \bibnamefont {Zibrov}}, \bibinfo {author} {\bibfnamefont {M.}~\bibnamefont
  {Endres}}, \bibinfo {author} {\bibfnamefont {M.}~\bibnamefont {Greiner}},
  \bibinfo {author} {\bibfnamefont {V.}~\bibnamefont {Vuleti$\acute{\text{
  c}}$}},\ and\ \bibinfo {author} {\bibfnamefont {M.~D.}\ \bibnamefont
  {Lukin}},\ }\bibfield  {title} {\bibinfo {title} {Probing many-body dynamics
  on a 51-atom quantum simulator},\ }\href@noop {} {\bibfield  {journal}
  {\bibinfo  {journal} {Nature}\ }\textbf {\bibinfo {volume} {551}},\ \bibinfo
  {pages} {579} (\bibinfo {year} {2017})}\BibitemShut {NoStop}%
\bibitem [{\citenamefont {Zhang}\ \emph
  {et~al.}(2017{\natexlab{b}})\citenamefont {Zhang}, \citenamefont {Pagano},
  \citenamefont {Hess}, \citenamefont {Kyprianidis}, \citenamefont {Becker},
  \citenamefont {Kaplan}, \citenamefont {Gorshkov}, \citenamefont {Gong},\ and\
  \citenamefont {Monroe}}]{DQPTN17b}%
  \BibitemOpen
  \bibfield  {author} {\bibinfo {author} {\bibfnamefont {J.}~\bibnamefont
  {Zhang}}, \bibinfo {author} {\bibfnamefont {G.}~\bibnamefont {Pagano}},
  \bibinfo {author} {\bibfnamefont {P.~W.}\ \bibnamefont {Hess}}, \bibinfo
  {author} {\bibfnamefont {A.}~\bibnamefont {Kyprianidis}}, \bibinfo {author}
  {\bibfnamefont {P.}~\bibnamefont {Becker}}, \bibinfo {author} {\bibfnamefont
  {H.}~\bibnamefont {Kaplan}}, \bibinfo {author} {\bibfnamefont {A.~V.}\
  \bibnamefont {Gorshkov}}, \bibinfo {author} {\bibfnamefont {Z.-X.}\
  \bibnamefont {Gong}},\ and\ \bibinfo {author} {\bibfnamefont
  {C.}~\bibnamefont {Monroe}},\ }\bibfield  {title} {\bibinfo {title}
  {Observation of a many-body dynamical phase transition with a 53-qubit
  quantum simulator},\ }\href@noop {} {\bibfield  {journal} {\bibinfo
  {journal} {Nature}\ }\textbf {\bibinfo {volume} {551}},\ \bibinfo {pages}
  {601} (\bibinfo {year} {2017}{\natexlab{b}})}\BibitemShut {NoStop}%
\bibitem [{\citenamefont {Guo}\ \emph {et~al.}(2019{\natexlab{b}})\citenamefont
  {Guo}, \citenamefont {Yang}, \citenamefont {Zeng}, \citenamefont {Peng},
  \citenamefont {Li}, \citenamefont {Deng}, \citenamefont {Jin}, \citenamefont
  {Chen}, \citenamefont {Zheng},\ and\ \citenamefont {Fan}}]{GuoApplied19}%
  \BibitemOpen
  \bibfield  {author} {\bibinfo {author} {\bibfnamefont {X.~Y.}\ \bibnamefont
  {Guo}}, \bibinfo {author} {\bibfnamefont {C.}~\bibnamefont {Yang}}, \bibinfo
  {author} {\bibfnamefont {Y.}~\bibnamefont {Zeng}}, \bibinfo {author}
  {\bibfnamefont {Y.}~\bibnamefont {Peng}}, \bibinfo {author} {\bibfnamefont
  {H.~K.}\ \bibnamefont {Li}}, \bibinfo {author} {\bibfnamefont
  {H.}~\bibnamefont {Deng}}, \bibinfo {author} {\bibfnamefont {Y.~R.}\
  \bibnamefont {Jin}}, \bibinfo {author} {\bibfnamefont {S.}~\bibnamefont
  {Chen}}, \bibinfo {author} {\bibfnamefont {D.}~\bibnamefont {Zheng}},\ and\
  \bibinfo {author} {\bibfnamefont {H.}~\bibnamefont {Fan}},\ }\bibfield
  {title} {\bibinfo {title} {Observation of a dynamical quantum phase
  transition by a superconducting qubit simulation},\ }\href@noop {} {\bibfield
   {journal} {\bibinfo  {journal} {Phys. Rev. Applied}\ }\textbf {\bibinfo
  {volume} {11}},\ \bibinfo {pages} {044080} (\bibinfo {year}
  {2019}{\natexlab{b}})}\BibitemShut {NoStop}%
\bibitem [{\citenamefont {Wang}\ \emph {et~al.}(2019)\citenamefont {Wang},
  \citenamefont {Qiu}, \citenamefont {Xiao}, \citenamefont {Zhan},
  \citenamefont {Bian}, \citenamefont {Yi},\ and\ \citenamefont
  {Xue}}]{WangPRL19}%
  \BibitemOpen
  \bibfield  {author} {\bibinfo {author} {\bibfnamefont {K.}~\bibnamefont
  {Wang}}, \bibinfo {author} {\bibfnamefont {X.}~\bibnamefont {Qiu}}, \bibinfo
  {author} {\bibfnamefont {L.}~\bibnamefont {Xiao}}, \bibinfo {author}
  {\bibfnamefont {X.}~\bibnamefont {Zhan}}, \bibinfo {author} {\bibfnamefont
  {Z.}~\bibnamefont {Bian}}, \bibinfo {author} {\bibfnamefont {W.}~\bibnamefont
  {Yi}},\ and\ \bibinfo {author} {\bibfnamefont {P.}~\bibnamefont {Xue}},\
  }\bibfield  {title} {\bibinfo {title} {Simulating dynamic quantum phase
  transitions in photonic quantum walks},\ }\href@noop {} {\bibfield  {journal}
  {\bibinfo  {journal} {Phys. Rev. Lett.}\ }\textbf {\bibinfo {volume} {122}},\
  \bibinfo {pages} {020501} (\bibinfo {year} {2019})}\BibitemShut {NoStop}%
\bibitem [{\citenamefont {Chen}\ \emph {et~al.}(2020)\citenamefont {Chen},
  \citenamefont {Hou}, \citenamefont {Zhou}, \citenamefont {Qian},
  \citenamefont {Shen},\ and\ \citenamefont {Xu}}]{ChenAPL20}%
  \BibitemOpen
  \bibfield  {author} {\bibinfo {author} {\bibfnamefont {B.}~\bibnamefont
  {Chen}}, \bibinfo {author} {\bibfnamefont {X.}~\bibnamefont {Hou}}, \bibinfo
  {author} {\bibfnamefont {F.}~\bibnamefont {Zhou}}, \bibinfo {author}
  {\bibfnamefont {P.}~\bibnamefont {Qian}}, \bibinfo {author} {\bibfnamefont
  {H.}~\bibnamefont {Shen}},\ and\ \bibinfo {author} {\bibfnamefont
  {N.}~\bibnamefont {Xu}},\ }\bibfield  {title} {\bibinfo {title} {Detecting
  the out-of-time-order correlations of dynamical quantum phase transitions in
  a solid-state quantum simulator},\ }\href@noop {} {\bibfield  {journal}
  {\bibinfo  {journal} {Appl. Phys. Lett.}\ }\textbf {\bibinfo {volume}
  {116}},\ \bibinfo {pages} {194002} (\bibinfo {year} {2020})}\BibitemShut
  {NoStop}%
\bibitem [{\citenamefont {Yang}\ \emph
  {et~al.}(2019{\natexlab{b}})\citenamefont {Yang}, \citenamefont {Tian},
  \citenamefont {Yang}, \citenamefont {Qiu}, \citenamefont {Liang},
  \citenamefont {Chu}, \citenamefont {Dag}, \citenamefont {Xu}, \citenamefont
  {Liu},\ and\ \citenamefont {Duan}}]{Yang19}%
  \BibitemOpen
  \bibfield  {author} {\bibinfo {author} {\bibfnamefont {H.~X.}\ \bibnamefont
  {Yang}}, \bibinfo {author} {\bibfnamefont {T.}~\bibnamefont {Tian}}, \bibinfo
  {author} {\bibfnamefont {Y.~B.}\ \bibnamefont {Yang}}, \bibinfo {author}
  {\bibfnamefont {L.~Y.}\ \bibnamefont {Qiu}}, \bibinfo {author} {\bibfnamefont
  {H.~Y.}\ \bibnamefont {Liang}}, \bibinfo {author} {\bibfnamefont {A.~J.}\
  \bibnamefont {Chu}}, \bibinfo {author} {\bibfnamefont {C.~B.}\ \bibnamefont
  {Dag}}, \bibinfo {author} {\bibfnamefont {Y.}~\bibnamefont {Xu}}, \bibinfo
  {author} {\bibfnamefont {Y.}~\bibnamefont {Liu}},\ and\ \bibinfo {author}
  {\bibfnamefont {L.~M.}\ \bibnamefont {Duan}},\ }\bibfield  {title} {\bibinfo
  {title} {Observation of dynamical quantum phase transitions in a spinor
  condensate},\ }\href@noop {} {\bibfield  {journal} {\bibinfo  {journal}
  {Phys. Rev. A}\ }\textbf {\bibinfo {volume} {100}},\ \bibinfo {pages}
  {013622} (\bibinfo {year} {2019}{\natexlab{b}})}\BibitemShut {NoStop}%
\bibitem [{\citenamefont {Tian}\ \emph {et~al.}(2020)\citenamefont {Tian},
  \citenamefont {Yang}, \citenamefont {Qiu}, \citenamefont {Liang},
  \citenamefont {Yang}, \citenamefont {Xu},\ and\ \citenamefont
  {Duan}}]{Duan20}%
  \BibitemOpen
  \bibfield  {author} {\bibinfo {author} {\bibfnamefont {T.}~\bibnamefont
  {Tian}}, \bibinfo {author} {\bibfnamefont {H.-X.}\ \bibnamefont {Yang}},
  \bibinfo {author} {\bibfnamefont {L.-Y.}\ \bibnamefont {Qiu}}, \bibinfo
  {author} {\bibfnamefont {H.-Y.}\ \bibnamefont {Liang}}, \bibinfo {author}
  {\bibfnamefont {Y.-B.}\ \bibnamefont {Yang}}, \bibinfo {author}
  {\bibfnamefont {Y.}~\bibnamefont {Xu}},\ and\ \bibinfo {author}
  {\bibfnamefont {L.-M.}\ \bibnamefont {Duan}},\ }\bibfield  {title} {\bibinfo
  {title} {Observation of dynamical quantum phase transitions with
  correspondence in an excited state phase diagram},\ }\href@noop {} {\bibfield
   {journal} {\bibinfo  {journal} {Phys. Rev. Lett.}\ }\textbf {\bibinfo
  {volume} {124}},\ \bibinfo {pages} {043001} (\bibinfo {year}
  {2020})}\BibitemShut {NoStop}%
\bibitem [{\citenamefont {Nie}\ \emph {et~al.}(2020)\citenamefont {Nie},
  \citenamefont {Wei}, \citenamefont {Chen}, \citenamefont {Zhang},
  \citenamefont {Zhao}, \citenamefont {Qiu}, \citenamefont {Tian},
  \citenamefont {Ji}, \citenamefont {Xin}, \citenamefont {Lu},\ and\
  \citenamefont {Li}}]{Nie20}%
  \BibitemOpen
  \bibfield  {author} {\bibinfo {author} {\bibfnamefont {X.}~\bibnamefont
  {Nie}}, \bibinfo {author} {\bibfnamefont {B.~B.}\ \bibnamefont {Wei}},
  \bibinfo {author} {\bibfnamefont {X.}~\bibnamefont {Chen}}, \bibinfo {author}
  {\bibfnamefont {Z.}~\bibnamefont {Zhang}}, \bibinfo {author} {\bibfnamefont
  {X.}~\bibnamefont {Zhao}}, \bibinfo {author} {\bibfnamefont {C.}~\bibnamefont
  {Qiu}}, \bibinfo {author} {\bibfnamefont {Y.}~\bibnamefont {Tian}}, \bibinfo
  {author} {\bibfnamefont {Y.}~\bibnamefont {Ji}}, \bibinfo {author}
  {\bibfnamefont {T.}~\bibnamefont {Xin}}, \bibinfo {author} {\bibfnamefont
  {D.}~\bibnamefont {Lu}},\ and\ \bibinfo {author} {\bibfnamefont
  {J.}~\bibnamefont {Li}},\ }\bibfield  {title} {\bibinfo {title} {Experimental
  observation of equilibrium and dynamical quantum phase transitions via
  out-of-time-ordered correlators},\ }\href@noop {} {\bibfield  {journal}
  {\bibinfo  {journal} {Phys. Rev. Lett.}\ }\textbf {\bibinfo {volume} {124}},\
  \bibinfo {pages} {250601} (\bibinfo {year} {2020})}\BibitemShut {NoStop}%
\bibitem [{\citenamefont {Budich}\ and\ \citenamefont
  {Heyl}(2016)}]{Budich201693}%
  \BibitemOpen
  \bibfield  {author} {\bibinfo {author} {\bibfnamefont {J.~C.}\ \bibnamefont
  {Budich}}\ and\ \bibinfo {author} {\bibfnamefont {M.}~\bibnamefont {Heyl}},\
  }\bibfield  {title} {\bibinfo {title} {Dynamical topological order parameters
  far from equilibrium},\ }\href {https://doi.org/10.1103/PhysRevB.93.085416}
  {\bibfield  {journal} {\bibinfo  {journal} {Phys. Rev. B}\ }\textbf {\bibinfo
  {volume} {93}},\ \bibinfo {pages} {085416} (\bibinfo {year}
  {2016})}\BibitemShut {NoStop}%
\bibitem [{\citenamefont {Pancharatnam}(1956)}]{Pancharatnam56}%
  \BibitemOpen
  \bibfield  {author} {\bibinfo {author} {\bibfnamefont {S.}~\bibnamefont
  {Pancharatnam}},\ }\bibfield  {title} {\bibinfo {title} {Generalized theory
  of interference, and its applications},\ }\href@noop {} {\bibfield  {journal}
  {\bibinfo  {journal} {Proc. Indian. Acad. Sci. A}\ }\textbf {\bibinfo
  {volume} {44}},\ \bibinfo {pages} {247} (\bibinfo {year} {1956})}\BibitemShut
  {NoStop}%
\bibitem [{\citenamefont {Sj\"oqvist}\ \emph {et~al.}(2000)\citenamefont
  {Sj\"oqvist}, \citenamefont {Pati}, \citenamefont {Ekert}, \citenamefont
  {Anandan}, \citenamefont {Ericsson}, \citenamefont {Oi},\ and\ \citenamefont
  {Vedral}}]{PhysRevLett.85.2845}%
  \BibitemOpen
  \bibfield  {author} {\bibinfo {author} {\bibfnamefont {E.}~\bibnamefont
  {Sj\"oqvist}}, \bibinfo {author} {\bibfnamefont {A.~K.}\ \bibnamefont
  {Pati}}, \bibinfo {author} {\bibfnamefont {A.}~\bibnamefont {Ekert}},
  \bibinfo {author} {\bibfnamefont {J.~S.}\ \bibnamefont {Anandan}}, \bibinfo
  {author} {\bibfnamefont {M.}~\bibnamefont {Ericsson}}, \bibinfo {author}
  {\bibfnamefont {D.~K.~L.}\ \bibnamefont {Oi}},\ and\ \bibinfo {author}
  {\bibfnamefont {V.}~\bibnamefont {Vedral}},\ }\bibfield  {title} {\bibinfo
  {title} {Geometric phases for mixed states in interferometry},\ }\href
  {https://doi.org/10.1103/PhysRevLett.85.2845} {\bibfield  {journal} {\bibinfo
   {journal} {Phys. Rev. Lett.}\ }\textbf {\bibinfo {volume} {85}},\ \bibinfo
  {pages} {2845} (\bibinfo {year} {2000})}\BibitemShut {NoStop}%
\bibitem [{\citenamefont {Uhlmann}(1986)}]{Uhlmann86}%
  \BibitemOpen
  \bibfield  {author} {\bibinfo {author} {\bibfnamefont {A.}~\bibnamefont
  {Uhlmann}},\ }\bibfield  {title} {\bibinfo {title} {Parallel transport and
  "quantum holonomy" along density operators},\ }\href@noop {} {\bibfield
  {journal} {\bibinfo  {journal} {Rep. Math. Phys.}\ }\textbf {\bibinfo
  {volume} {24}},\ \bibinfo {pages} {229} (\bibinfo {year} {1986})}\BibitemShut
  {NoStop}%
\bibitem [{\citenamefont {Tang}\ \emph {et~al.}(2024)\citenamefont {Tang},
  \citenamefont {Hou}, \citenamefont {Zhou}, \citenamefont {Guo},\ and\
  \citenamefont {Chien}}]{PhysRevB.110.134319}%
  \BibitemOpen
  \bibfield  {author} {\bibinfo {author} {\bibfnamefont {J.-C.}\ \bibnamefont
  {Tang}}, \bibinfo {author} {\bibfnamefont {X.-Y.}\ \bibnamefont {Hou}},
  \bibinfo {author} {\bibfnamefont {Z.}~\bibnamefont {Zhou}}, \bibinfo {author}
  {\bibfnamefont {H.}~\bibnamefont {Guo}},\ and\ \bibinfo {author}
  {\bibfnamefont {C.-C.}\ \bibnamefont {Chien}},\ }\bibfield  {title} {\bibinfo
  {title} {Uhlmann quench and geometric dynamic quantum phase transition of
  mixed states},\ }\href {https://doi.org/10.1103/PhysRevB.110.134319}
  {\bibfield  {journal} {\bibinfo  {journal} {Phys. Rev. B}\ }\textbf {\bibinfo
  {volume} {110}},\ \bibinfo {pages} {134319} (\bibinfo {year}
  {2024})}\BibitemShut {NoStop}%
\bibitem [{\citenamefont {Hou}\ \emph {et~al.}(2022)\citenamefont {Hou},
  \citenamefont {Gao}, \citenamefont {Guo},\ and\ \citenamefont
  {Chien}}]{PhysRevB.106.014301}%
  \BibitemOpen
  \bibfield  {author} {\bibinfo {author} {\bibfnamefont {X.-Y.}\ \bibnamefont
  {Hou}}, \bibinfo {author} {\bibfnamefont {Q.-C.}\ \bibnamefont {Gao}},
  \bibinfo {author} {\bibfnamefont {H.}~\bibnamefont {Guo}},\ and\ \bibinfo
  {author} {\bibfnamefont {C.-C.}\ \bibnamefont {Chien}},\ }\bibfield  {title}
  {\bibinfo {title} {Metamorphic dynamical quantum phase transition in
  double-quench processes at finite temperatures},\ }\href
  {https://doi.org/10.1103/PhysRevB.106.014301} {\bibfield  {journal} {\bibinfo
   {journal} {Phys. Rev. B}\ }\textbf {\bibinfo {volume} {106}},\ \bibinfo
  {pages} {014301} (\bibinfo {year} {2022})}\BibitemShut {NoStop}%
\bibitem [{\citenamefont {Heyl}\ and\ \citenamefont
  {Budich}(2017{\natexlab{b}})}]{PhysRevB.96.180304}%
  \BibitemOpen
  \bibfield  {author} {\bibinfo {author} {\bibfnamefont {M.}~\bibnamefont
  {Heyl}}\ and\ \bibinfo {author} {\bibfnamefont {J.~C.}\ \bibnamefont
  {Budich}},\ }\bibfield  {title} {\bibinfo {title} {Dynamical topological
  quantum phase transitions for mixed states},\ }\href
  {https://doi.org/10.1103/PhysRevB.96.180304} {\bibfield  {journal} {\bibinfo
  {journal} {Phys. Rev. B}\ }\textbf {\bibinfo {volume} {96}},\ \bibinfo
  {pages} {180304} (\bibinfo {year} {2017}{\natexlab{b}})}\BibitemShut
  {NoStop}%
\bibitem [{\citenamefont {Mukunda}\ and\ \citenamefont
  {Simon}(1993)}]{MUKUNDA1993205}%
  \BibitemOpen
  \bibfield  {author} {\bibinfo {author} {\bibfnamefont {N.}~\bibnamefont
  {Mukunda}}\ and\ \bibinfo {author} {\bibfnamefont {R.}~\bibnamefont
  {Simon}},\ }\bibfield  {title} {\bibinfo {title} {Quantum kinematic approach
  to the geometric phase. i. general formalism},\ }\href
  {https://doi.org/https://doi.org/10.1006/aphy.1993.1093} {\bibfield
  {journal} {\bibinfo  {journal} {Ann. Phys.}\ }\textbf {\bibinfo {volume}
  {228}},\ \bibinfo {pages} {205} (\bibinfo {year} {1993})}\BibitemShut
  {NoStop}%
\bibitem [{\citenamefont {Andersson}\ \emph {et~al.}(2016)\citenamefont
  {Andersson}, \citenamefont {Bengtsson}, \citenamefont {Ericsson},\ and\
  \citenamefont {Sj{\"o}qvist}}]{andersson2016geometric}%
  \BibitemOpen
  \bibfield  {author} {\bibinfo {author} {\bibfnamefont {O.}~\bibnamefont
  {Andersson}}, \bibinfo {author} {\bibfnamefont {I.}~\bibnamefont
  {Bengtsson}}, \bibinfo {author} {\bibfnamefont {M.}~\bibnamefont
  {Ericsson}},\ and\ \bibinfo {author} {\bibfnamefont {E.}~\bibnamefont
  {Sj{\"o}qvist}},\ }\bibfield  {title} {\bibinfo {title} {Geometric phases for
  mixed states of the kitaev chain},\ }\href@noop {} {\bibfield  {journal}
  {\bibinfo  {journal} {Philosophical Transactions of the Royal Society A:
  Mathematical, Physical and Engineering Sciences}\ }\textbf {\bibinfo {volume}
  {374}},\ \bibinfo {pages} {20150231} (\bibinfo {year} {2016})}\BibitemShut
  {NoStop}%
\bibitem [{\citenamefont {Jeevanjee}(2015)}]{Jeevanjeebook}%
  \BibitemOpen
  \bibfield  {author} {\bibinfo {author} {\bibfnamefont {N.}~\bibnamefont
  {Jeevanjee}},\ }\href@noop {} {\emph {\bibinfo {title} {An Introduction to
  Tensors and Group Theory for Physicists, 2nd ed.}}}\ (\bibinfo  {publisher}
  {Birkh\"auser, Basel},\ \bibinfo {year} {2015})\BibitemShut {NoStop}%
\bibitem [{\citenamefont {Foot}(2005)}]{FootBook}%
  \BibitemOpen
  \bibfield  {author} {\bibinfo {author} {\bibfnamefont {C.~J.}\ \bibnamefont
  {Foot}},\ }\href@noop {} {\emph {\bibinfo {title} {Atomic physics}}}\
  (\bibinfo  {publisher} {Oxford University Press},\ \bibinfo {address}
  {Oxford, UK},\ \bibinfo {year} {2005})\BibitemShut {NoStop}%
\bibitem [{\citenamefont {Pethick}\ and\ \citenamefont
  {Smith}(2008)}]{PethickBook}%
  \BibitemOpen
  \bibfield  {author} {\bibinfo {author} {\bibfnamefont {C.~J.}\ \bibnamefont
  {Pethick}}\ and\ \bibinfo {author} {\bibfnamefont {H.}~\bibnamefont
  {Smith}},\ }\href@noop {} {\emph {\bibinfo {title} {Bose-Einstein
  condensation in dilute gases}}},\ \bibinfo {edition} {2nd}\ ed.\ (\bibinfo
  {publisher} {Cambridge University Press},\ \bibinfo {address} {Cambridge,
  UK},\ \bibinfo {year} {2008})\BibitemShut {NoStop}%
\bibitem [{\citenamefont {Du}\ \emph {et~al.}(2003)\citenamefont {Du},
  \citenamefont {Zou}, \citenamefont {Shi}, \citenamefont {Kwek}, \citenamefont
  {Pan}, \citenamefont {Oh}, \citenamefont {Ekert}, \citenamefont {Oi},\ and\
  \citenamefont {Ericsson}}]{PhysRevLett.91.100403}%
  \BibitemOpen
  \bibfield  {author} {\bibinfo {author} {\bibfnamefont {J.}~\bibnamefont
  {Du}}, \bibinfo {author} {\bibfnamefont {P.}~\bibnamefont {Zou}}, \bibinfo
  {author} {\bibfnamefont {M.}~\bibnamefont {Shi}}, \bibinfo {author}
  {\bibfnamefont {L.~C.}\ \bibnamefont {Kwek}}, \bibinfo {author}
  {\bibfnamefont {J.-W.}\ \bibnamefont {Pan}}, \bibinfo {author} {\bibfnamefont
  {C.~H.}\ \bibnamefont {Oh}}, \bibinfo {author} {\bibfnamefont
  {A.}~\bibnamefont {Ekert}}, \bibinfo {author} {\bibfnamefont {D.~K.~L.}\
  \bibnamefont {Oi}},\ and\ \bibinfo {author} {\bibfnamefont {M.}~\bibnamefont
  {Ericsson}},\ }\bibfield  {title} {\bibinfo {title} {Observation of geometric
  phases for mixed states using nmr interferometry},\ }\href
  {https://doi.org/10.1103/PhysRevLett.91.100403} {\bibfield  {journal}
  {\bibinfo  {journal} {Phys. Rev. Lett.}\ }\textbf {\bibinfo {volume} {91}},\
  \bibinfo {pages} {100403} (\bibinfo {year} {2003})}\BibitemShut {NoStop}%
\bibitem [{\citenamefont {Ericsson}\ \emph {et~al.}(2005)\citenamefont
  {Ericsson}, \citenamefont {Achilles}, \citenamefont {Barreiro}, \citenamefont
  {Branning}, \citenamefont {Peters},\ and\ \citenamefont
  {Kwiat}}]{PhysRevLett.94.050401}%
  \BibitemOpen
  \bibfield  {author} {\bibinfo {author} {\bibfnamefont {M.}~\bibnamefont
  {Ericsson}}, \bibinfo {author} {\bibfnamefont {D.}~\bibnamefont {Achilles}},
  \bibinfo {author} {\bibfnamefont {J.~T.}\ \bibnamefont {Barreiro}}, \bibinfo
  {author} {\bibfnamefont {D.}~\bibnamefont {Branning}}, \bibinfo {author}
  {\bibfnamefont {N.~A.}\ \bibnamefont {Peters}},\ and\ \bibinfo {author}
  {\bibfnamefont {P.~G.}\ \bibnamefont {Kwiat}},\ }\bibfield  {title} {\bibinfo
  {title} {Measurement of geometric phase for mixed states using single photon
  interferometry},\ }\href {https://doi.org/10.1103/PhysRevLett.94.050401}
  {\bibfield  {journal} {\bibinfo  {journal} {Phys. Rev. Lett.}\ }\textbf
  {\bibinfo {volume} {94}},\ \bibinfo {pages} {050401} (\bibinfo {year}
  {2005})}\BibitemShut {NoStop}%
\bibitem [{\citenamefont {Klepp}\ \emph {et~al.}(2008)\citenamefont {Klepp},
  \citenamefont {Sponar}, \citenamefont {Filipp}, \citenamefont {Lettner},
  \citenamefont {Badurek},\ and\ \citenamefont
  {Hasegawa}}]{PhysRevLett.101.150404}%
  \BibitemOpen
  \bibfield  {author} {\bibinfo {author} {\bibfnamefont {J.}~\bibnamefont
  {Klepp}}, \bibinfo {author} {\bibfnamefont {S.}~\bibnamefont {Sponar}},
  \bibinfo {author} {\bibfnamefont {S.}~\bibnamefont {Filipp}}, \bibinfo
  {author} {\bibfnamefont {M.}~\bibnamefont {Lettner}}, \bibinfo {author}
  {\bibfnamefont {G.}~\bibnamefont {Badurek}},\ and\ \bibinfo {author}
  {\bibfnamefont {Y.}~\bibnamefont {Hasegawa}},\ }\bibfield  {title} {\bibinfo
  {title} {Observation of nonadditive mixed-state phases with polarized
  neutrons},\ }\href {https://doi.org/10.1103/PhysRevLett.101.150404}
  {\bibfield  {journal} {\bibinfo  {journal} {Phys. Rev. Lett.}\ }\textbf
  {\bibinfo {volume} {101}},\ \bibinfo {pages} {150404} (\bibinfo {year}
  {2008})}\BibitemShut {NoStop}%
\bibitem [{\citenamefont {Ghosh}\ and\ \citenamefont
  {Kumar}(2006)}]{GHOSH200627}%
  \BibitemOpen
  \bibfield  {author} {\bibinfo {author} {\bibfnamefont {A.}~\bibnamefont
  {Ghosh}}\ and\ \bibinfo {author} {\bibfnamefont {A.}~\bibnamefont {Kumar}},\
  }\bibfield  {title} {\bibinfo {title} {Experimental measurement of mixed
  state geometric phase by quantum interferometry using nmr},\ }\href
  {https://doi.org/https://doi.org/10.1016/j.physleta.2005.08.092} {\bibfield
  {journal} {\bibinfo  {journal} {Physics Letters A}\ }\textbf {\bibinfo
  {volume} {349}},\ \bibinfo {pages} {27} (\bibinfo {year} {2006})}\BibitemShut
  {NoStop}%
\end{thebibliography}%

\end{document}